\documentclass[times,twocolumn]{aastex631} 

\usepackage[figure,figure*]{hypcap}

\usepackage{natbib}

\usepackage{amsmath}

\usepackage{graphics,graphicx}
\usepackage{epsfig}

\newcommand\subs[1]{\textsubscript{#1}}

\newcommand{\chemone}{\raisebox{0.03cm}{$-$}} 

\newcommand{\ltsimeq}{\raisebox{-0.6ex}{$\,\stackrel
	{\raisebox{-.2ex}{$\textstyle <$}}{\sim}\,$}}
\newcommand{\gtsimeq}{\raisebox{-0.6ex}{$\,\stackrel
	{\raisebox{-.2ex}{$\textstyle >$}}{\sim}\,$}}

\usepackage{color,soul}

\shorttitle{JWST and Comet C/2017 K2 PS}
\shortauthors{Woodward, C.E. et al.}

\begin{document}
\title{A JWST Study of the Remarkable Oort Cloud Comet C/2017 K2 (PanSTARRS)} 

\author[0000-0001-6567-627X]{Charles E.\ Woodward}
\affiliation{Minnesota Institute for Astrophysics, School of
Physics and Astronomy, 116 Church Street, S.E., University of
Minnesota, Minneapolis, MN 55455, USA}

\author[0000-0001-6567-627X]{Dominique Bock\'{e}lee-Morvan}
\affiliation{Observatoire de Paris, France}

\author[0000-0001-6397-9082]{David E.\ Harker}
\affiliation{Department of Astronomy and Astrophysics
University of California, San Diego
9500 Gilman Drive, MC 0424
La Jolla, CA 92093-0424 USA}

\author[0000-0002-6702-7676]{Michael S.~P.\ Kelley}
\affiliation{University of Maryland, Department of Astronomy,
4254 Stadium Dr, College Park, MD 20742-2421, USA}

\author[0000-0002-6006-9574]{Nathan X.\ Roth}
\affiliation{Solar System Exploration Division, Astrochemistry Laboratory Code 691, NASA 
Goddard Space Flight Center, 8800 Greenbelt Rd, Greenbelt, MD 20771, USA}
\affiliation{Department of Physics, American University, 4400 Massachusetts Ave NW, Washington, DC 20016, USA}

\author[0000-0002-1888-7258]{Diane H.\ Wooden}
\affiliation{NASA Ames Research Center, Space
Science Division, MS~245-1, Moffett Field, CA 94035-1000, USA}

\author[0000-0001-6567-627X]{Stefanie N.\ Milam}
\affiliation{NASA Goddard Space Flight Center, USA}

\correspondingauthor{C.E. Woodward}
\email{mailto:chickw024@gmail.com}
\received{2025 Feb 13}
\revised{2025 Apr 18}
\accepted{2025 Apr 26}
\submitjournal{Planetary Science Journal}

\begin{abstract}
Comets, during their journeys into the inner solar system, deliver volatile gases, organics, 
and particulates into their comae that provide crucial information for assessing the 
physico-chemical conditions in the outer disk from which they formed. Here we present 
observational and modeling results of a JWST NIRSpec and MIRI MRS integral-field-unit 
spatial-spectral study of the inner coma of the Oort Cloud comet C/2017 K2 (PanSTARRS) at a 
heliocentric distance of 2.35~au. We find the comet is hyperactive (water ice active fraction $\gtsimeq 86$\%),
with a nucleus radius of $<$4.2~km, exhibiting strong emission from H$_{2}$O, $^{12}$CO, $^{13}$CO, and
CO$_{2}$ as well as CN, H$_2$CO, CH$_3$OH, CH$_4$, C$_2$H$_6$, HCN, NH$_2$, and OH prompt emission. 
The water ortho-to-para ratio is $\gtsimeq 2.75.$ The modeled dust composition (relative 
mass fraction of the sub-micron grains) in the coma is dominated by amorphous carbon ($\simeq 25$\%), 
amorphous Mg:Fe olivine ($\simeq 19$\%), amorphous Mg:Fe pyroxene ($\simeq 16$\%), and 
Mg-rich crystalline olivine ($\simeq 39$\%) and the crystalline mass fraction of the sub-micron grains 
in the coma is, $f_{cryst} \simeq 0.384 \pm 0.065$.  Analysis of residuals in 3 to 8~\micron{} region of 
the spectral energy distribution strongly suggests the presence of polycyclic aromatic hydrocarbon (PAHs) 
species in the coma. 

\end{abstract}

\keywords{Long period comets (933),  Comet origins (2203), 
Dust composition (2271), Infrared spectroscopy (2285)}

\section{Introduction}
\label{sec-intro}

Comets are fossils of the solar system. They formed in the cold disk mid-plane, preserving a mixture 
of ices from the cold outer disk, while also capturing refractory dust particles including 
crystalline silicates, calcium-aluminum-rich inclusions (CAIs), and in some cases micro-chondrules 
from the hotter inner disk. Prior to being perturbed into the inner solar system, comet nuclei 
have been cryogenically preserved in the outer solar system at temperatures below $\sim$30~K
\citep{2019ARA&A..57..113A, 2020SSRv..216..102R} for the past 4.5~Gyr, hence comets 
contain materials from the earliest stage of solar system formation. As comets journey into 
the inner solar system, they deliver volatile gases, organics, and particulates into their comae that 
provide crucial information for assessing the physico-chemical conditions in the outer disk from 
which they formed \cite[e.g.,][]{2021PSJ.....2...25W}. 

Here we report results of our JWST \citep{2023PASP..135d8001R, 2023PASP..135f8001G}
spectroscopic study of comet C/2017 K2 (PanSTARRS) near a heliocentric distance, $r_{h}$, of 2.35~au, 
pre-perihelion. 

Comet C/2017 K2 (PanSTARRS) was selected as a JWST Cycle 1 target due to its expected IR brightness,
a figure of Merit \citep[FoM,][]{2017ApJ...849L...8M} conducive to the detection of
strong molecular emission lines (estimated FoM \gtsimeq 0.1), its modest non-sidereal 
track rates that were $\ltsimeq 15$~mas~s$^{-1}$, and its observability within the JWST field-of-regard. 
C/2017 K2 (PanSTARRS) , discovered by the PanSTARRS1 
(PS1) survey \citep{2017CBET.4393....1W, 2017ApJ...849L...8M}, is a comet of interest
given its activity at large $r_{h} \simeq 23.7$~au \citep{2017ApJ...847L..19J} attributed to the sublimation of
super-volatile ices. The dust properties of material in the coma was studied by \citet{2022PSJ.....3..135Z} with HST
polarimetric observations (r$_{h} \simeq 6$~au in-bound) who concluded that the coma was in part populated
with materials released by sublimation of large (micron-sized), carbonaceous-laden water ice grains.
Post-perihelion gas productions rates  \citep[measured in the narrowband ``Hale-Bopp'' filters,][]{2000Icar..147..180F} 
near r$_{h} = 2.2$~au were of the order $\simeq 4 \times 10^{28}$~molecules~s$^{-1}$ for OH, 
1.4 $\times 10^{26}$~molecules~s$^{-1}$ for CN and C$_2$, and $\simeq 3.5 \times 10^{25}$~molecules~s$^{-1}$ 
for C$_{3}$, with an Af$\rho$ \citep[a dust production rate proxy,][]{1995Icar..118..223A} 
of $\simeq 6300$~cm in blue filter \citep{2023ATel15973....1J}.  \citet{2023A&A...674A.206K} find an  upper limit 
on H$_{2}$O) production rate of $\simeq 7 \times 10^{28}$~molecules~s$^{-1}$ near 2.53~au pre-perihelion.
Recently, \citep{2025AJ....169..102E}  report that all volatiles observed with high resolution ground-base
observations are enriched relative to H$_{2}$O compared to the mean of other Oort Cloud comets (OCCs).

\citet{2018A&A...615A.170K} suggest that C/2017 K2 (PanSTARRS) is a dynamically ``old''  OCC
with at least one prior perihelion passage limited to the outer 
solar system, perihelion distance $\simeq 5$~au,  based on dynamical modeling. 
Its latest perihelion passage at $r_{h} = 1.79$~au was on 2022 December 19 UT
interior to the H$_{2}$O sublimation distance which drives vigorous nucleus sublimation producing a
significant extended coma.

JWST is a space-based 6.5m telescope with a variety of instrumental capabilities that are described
in detail by \citet{2023PASP..135f8001G}, \citet{2023PASP..135d8001R}, and \citet{2023PASP..135e8001M}.
Its unprecedented sensitivity, infrared (IR) wavelength coverage, and aperture size combined
with its spectrographs opens new discovery space for the study of comets.  In this manuscript, 
we discuss the analysis of JWST spectra highlighting
the detection of numerous molecular and gas phase emission features and their 
spatial distribution in the inner coma, as well as the analysis of coma dust constituents derived from 
thermal model analysis of the IR spectral energy distribution (SED).  

In Section~\ref{sec-obspipe} we describe our JWST reduction of the raw MAST data sets, 
and Section~\ref{sec-nucleus} is a discussion of the analysis and results of using the high-spatial resolution
of JWST to ascertain the photometric signature of the nucleus of comet C/2017 K2 (PanSTARRS).
In Section~\ref{sec-volatiles}, the spatial distribution of major gases 
(H$_{2}$O, CO$_{2}$, and CO) as well as trace volatile species, including production rates
and relative abundances is discussed. Section~\ref{sec-trace-volatiles} discusses
analysis of the detected volatile trace species. The activity of the comet is presented in Section~\ref{sec:activity}.  
In Section~\ref{sec:sec-dust} the observations and characteristics of the dust species detected in 
the coma are presented and outcomes described based on thermal model analysis, including 
comparison of the dust found in comet C/2017 K2 (PanSTARRS) with other comets studied by 
remote sensing techniques at wavelengths similar to that accessible with JWST.  The presence of 
polycyclic aromatic carbons (PAHs) in the coma of C/2017 K2 (PanSTARRS) is explored in 
Section~\ref{sec-pah-models-dw}. 

In each Section the observed characteristics of comet C/2017 K2 (PanSTARRS) is compared 
to other comets observed with a variety of ground-based as well 
as in situ studies of comets, including the broad ensemble of comets studied with 
Spitzer \citep{2023PSJ.....4..242H}.  The insights gained from solar system 
comet investigations with JWST and their impact on our understanding of 
proto-planetary disk processes is presented. We conclude in Section~\ref{sec-summary}, 
summarizing key and unexpected returns from JWST observations of comet
C/2017 K2 (PanSTARRS) stemming from our analyses. More detailed analysis 
and modeling efforts of specific aspects of these JWST observations will follow in a series of 
follow-on papers.


\begin{deluxetable*}{@{\extracolsep{0pt}}lccccccccc}
\tablenum{1}
\setlength{\tabcolsep}{3pt} 
\tablecaption{JWST Observational Log Comet C/2017 K2 (PanSTARRS)\label{tab:table-1}}
\tablewidth{0pt}
\tablehead{
\colhead{JWST Archive}            &\colhead{IFU}                  &\colhead{Number}   &\colhead{} &&&&&& \colhead{Phase}\\ 
\colhead{Observation}               &\colhead{Instrument}       &\colhead{Dither}  &\colhead{Exposure}      &\colhead{Exposure}   &&&&& \colhead{Angle}   \\
\colhead{Number\,\tablenotemark{a} } &\colhead{Configuration}  &\colhead{Positions}  &\colhead{Mid-point } &\colhead{Time}  &\colhead{$r_h$}  &\colhead{$\Delta$} &\colhead{PSAng\,\tablenotemark{b} }&\colhead{PSAMV\,\tablenotemark{b} } &\colhead{$\phi$} \\
&& &\colhead{(date | time)} &\colhead{(sec)} &\colhead{(au)}    &\colhead{(au)} &\colhead{(deg)} &\colhead{(deg)} &\colhead{(deg)} 
}
\startdata
12 (comet)      & MRS SHORT\_A & 4  & 2022 Aug 21 TT 09:28:36.8 & 4300 & 2.35 & 2.02 & 101.42 & 21.8 & 25.67 \\
13 (background) & MRS SHORT\_A & 4 \\ 
14 (comet)      & MRS MEDIUM\_B & 4 & 2022 Aug 21 TT 15:05:28.1 & 2867 & $\ldots$ & $\ldots$ & $\ldots$ & $\ldots$ & $\ldots$ \\ 
15 (background) & MRS MEDIUM\_B & 4 \\ 
16 (comet)      & MRS LONG\_C & 4  & 2022 Aug 21 TT 15:04:29.2 & 2389 & $\ldots$ & $\ldots$ & $\ldots$ & $\ldots$ & $\ldots$\\ 
17 (background) & MRS LONG\_C & 4 \\ 
18 (comet)      & NIRSpec G395M & 1 & 2022 Aug 21 TT 21:40:46.3 & 322 & $\ldots$ & $\ldots$ & $\ldots$ & $\ldots$ & $\ldots$\\      
19 (background) & NIRSpec G395M & 4\\
\enddata
\tablenotetext{a}{\, Filenames are: jw015660\_\{observation number\}001\_\{31001\}, i.e., \{observation number\}\_\{visit number\}\_\{visit group\}.}
\tablenotetext{b}{\, The position angles of the extended Sun-to-target radius vector (PsAng)
and the negative of the target's heliocentric velocity vector (PsAMV), as
seen in the observer's plane-of-sky, measured counter-clockwise (east) from
reference-frame north-pole. For comets like C/2017 K2 (PanSTARRS), PsAng
is an indicator of the comet's gas-tail orientation in the sky (being in the
anti-sunward direction) while PsAMV is an indicator of dust-tail orientation. }
\end{deluxetable*}


\section{Observations and Reductions} 
\label{sec-obspipe}

Observations of comet C/2017 K2 (PanSTARRS) were formulated using the best available knowledge
and performance characteristics of JWST prior to the observatory completing its commissioning phase.
We will discuss data obtained with both NIRSpec 
\citep[for a complete instrumental description see][]{2022A&A...661A..80J, 2022A&A...661A..82B}
integral field unit (IFU; 2.9 -- 5.3~\micron,  $\lambda/\Delta\lambda \simeq 1000$, G395M/F290LP 
disperser/filter configuration) and MIRI Medium Resolution \citep[henceforth MRS, see][]{2015PASP..127..584R, 2021A&A...656A..57L} 
IFU (4.9 -- 28.1~\micron, $\lambda/\Delta\lambda \simeq 3000$) data sets (Table~\ref{tab:table-1}).

The original JWST astronomical observing templates used the NASA JPL Horizons 
System\footnote{\url{https://ssd.jpl.nasa.gov/horizons/app.html\#/}} 
generated ephemerides (which had high astrometric accuracy) to acquire the comet 
directly with no target acquisition. However, the initial attempt to execute the program (observations 
01 through 08) on MJD $\simeq 2459807$ (2022 August 16) encountered a variety of failures 
that caused background sequences to be completely skipped by the observatory (exception reports 
generated on-board software) and/or the commanded dither patterns failed to complete. The cause 
of these exceptions were eventually traced to issues with the fine guide star bad/hot pixel mask 
that were eventually rectified.  As a result, some of the program's observations were rescheduled 
and repeated (although the proposed NIRCam imagery was dropped due to overall programmatic 
time constraints). Herein, we address only the more robust MRS and NIRSpec data sets obtained 
on MJD $\simeq 2459812$ (2022 August 21 UT).

The observational circumstances of comet C/2017 K2 (PanSTARRS) at the epoch of 
JWST observations are summarized in Table~\ref{tab:table-1}. The JWST level~0 data 
files were downloaded from the Mikulski Archive for Space Telescopes (MAST) and 
re-reduced using a local instance of JWST pipeline version 1.14.0 through all three 
stages of pipeline processing to produce three-dimensional spatial-spectral data 
cubes (s3d) for analysis. The MRS data discussed in this manuscript 
(summarized in Table~\ref{tab:table-1}) are those in the archive for which 
both the MRS on-source (comet in the field of view in each channel) and 
the off-source nod position (measurements of the background toward the sky 
position of the target) and associated 4-point dither pattern 
were successfully completed. However, only one of four NIRSpec on-source
dither positions were executed.  These data are sufficient for science analysis 
as all four of the background observations were completed. The remainder 
of the NIRSpec observations, and subsequently scheduled NIRCam observations 
were skipped due to a guide star re-acquisition failure.  A second repeat of the 
observational sequence was not possible due to solar elongation constraints.

The CDRS files (2022 August, MRS jwst\_0994.pmap and NIRSpec jwst\_1105.pmap) 
called in the reductions included improved (over the initial commissioning release) 
MRS spectral photometric calibration files as well as corrections for the time dependent 
sensitivity of the longest wavelengths of the MRS detectors.  An MRS defringing algorithm 
was run in Stage 1 of the pipeline (calwebb1), and background subtraction 
was enabled for NIRSpec in Stage 2 (calwebb2) of the pipeline processing and 
MRS in Stage 3 (callwebb3) of the processing using the observed 
backgrounds (Table~\ref{tab:table-1}).  Due to pointing errors, the position of the photocenter of 
the comet moved in the respective fields of view for each MRS band. 
Hence separate MRS s3d data cubes were generated for each band after 
dither position combination and spaxel image reconstruction with drizzle (in pixel coordinates).

Extended spatial emission was clearly evident in the NIRSpec and MRS s3d data cubes.
The position of the comet photocenter was determined in each band, and aperture photometry
using python package photutils (aperture\_photometry invoking the 
method = `subpixel', subpixels = 5 options) was performed by ``drilling'' each 
spatial spectral cube along the wavelength axis with an effective circular 
aperture using a 1\farcs0 diameter (subtending $\simeq 1465$~km of the coma). Choice of
a 1\farcs0 diameter effective circular aperture was driven by the desire to study the broad wavelength 
coverage spectral features (3 to 28~\micron) across the spatial-extent of the comet 
C/2017 K2 (PanSTARRS) as captured in the NIRSpec and MRS fields-of-view, excluding 
regions of low signal to noise and/or NAN values at the edges and properly sample the point-spred function
(PSF). The effective CDELT1 of the longest wavelength MRS channel 4 
(9.722222e-05$^\circ$) yields a 2.86~spaxel diameter (approximately satisfying a Nyquist sampling 
criteria) effective circular radius for a 1\farcs0 diameter beam. This constraint provided the 
choice for the beam size diameter in arcseconds for the other channels. 
Recently \citep{2024NatAs...8.1237F} have investigated the distribution of NIRSpec emission
from the active Centaur 29P/Schwassmann-Wachmann~1 of on spatial scale $\simeq 398$~km using a 
spaxel-by-spaxel approach (which we employ to study the H$_{2}$O 5.66-6.63~\micron\ emission,
section~\ref{sec:results_waterband_analysis}); however, here we are investigating emission in both NIRSpec 
and MRS toward the same lines-of-sight in the coma over a large spatial size-scales. 

\begin{figure}[ht!]
\figurenum{1}
\begin{center}
\includegraphics[trim=0.07cm 0.05cm 0.07cm 0.05cm, clip, width=0.450\textwidth]{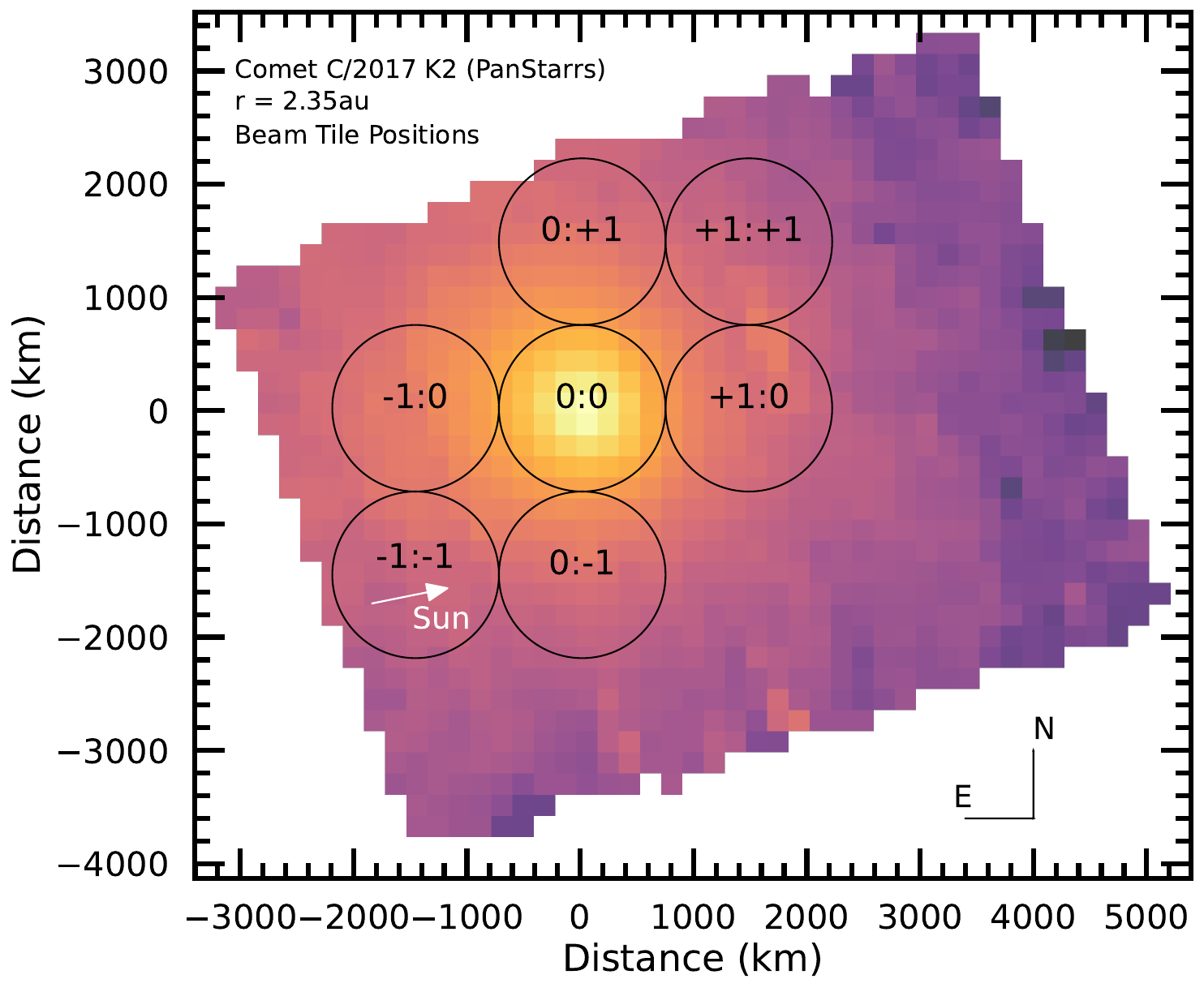}
\caption{The tile positions of 1\farcs0 diameter circular aperture beams used to study the spatial distribution
of coma emission from the  2.9 - 27.9~\micron{} JWST spectra of comet C/20217 K2 (PanSTARRS) derived
from NIRSpec and MRS s3d data cubes. The background is the wavelength collapsed
MRS Ch1 MEDIUM(B) s3d data cube (arbitrary logNormal scaling), the direction of the Sun vector is 
indicated for the date of the MRS observations, MJD 2459812 (2022 August 21 UT). North is up and East to the left.
}
\label{fig:mrs-beamtile-positions}
\end{center}
\end{figure}

The MRS s3d data cubes were drilled in seven beams 
(both the flux plane and the corresponding variance plane) whose  spatial positions across the 
image (spatial extent) were defined by that of MRS Ch1~SHORT(A), which has the smallest 
field-of-view  ($\simeq$ 3\farcs0 $\times$ 3\farcs0). Selected beam positions in the beam grid 
avoided regions where there are extensive ``white pixels'' (indicative of NANs and/or other 
data artifacts in the dither combined background subtracted dithered images). The
beam positions and their labeling are shown in Figure~\ref{fig:mrs-beamtile-positions}.

The MRS spectral segments were then channel-band merged (i.e., Ch1 SHORT(A)+MEDIUM(B)+LONG(C))
and then channel-merged together: 1~($4.9004 \ltsimeq \lambda (\mu\rm{m})$ \linebreak $\ltsimeq 7.6496$) +  
2~($7.5107 \ltsimeq \lambda (\mu\rm{m}) $$\ltsimeq 11.7003$) + 
3 ($11.5513 \ltsimeq \lambda (\mu\rm{m})$$\ltsimeq 17.9787$) + 
4 ($17.7030 \ltsimeq \lambda (\mu\rm{m})$ \linebreak $\ltsimeq 27.9010$)
using local post-JWST pipeline Python code to produce complete 4.9 to 28.0~\micron{} spectra. 
Given residual fringing effects, Channels 2, 3, and 4 segments were smoothed by convolving the 
spectral segments with an astropy scipy Box1DKernel (width = 21 pixels, mode = oversample, factor = 12) function.
Use of this smoothing kernel technique for early cycle JWST pipeline products ($\ltsimeq$ v1.140) has 
been demonstrated by \citet[][]{2022NatAs...6.1308L} to produce reasonable results. However, exploration of 
Savitzky-Golay filtering convolution (scipy.signal.savgol\_filter) as well as a Gassian1DKernal (order 9) convolution 
smoothing was done to ensure that subtle features were not eliminated nor artifacts introduced into these 
MRS spectral segments. All MRS spectra discussed and analyzed herein are those with Box1DKernel smoothing.
The Channel 1 spectral segment was not  smoothed due to the presence of a `forrest' of water $\nu_{2}$ band
emission features. 

\begin{figure*}[h!]
\figurenum{2}
\begin{center}
\includegraphics[trim=0.05cm 0.05cm 0.05cm 0.05cm, clip, width=1.0\textwidth]{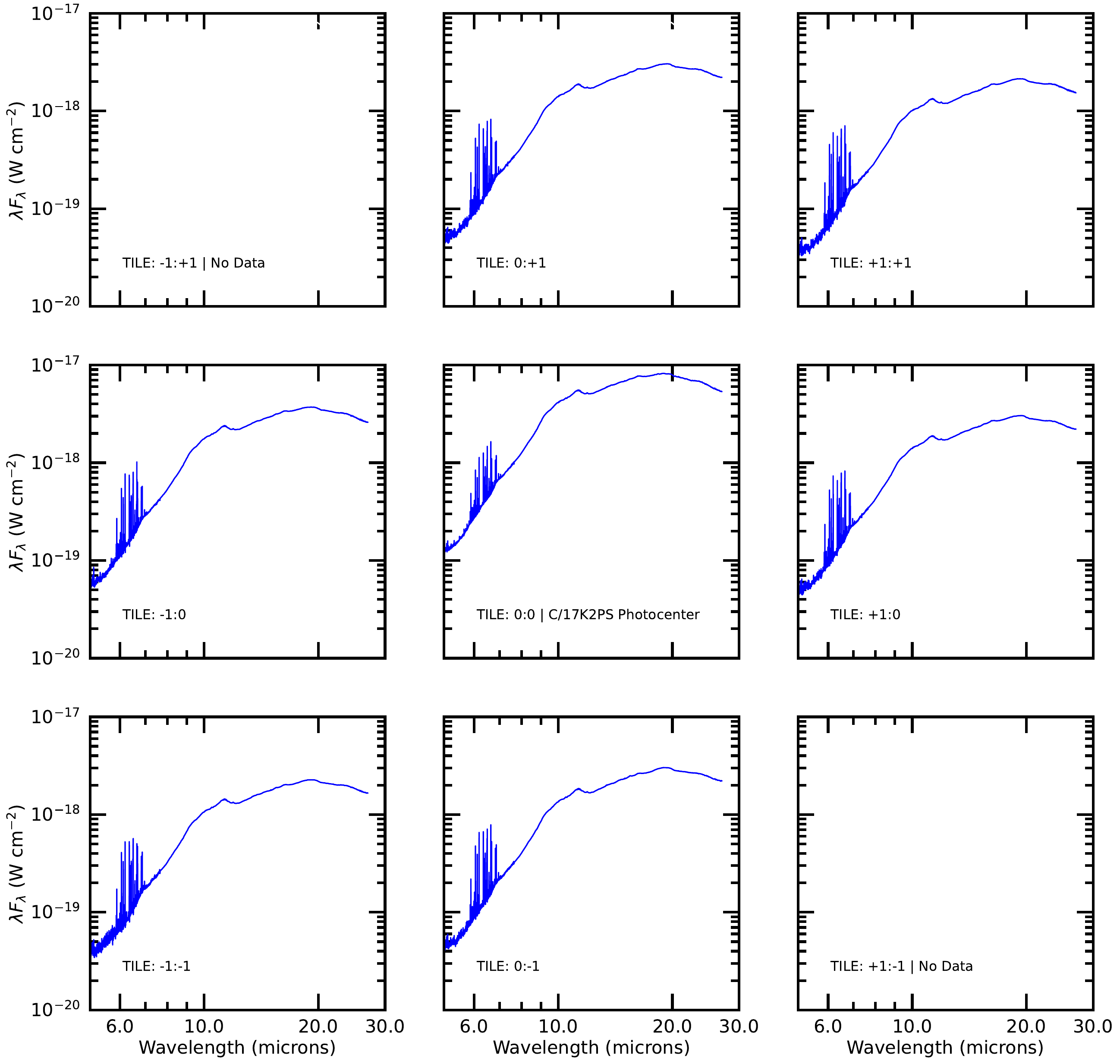}
\caption{The spatial distribution of the 4.9 to 28.0~\micron{} SEDs of comet C/2017 K2 (PanSTARRS) 
obtained on 2022-08-21TT14:42:00 (JWST MRS observation date) when the comet was at a 
heliocentric distance $r_{h} =$ 2.35~au and $\Delta =$ 2.02~au in 1\farcs0 diameter circular 
beams that tile the coma (see section~\ref{sec-obspipe}). Tiles indicating no data are places in the beam 
grid where the MRS Ch1 SHORT(A) cube had no of valid data. The center of the beam grid is a beam placed 
on the approximate centroid position of the peak in the surface brightness of the comet (as measured 
in all the channels drilled) as given in Figure~\ref{fig:mrs-beamtile-positions}.
The 1\farcs0 diameter beam accounts for the values of PIX\_SAR variations 
in each a channel. 
}
\label{fig:fig-three}
\end{center}
\end{figure*}

Channel 2 was selected as the anchor order to which all the other 
channels (1, 3, and 4) were scaled (normalizing to the average flux density computed from the flux density
average within overlapping wavelength regions) and trimmed to generate a piecewise smooth, 
contiguous SEDs. The scaling factors for Channels 1 and 3 were of the order of 2\% to 3\%, and 
that of Channel 4 of the order of $\ltsimeq 14$\%.  The MRS spectra at each spatial beam position across the 
inner coma of comet C/2017 K2 (PanSTARRS) are presented in Figure~\ref{fig:fig-three}.

The NIRSpec IFU s3d cube was generated from a level~0 data reprocessed through the JWST 
v1.11.3 version of the pipeline from single observation at only one dither position. 
Because bad, hot, and/or noisy pixels could not be masked and replaced by valid spaxel data 
from other dither positions, the final pipeline processed s3d cube has more regions with 
NAN-values. A central rectangular region of interest was cut 
from the larger s3d cube and the spectrum was generated using a single 1\farcs0 diameter 
beam centered on the comet photocenter summing along the cube-spectral dimension
as described above. The extracted spectrum was then sigma-clipped ($10\sigma$ threshold) to
remove remaining outliers. 

\begin{figure}[ht!]
\figurenum{3}
\begin{center}
\includegraphics[trim=0.07cm 0.05cm 0.07cm 0.05cm, clip, width=0.480\textwidth]{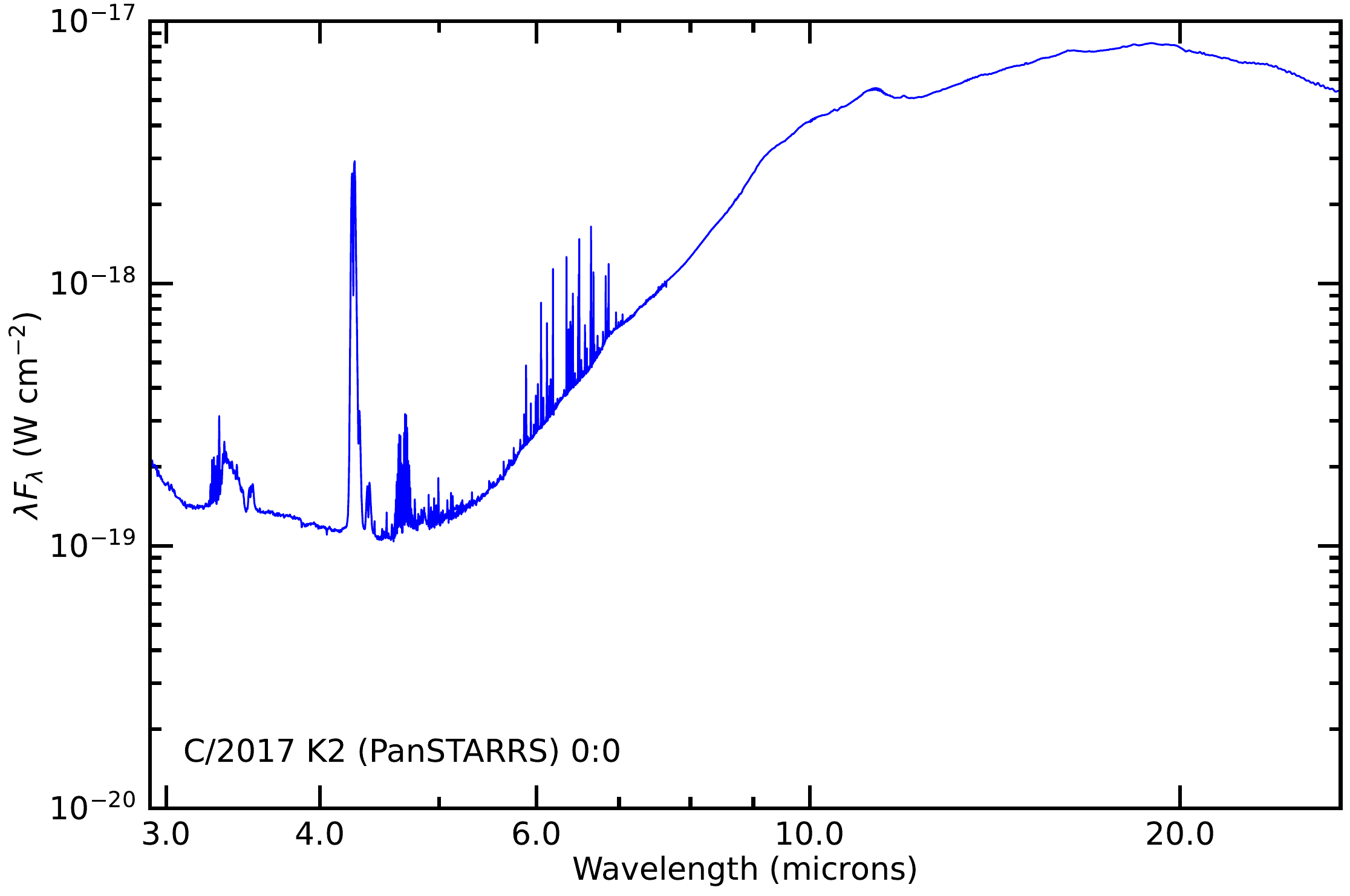}
\caption{The composite 2.9 - 27.9~\micron{} JWST spectrum of comet C/20217 K2 (PanSTARRS) derived
from drilling the NIRSpec and MRS s3d data cubes using a 1\farcs0 diameter circular aperture centered on the 
comet photocenter (position 0:0, see Fig.~\ref{fig:mrs-beamtile-positions}) along their wavelength dimension 
(hence a cylinder in the s3d data cube space) on MJD 2459812 (2022 August 21 UT).
}
\label{fig:fig-composite-nirspec-mrs-00}
\end{center}
\end{figure}

The complete 2.9 to 27.9~\micron{} JWST spectrum of comet C/20217 K2 (PanSTARRS) in
a 1\farcs0 diameter aperture centered on the peak of the photocenter emission is
presented in Figure~\ref{fig:fig-composite-nirspec-mrs-00}.

\section{The Nucleus} 
\label{sec-nucleus}

\begin{figure*}[ht!]
\figurenum{4}
\begin{center}
\gridline{
    \fig{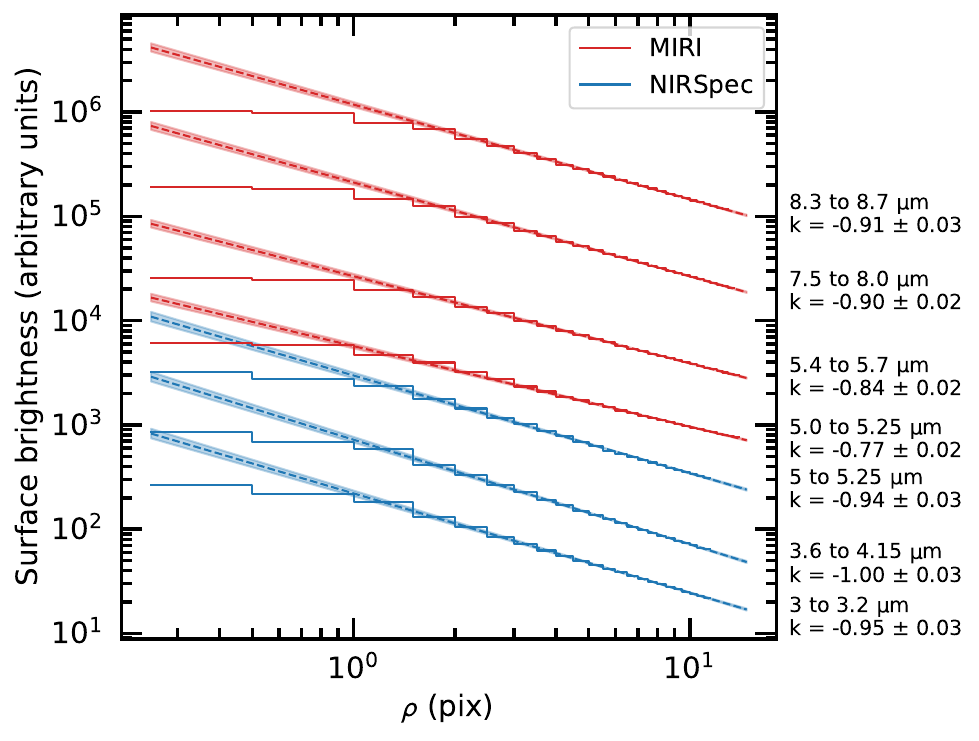}{0.48\textwidth}{(a)}
    \fig{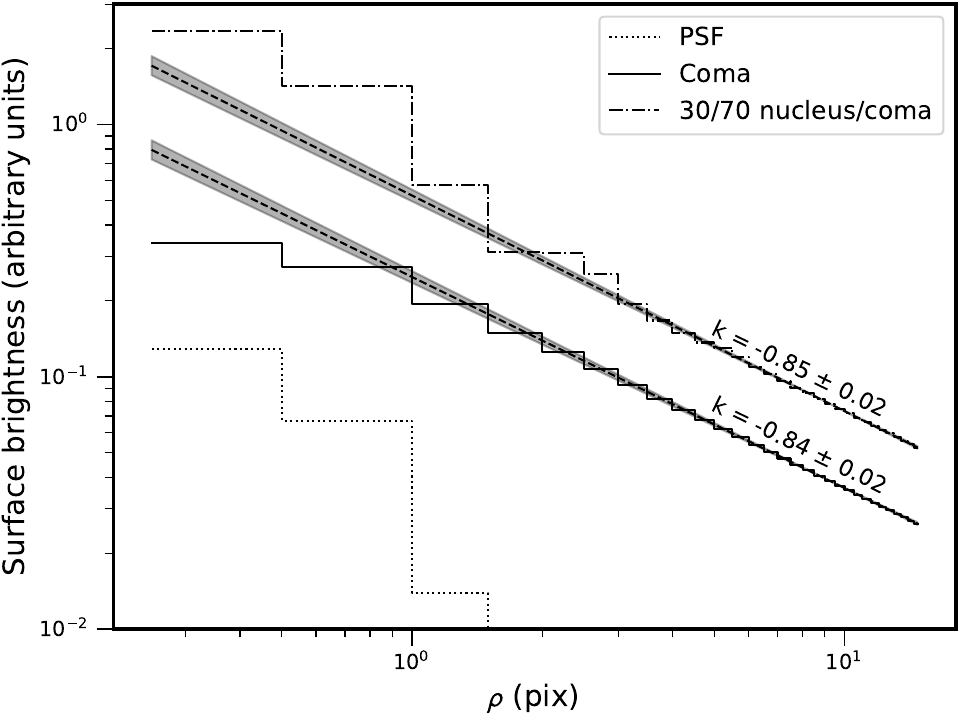}{0.48\textwidth}{(b)}
}
\caption{(a) Radial surface brightness profiles (solid lines) for six selected wavelength 
ranges targeting the spectral continuum.  Wavelength range and best-fit power-law 
slope ($k$, dashed lines) are as indicated.  The bottom three profiles are 
based on the NIRSpec data, and the remaining profiles are based on the MRS data. 
The profiles are offset for clarity. (b) Model radial profiles for 
the MRS and a monochromatic wavelength of 5.55~\micron.  Three 
models are provided: a pure point-source (PSF), a pure coma with a 
radial surface brightness profile equivalent to that observed at 5.4 to 
5.7~\micron{} (coma), and a combined coma and nucleus model, based on a 
nucleus-to-coma ratio of 30/70 within a 1\arcsec{} diameter aperture. The profiles 
are offset for clarity.
}
\label{fig:profiles}
\end{center}
\end{figure*}

The high spatial resolution of the NIRSpec and MRS instruments (150 
and 290 km~pixel$^{-1}$, respectively at the comet's geocentric distance, $\Delta$) 
provides an opportunity to search for the photometric signature of the nucleus.  We averaged the 
NIRSpec and MRS (short) spectral data cubes over select wavelength ranges and 
computed azimuthally averaged radial profiles.  The wavelength ranges were 
chosen to avoid significant emission from gaseous species.  The radial 
profiles were fit at $\geq4$~pixels from the photometric centroid with a constant 
power-law slope.  The results are shown in Figure~\ref{fig:profiles}a.  
A clear nucleus signal would be apparent as excess surface brightness above the 
power-law trend lines.  However, no obvious nuclear contribution is seen.  The trend 
lines are in agreement with the surface brightness profiles down to $\sim2$~pixels. Data 
interior to this point are affected by the instrumental spatial resolving power.
The best-fit NIRSpec slopes are: 
--0.95$\pm$0.03 at 3.0--3.25~\micron, --1.00$\pm$0.03 at 3.6--4.15~\micron, 
and --0.94$\pm$0.03 at 5.0 to 5.25~\micron; for MRS: --0.77$\pm$0.02 at 5.0 to 
5.25~\micron{}, --0.84$\pm$0.02 at 5.4--5.7~\micron, --0.90$\pm$0.02 at 7.5--8.0~\micron, 
and --0.91$\pm$0.03 at 8.3--8.7~\micron. The disagreement between the NIRSpec and 
MRS profiles at 5.0 to 5.25~\micron{}, yet the better agreement between the radial profiles for 
MRS's longer wavelengths suggests significant residual background is present in the 
MRS short wavelengths ($\simeq 8$~MJy sr$^{-1}$).

\begin{figure}[ht!]
\figurenum{5}
\begin{center}
\includegraphics[width=0.450\textwidth]{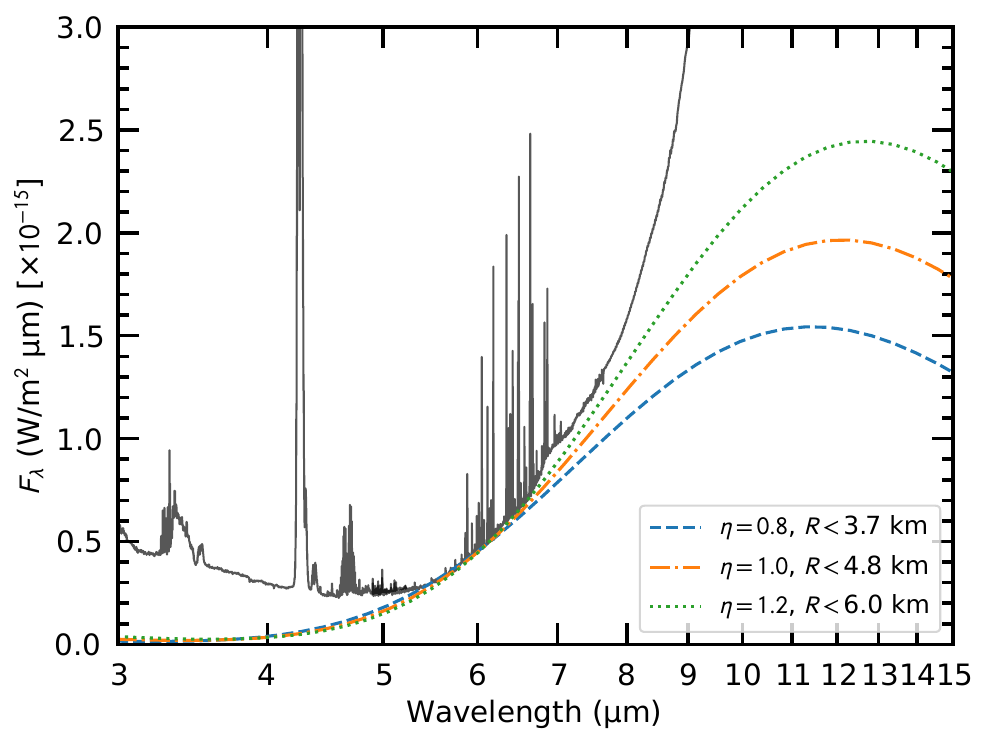}
\caption{Upper-limit nucleus models (dashed, dashed-dotted, and dotted lines) compared 
to the combined NIRSpec and MRS spectrum (solid line).  The 5 to 7~\micron{} range provides 
the strongest constraint on the model nucleus radius ($R$) for our assumed IR beaming 
parameter values ($\eta$).
}
\label{fig:nucleus}
\end{center}
\end{figure}

Given that there is no clear spatial signature of emission from the nucleus, 
we next examine the NIRSpec and MRS spectra to estimate an upper-limit 
to the nucleus effective radius.  A strict upper-limit can be given assuming 
that the spectrum is 100\% nucleus, even though it is clearly coma dominated. 
We fitted the combined NIRSpec and MRS central aperture spectrum 
(1\farcs0 diameter) with example models based on the near-Earth asteroid 
thermal model of \citet{1998Icar..131..291H}.  The model is based on the 
thermal emission from a sphere in instantaneous thermal equilibrium with 
insolation.  The model parameters used are similar to those of  
\citet{2013Icar..226.1138F}, who studied the 16 to 24~\micron{} emission 
from cometary nuclei: an infrared emissivity of 0.95, a low optical geometric 
albedo of 6\%, and an infrared beaming parameter, $\eta\sim1$. The choice 
of beaming parameter affects the model temperature, and mimics the effects 
of surface roughness and non-zero thermal inertia.  Specifically, 
\citet{2013Icar..226.1138F} derived $\eta=1.03\pm0.11$ for the cometary 
population, therefore we chose values over the $\pm2\sigma$ range: 0.8, 1.0, 
and 1.2.  We find that the spectrum at 5 to 7~\micron{} yields the strictest 
limits on the nucleus brightness, and therefore its effective radius. The 
nucleus cannot be larger than 6.0~km in radius, or else the $\eta=1.2$ model 
would be brighter than the observed spectrum at $\sim6~\micron$. Alternatively,
if one assumes $\eta=0.8$, the nucleus cannot be larger than 3.7~km in effective
radius. These models are show in Figure~\ref{fig:nucleus}. 

We further investigate the radial profile using the WebbPSF tool 
v1.2 \citep{2012SPIE.8442E..3DP} to estimate the amount of nucleus that 
could be present in the MRS 5.4 to 5.7~\micron{} radial profile, based 
on the prediction that nucleus-coma contrast is strongest in this wavelength 
regime.  WebbPSF can produce simulated PSF models for a given telescope and 
instrument optical state.  Although the details of the MIRI MRS instrument 
are not modeled in this version of WebbPSF, we can still use the WebbPSF results 
by proxy.  To that end, we produced a model coma at 5.55~\micron{} through 
convolution of the PSF with a power-law coma profile.  The radial profile of 
the resulting model coma behaves similarly to the observed radial profile at 
5.4 to 5.7~\micron, i.e., the profile is consistent with a power-law 
distribution, except inside the inner $\sim$2~pix, where the power-law 
extrapolation over-estimates the profile by a factor of $\lesssim$2 (Figure~\ref{fig:profiles}b).  
We next produced a combined nucleus and 
coma model, with a 30/70 respective flux ratio within a 1\arcsec{} diameter 
aperture. The radial profile for this model is significantly 
different, and the power-law fit to the coma now under-estimates 
the profile.  Given the lack of a precise PSF model for the reconstructed 
MRS data cubes, we take this nucleus-to-coma ratio, 30/70, as a 
conservative upper-limit to the nucleus contribution.  The average flux 
density of the comet between 5.4 and 5.7~\micron{} in a 1\arcsec{} diameter 
is 3.2$\times$10$^{-16}$~W~m$^{-2}$~\micron$^{-1}$.  Therefore, the nucleus 
must be less than 1.4$\times$10$^{-16}$~W~m$^{-2}$~\micron$^{-1}$.  Returning 
to our $\eta=1.2$ model, the nucleus is smaller than 4.2~km in radius.
 
\section{Volatiles and Molecular Inventory} 
\label{sec-volatiles}

\subsection{Overview} \label{sec:sec-summary-volatiles}

\begin{figure}[hbt]
\figurenum{6}
\begin{center}
\vspace{-0.1cm}
\includegraphics[trim=0.15cm 0.10cm 0.15cm 0.01cm, clip, width=0.460\textwidth]{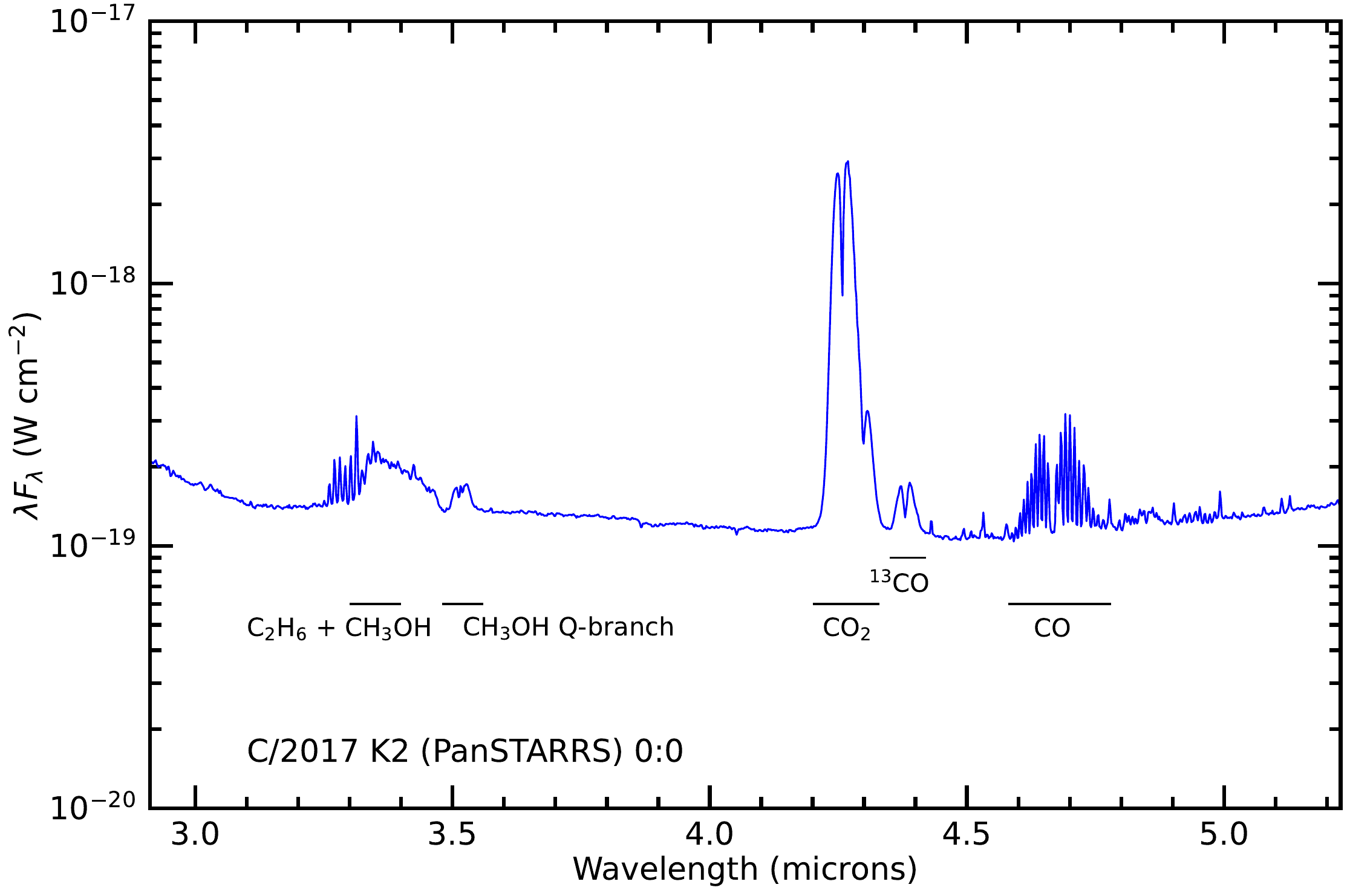}
\caption{The NIRSpec spectrum of comet C/2017 K2 (PanSTARRS) derived 
from the G395M IFU data cube in a 1\farcs0 diameter circular aperture centered on the comet photocenter. 
Emission from CO$_{2}$, $^{13}$CO$_{2}$, CO, OCS (carbonyl sulfide), and H$_{2}$O are 
present. The complex near 3.35~\micron{} is comprised of C$_{2}$H$_{6}$ and CH$_{3}$OH (ethane 
and methanol) and the CH$_{3}$OH Q-branch at 3.52~\micron. 
}
\label{fig:fig-nirspec-three}
\end{center}
\end{figure}

The inventory of volatiles commonly detected in comets include 
H$_{2}$O, $^{12}$CO,  as well as other hyper-volatiles
CH$_4$ and C$_2$H$_6$, together with HCN, CH$_3$OH and other species 
\citep[for a review see][]{Biver2022}.
The most volatile molecules in comets accessible in the infrared via remote sensing are  
the species CO, CH$_{4}$, and C$_{2}$H$_{6}$ \citep{2016Icar..278..301D}. 
The symmetric hydrocarbons, such as CH$_{4}$ and C$_{2}$H$_{6}$, are uniquely 
sampled in the near-IR, as they lack a rotational dipole. CO and CH$_4$ in particular  
are difficult species to adequately measure from the ground unless there is sufficient comet 
geocentric velocity to Doppler shift the emission from spectral wavelength regions were 
opaque telluric counterparts dominate, whereas CO$_2$ can only be accessed from space. 
JWST therefore has a distinct advantage. Figure~\ref{fig:fig-nirspec-three} shows the emission from
some of the stronger volatiles emitting in the coma of comet C/2017 K2 (PanSTARRS).

The volatile inventory provide insights into proto-planetary disk chemical pathways and evolution
\citep{2019A&A...629A..84E, 2020ApJ...898...97B, 2022ApJ...936...40E, 2023FaDi..245...52V}. 
Rosetta's ROSINA mass spectrometer demonstrated 
that CH$_4$ and C$_2$H$_6$ correlate better with CO$_2$ than with H$_2$O 
\citep{2017MNRAS.469S.108G} and similarly with hydrocarbons \citep{2019A&A...630A..31S},
indicating different species are embedded in the two different ice phases of H$_2$O and 
CO$_2$ and that these two ices are not intimately mixed. citet{2011Icar..216..227V}
also argue that distinct moieties of ice exist within the nucleus of comet C/2007 W1 (Boattin)
to account for the observed variation volatile emission.
Whether this is due to primordial heterogeneity or differentiation caused by to thermal 
evolution of the nucleus via its multiple perihelion passages as a Jupiter-family comet (JFC) is not 
clearly discernible for comet 67P/Churyumov-Gerasimenko \citep{2019A&A...630A..31S}.  However, these connection to the 
evolution in the early solar system are still being explored \citep[e.g.,][]{2021AJ....162...74L, 2021ApJ...919...45B}. 

The comet population displays a strong compositional diversity, in particular regarding the 
CO$_{2}$/H$_{2}$O and CO/H$_{2}$O abundance ratios \citep{2017RSPTA.37560252B, 2022PSJ.....3..247H} 
which typically range from 1\%-20\%. This diversity may result from evolutionary processes, affecting
mainly short-period comets, or indicate various formation conditions of comet nuclei in the solar nebula. 
Recent solar nebula models predict a CO$_{2}$/H$_{2}$O enhanced region in the outer disk that results from 
a combination of the inward flow of rock- and H$_{2}$O-rich bodies \citep{2022ApJ...936...40E, 2022NatAs...6..357I}.
For example, comet C/2016 R2 (PanSTARRS) reveals a rare nuclear composition with activity dominated 
by CO sublimation and lacking H$_2$O sublimation \citep{2018A&A...619A.127B, 2019AJ....158..128M, 2022ApJ...929...38C}. 
This composition provides evidence for disk models that predict water-deficient and CO-enriched outer 
disk regions due to the efficient inward flux of H$_2$O-rich pebbles coupled with enhanced CO 
condensation outside the CO snow line in the pebble accretion scenario \citep[e.g.,][]{2023A&A...670A..28S}.
In addition, molecules produced by extended sources, such as subliming icy grains or 
radiative-dissociation of organics \citep{2020Icar..33513411D} or of ammonium 
salts \citep{2020Sci...367.7462P} associated with the dust,  can be identified from column density 
radial profiles that are shallower than for species released from the nucleus. 

JWST also opens new discovery space for the study of 
carbonaceous species, including organics and polyaromatic hydrocarbons (PAHs, carbon ring-chains), 
Mg-carbonates, or carbonyl sulfide (OCS) in cometary coma as all have identifiable features between 
$3.0 \ltsimeq \lambda$(\micron)$\ltsimeq 16.0$. Detection of these species can constrain
models of gas-phase chemistry in the early disk, the survival or formation of PAHs in the disk, and 
the late-time condensation of  Mg-carbonates at high disk altitudes 
\citep[e.g.,][]{2008SSRv..138...75W, 2017RSPTA.37560260W, 2019M&PS...54.1632K}.  
 All are important discriminators for the disk age and radial regimes of cometary 
nuclei formation \citep{2008M&PS...43..261Z, 2012ApJ...745L..19O}.

\vspace{-0.5cm}
\subsection{Water} 
\label{sec:sec-watermodelling}

Emission lines from H$_2$O are detected in both NIRSPec and MRS spectra of C/2017 K2 (PanSTARRS). 
In the 4.5--5.2~$\mu$m range, faint H$_2$O ro-vibrational lines are detected from the $\nu_3$-$\nu_2$ 
and $\nu_1$-$\nu_2$ hot bands. The spectral region 5.5--7.25~$\mu$m, covered by MRS Ch1 SHORT(A), 
shows strong lines from the $\nu_2$ fundamental vibrational band, with some minor contribution from 
hot bands (e.g., $\nu_3+\nu_2-\nu_3$, 2$\nu_2-\nu_2$). The high spectral resolution of MRS 
offers the opportunity to investigate spectrally and spatially 
the ro-vibrational structure of the H$_2$O $\nu_2$ 6.3~$\mu$m band. This is of key interest for H$_2$O 
production rate measurements, because this band is optically thick, especially in the denser parts of the 
inner coma. Lines from $\nu_3$-$\nu_2$ and $\nu_1$-$\nu_2$ hot bands near 5~$\mu$m might also 
be affected by opacity effects, since, as for $\nu_2$, these bands are pumped by solar radiation 
through excitation of intrinsically strong fundamental bands, namely $\nu_3$ and $\nu_1$. However, 
the emitted photons from these hot bands should be poorly absorbed along the ray path towards JWST.

\subsection{Analysis of the H$_2$O bands}
State-of-the-art radiative transfer models with detailed, precise consideration of optical thickness effects in a wide
range of geometric observing conditions are not yet readily available for the $\nu_2$ H$_2$O band 
and hot bands in cometary atmospheres. The NASA Planetary Spectrum 
Generator \citep[NASA PSG;][]{2018JQSRT.217...86V} provides some guidance to this complex
problem for low Sun-Target-Observer phase angles using a first order approximation. Since. the phase angle
for comet C/2017 K2 (PanSTARRS) at the JWST observational epoch was $25.67 \degr$ 
and the NASA PSG opacity corrections are not yet benchmarked,  we used a simplistic approach to analyze the
$\nu_2$ 6.3~$\mu$m band. It consisted in fitting the observed spectrum with an optically thin fluorescence 
model parameterized by the water production rate $Q$(H$_2$O), the rotational temperature $T_{\rm rot}$, 
and the ortho-para ratio, OPR. The H$_2$O number density is described by the Haser model assuming 
an expansion velocity of 0.52~km~s$^{-1}$. The latter expansion velocity is derived
from the standard relation that v$_{\rm{expan}}$ (km~s$^{-1}$) $ = 0.8\, (r_{h})^{-0.5}$ and 
is used in the analysis of the NIRSpec data. This approach 
was used to analyze a series of H$_2$O 2.7~\micron{}  Rosetta Visible and Infrared Thermal 
Imaging Spectrometer (VIRTIS) experiment  spectra \citep{Debout2016, Cheng2022}. 
\citet{Cheng2022} showed that, under optically thick conditions, the inferred OPR and 
$T_{\rm rot}$ are respectively lower and higher than the effective 
values because fainter lines are generally less affected by optical depth effects.  Outcomes from
this approach can be compared to NASA PSG treatments.

For the 6.3~$\mu$m band analysis, we used the H$_2$O fluorescence model of \citet{Crovisier2009}, 
which includes fundamental bands and hot bands and uses the comprehensive H$_2$O ab initio 
database of \citet{Schwenke2000}. For analyzing hot-band H$_2$O lines in the 4.88 to 5.2~$\mu$m range 
covered by MRS and NIRSpec, we used the broader abilities of the NASA PSG because 
some water lines are blended with CN lines. The 6.3~\micron{} 
synthetic spectra from \citet{Crovisier2009} closely resemble those obtained by NASA PSG, 
which uses the BT2 ab initio database of \cite{Barber2006}. The total emission rates in the 
5.5 to 7.25~$\mu$m and 4.8 to 5.2~$\mu$m ranges are $\sim$15\% lower and $\sim$11\% higher, 
respectively, in the NASA PSG model.

The fitting procedure consisted in fitting simultaneously the H$_2$O lines and the underlying continuum, 
described as a cubic spline. The procedure also allows to correct for small deviations in the frequency 
calibration. 

%
\begin{figure}[h]
\begin{centering}
\figurenum{7}
\includegraphics[scale=0.38]{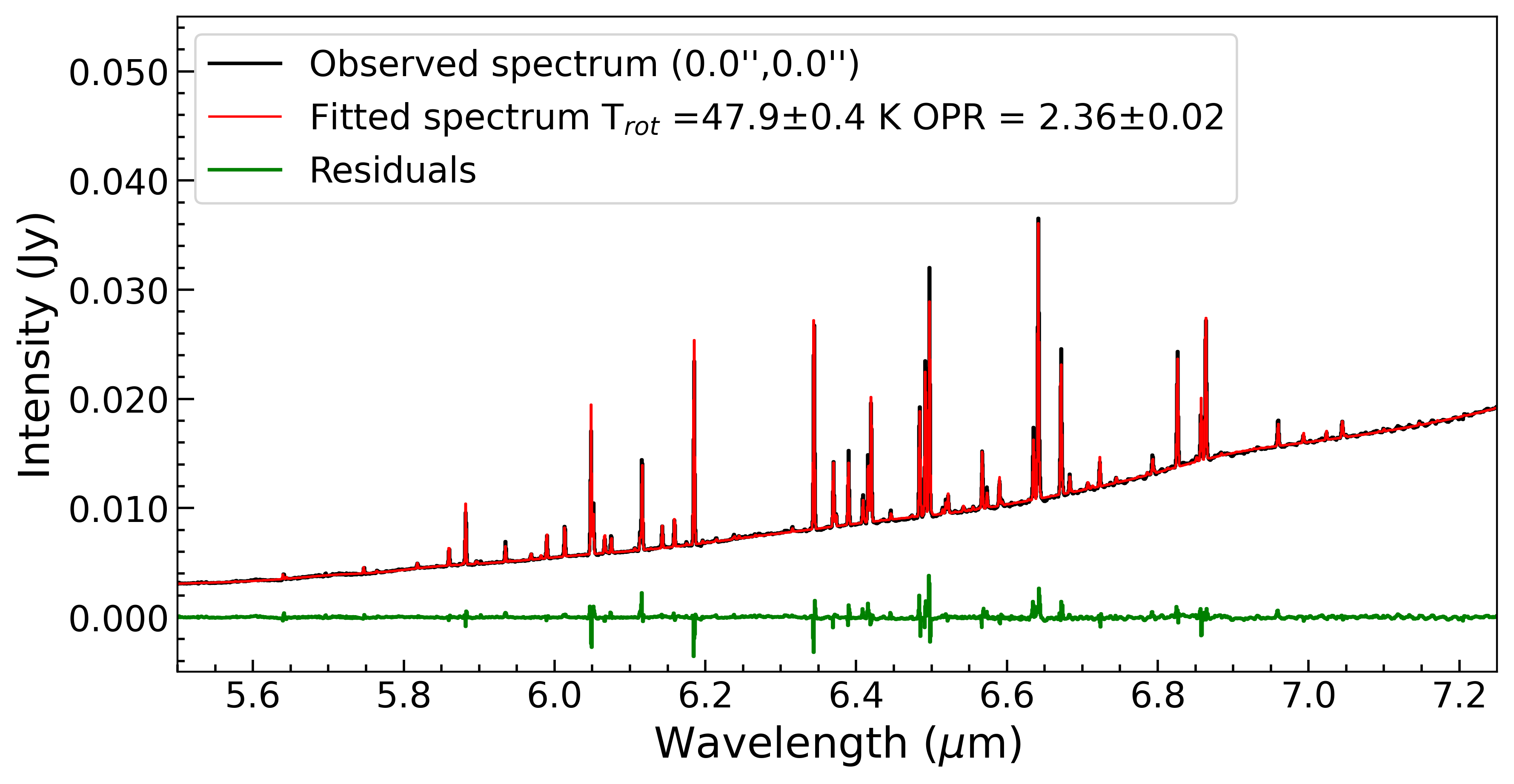}\\
\includegraphics[scale=0.38]{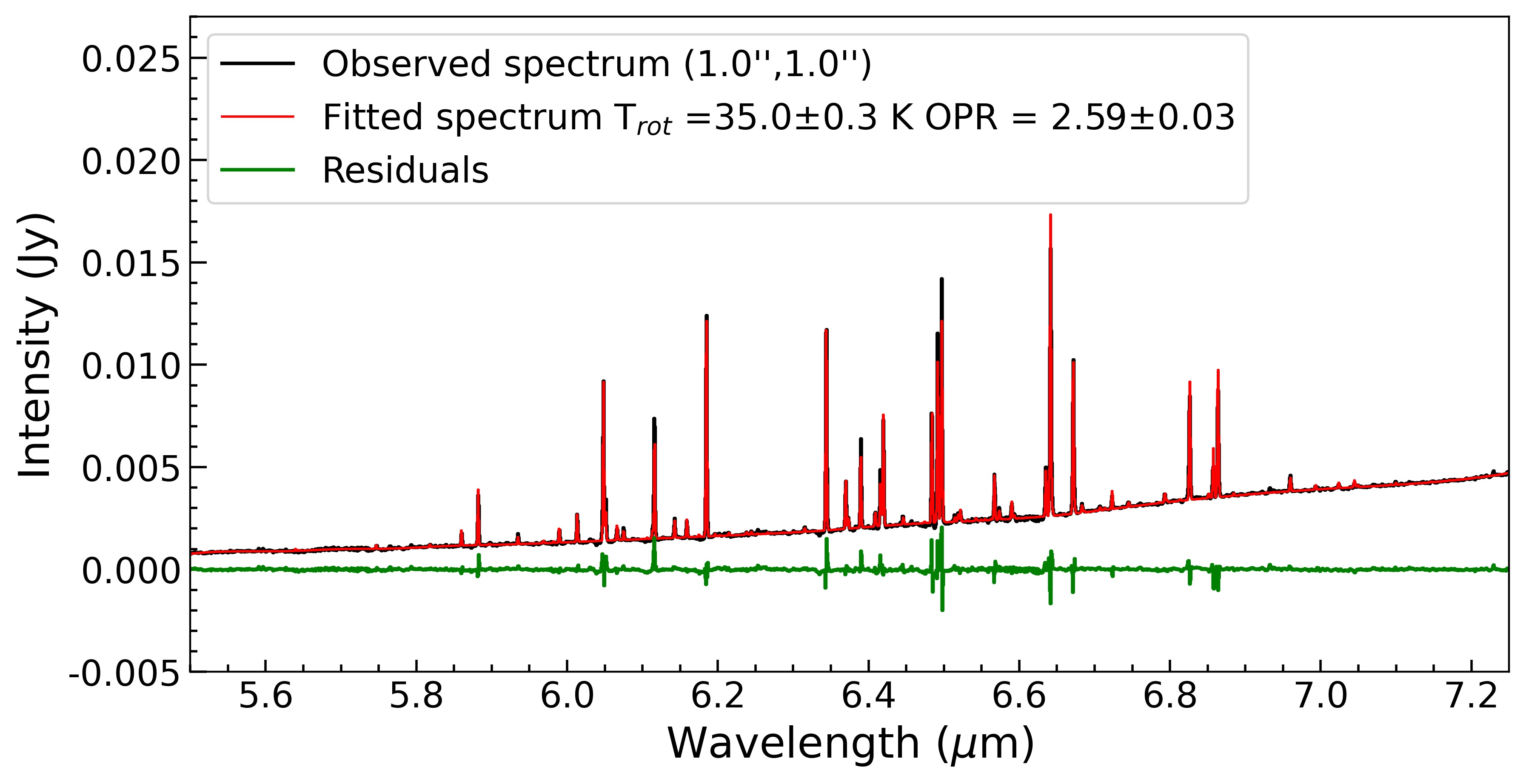}
\caption{Model fits to H$_2$O MRS spectra of C/2017 K2 (PanSTARRS). Measurements and 
model fits are shown in black and red, respectively. Residuals are in green. Top: Extracted on-nucleus spectrum.  
Bottom: extracted off-nucleus spectrum at the (+1:+1)  position
(see Figure~\ref{fig:mrs-beamtile-positions}). A 1\farcs0-diameter 
circular aperture is used. Fitted parameters are given in the inset. Derived apparent production 
rates are (0.967$\pm$0.005) 10$^{28}$~molecules~s$^{-1}$ and (2.64$\pm$0.01) 10$^{28}$~molecules~s$^{-1}$, for the 
on-nucleus and off-nucleus spectra, respectively. \label{fig:H2O-fit}}
\end{centering}
\end{figure}   

\begin{figure}[h]
\figurenum{8}
\includegraphics[trim=0.45cm 1.85cm 0.25cm 2.25cm, clip, width=0.50\textwidth]{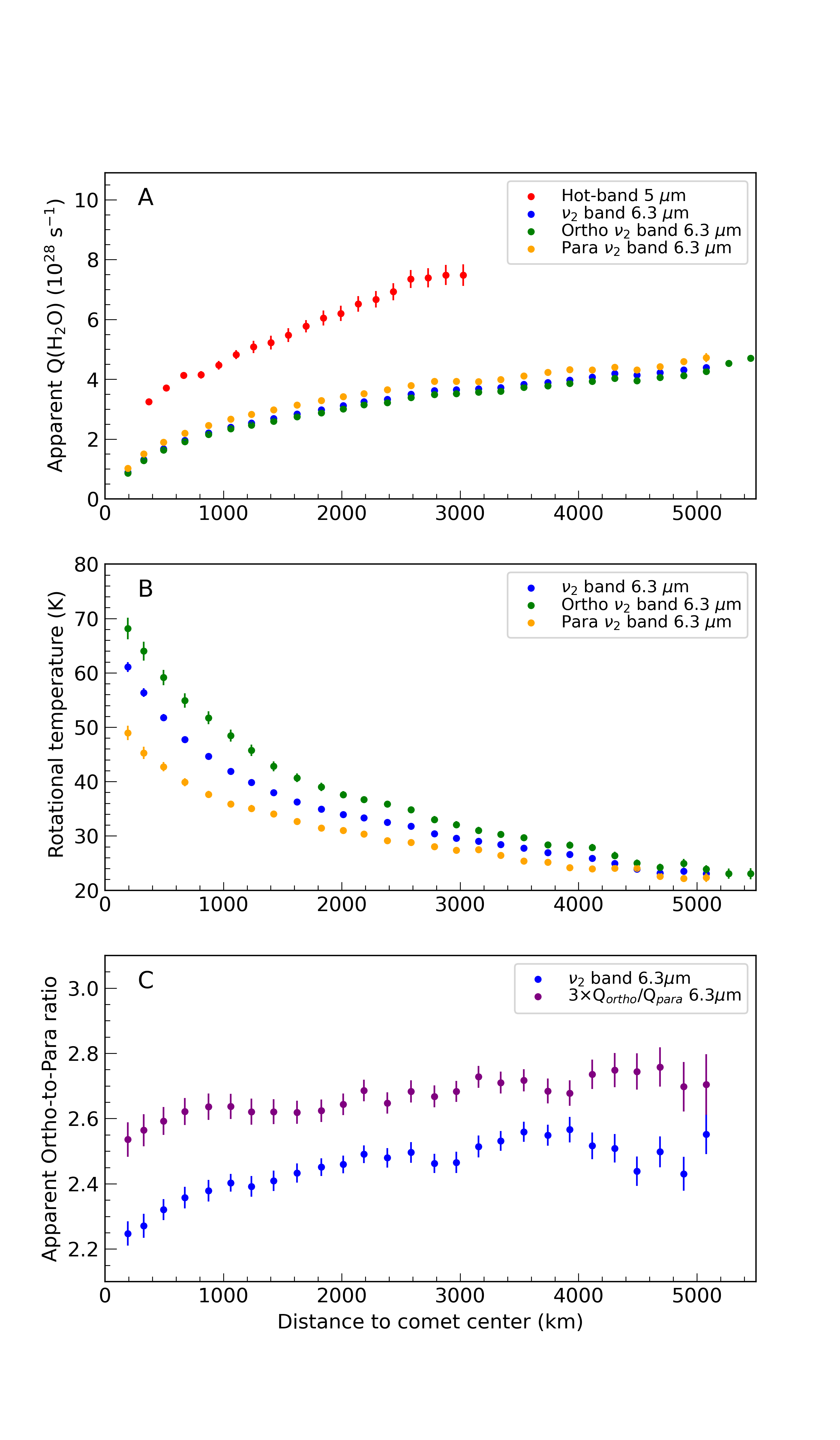}
\caption{Model derived properties of water band observed in comet C/2017 K2 (PanSTARRS) versus 
projected distance from the photocenter. Panel (a)~The water production rate; Panel (b)~The rotational 
temperature; Panel (c)~The ortho-para ratio (C) as a function 
of cometocentric distance derived from model fits. Considered spectra are MRS Channel 1 MED 
($\nu_2$ band from 5.66 to 6.63~$\mu$m, results with blue symbols) and NIRSpec (5~$\mu$m hot-band, 
red symbols) spectra extracted over annuli of 0.13'' (MRS) and 0\farcs1 (NIRSpec) diameter. Results 
from model fits to unblended ortho and para lines are shown with green and orange symbols, 
respectively. In panel A, $Q$(H$_2$O) values derived from the 6.3~$\mu$m band were multiplied 
by 1.17 for consistency with NASA PSG models (see text), and  those inferred from ortho and para lines assume 
OPR = 3. Panel (c) displays OPR values that provide consistent ortho and para-derived 
$Q$(H$_2$O) values (purple symbols). \label{fig:H2O-annuli}  }
\end{figure}  

\subsection{Results on H$_{2}$O band analysis} \label{sec:results_waterband_analysis}
Figure~\ref{fig:H2O-fit} shows two 5.5--7.25~$\mu$m MRS spectra extracted over a 1\farcs0-diameter 
circular aperture, either centered on the nucleus or at the position offset (1\farcs0, 1\farcs0) (cometocentric 
distance $\rho$ $\sim$ 2100~km), with the model fits superimposed and the retrieved model 
parameters given in the legend. The model is not able to correctly fit the observed spectra. Since 
ortho-to-para transitions are highly unlikely in the coma, the increase of the OPR (from 2.36 to 2.6) 
with increasing cometocentric distance is a clear indication of optical depth effects. Therefore, the 
inferred apparent $Q$(H$_2$O) increases from 1$\times$10$^{28}$~molecules~s$^{-1}$ (central position) 
to 2.6$\times$10$^{28}$~molecules~s$^{-1}$ (offset position) should be at least partly affected by opacity. 

\begin{figure*}[ht!]
\figurenum{9}
\begin{centering}
\includegraphics[trim=0.25cm 0.25cm 0.25cm 0.25cm, clip, width=0.95\textwidth]{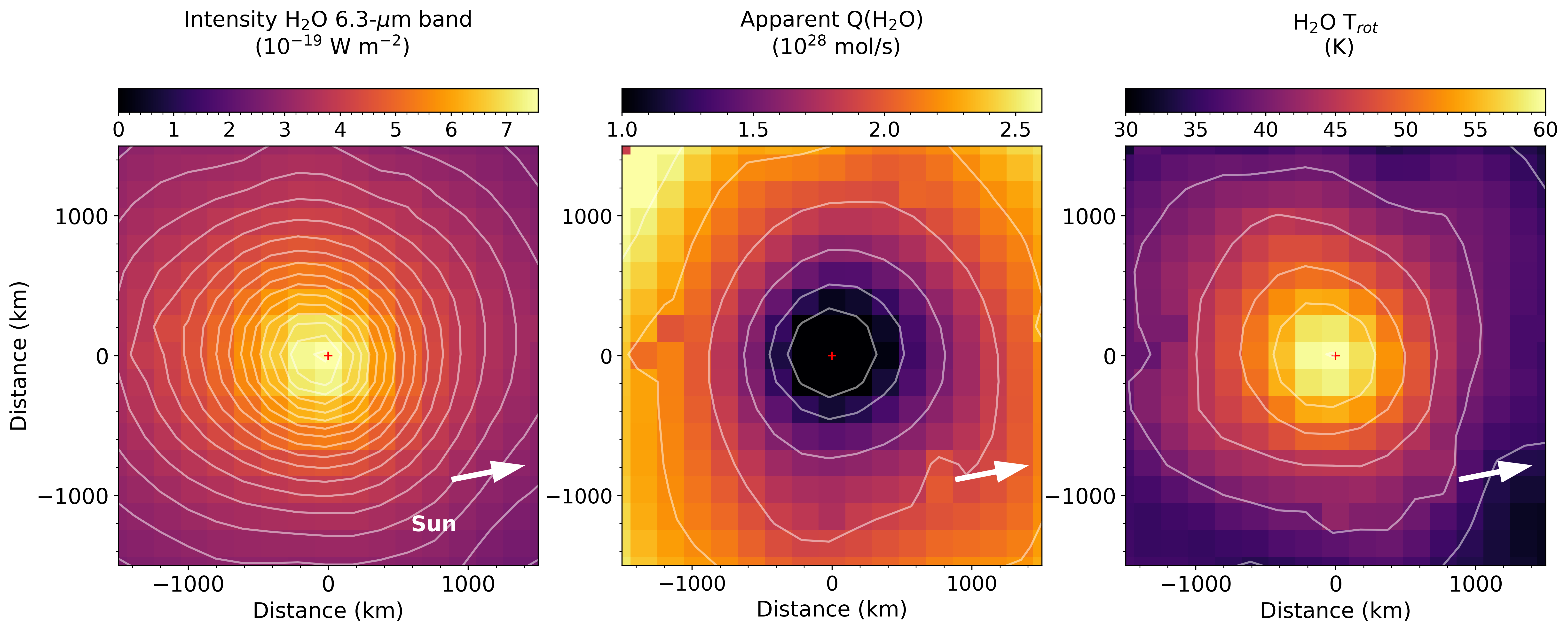}
\caption{Spaxel-by-spaxel spatial distribution of H$_2$O-related physical parameters deduced from 
MRS Channel 1 MED data. From left to right: band intensity in the range 5.66--6.63 $\mu$m, 
water production rate $Q$(H$_2$O) and rotational temperature $T_{\rm rot}$ (K) from model fits. 
A 3$\times$3 boxcar spatial smoothing was applied to the original data cube.  The inset white arrow 
indicates the Sunward direction. \label{fig:H2O-maps}  }
\end{centering}
\end{figure*}

The H$_2$O 5.66--6.63~$\mu$m spectra (MRS Ch1 MEDIUM(B)) were extracted over 0\farcs13--width 
annuli of increasing radius up to 5300~km. Despite this spectral window only partially covers the 
water $\nu_2$ band, larger cometocentric distances can be investigated, since the comet peak 
intensity is significantly offset from the center of MRS Ch1 MEDIUM(B) field of view.  The results of the model 
fits are shown in Figure~\ref{fig:H2O-annuli}. Maps from model fits to individual spaxels are shown
in Figure~\ref{fig:H2O-maps}. The rotational temperature continuously decreases with increasing 
$\rho$, from 60~K in the innermost annulus to $\sim$25 K at $\rho$ = 5000~km, whereas the opposite 
trend is obtained for both $Q$(H$_2$O) and OPR. Figure~\ref{fig:H2O-annuli} also shows results from 
spectral fits to ortho lines (masking para lines and para/ortho blends) and from spectral fits to para 
lines. 

Retrieved $Q$(H$_2$O) from these ortho and para spectra shown in Figure~\ref{fig:H2O-annuli}a 
make the assumption that the effective OPR is equal to 3. Striking in Figure~\ref{fig:H2O-annuli}b is the much 
lower rotational temperature deduced from the para lines (orange dots), with a 10 to 20~K difference 
with ortho lines (green dots) for $\rho$ $<$ 1500~km. Rotational temperatures of ortho and para lines 
become similar at large $\rho$. Since para levels are less populated than ortho levels and para lines 
are weaker than ortho lines, para lines are less optically thick, and $T_{\rm rot}$ values deduced from 
para lines are more representative of the population of H$_2$O rotational levels in the ground vibrational 
state. The fact that para $T_{\rm rot}$ values in the central coma are consistent with values derived 
from CO and CH$_4$ lines (see Figure~\ref{fig:fig-nirspec-temps} and discussion
in Section~\ref{sec:subsec-trace-results-q-and-t})  is supporting this interpretation. In addition, the  trend for 
higher $T_{\rm rot}$ in optically thick conditions was observed on 
67P/Churyumov-Gerasimeko H$_2$O 2.67~$\mu$m spectra \citep{Cheng2022}. 

Figure~\ref{fig:H2O-annuli}c shows with purple dots the OPR values that provide consistent ortho-line 
and para-line derived $Q$(H$_2$O) values. These OPR values, in the range 2.55 to 2.75, are higher 
than those derived from the simultaneous fitting of ortho and para lines (blue dots), and only slowly 
increase with increasing cometocentric distance. This suggests that the $\nu_2$ band is still 
optically thick at the edge of the MRS field-of-view despite the H$_2$O number density is lower. Optical 
depth effects do not significantly decrease as $\rho$ increases, likely because the decrease of 
$T_{\rm rot}$ with increasing $\rho$ strengthens the opacity of the strongest lines \citep{Cheng2022}. 
Since ortho lines are optically thick and thicker than the para lines, as discussed earlier, we can conclude 
that the effective H$_2$O OPR of C/2017 K2 (PanSTARRS) is larger than 2.75. Determining precisely 
the effective H$_2$O OPR would require radiative transfer modeling. 

Water production rates derived from the analysis of hot-band lines near 5~$\mu$m are significantly 
higher (by a factor $\sim$ 2) than those derived from the 6.3 $\mu$m band (Figure~\ref{fig:H2O-annuli}a). 
This is consistent with lesser opacity effects expected for these faint lines. 
We examined opacity corrections estimated by NASA PSG, making NASA PSG simulations 
for NIRSpec and MRS pixels close ($\rho = 600$~km, i.e., 0\farcs4 from comet center)  and significantly 
offset from the nucleus ($\rho = 3000$~km). From these simulations, the total intensity of the 6.3~\micron{}
band is attenuated by less than $\sim 15$\% at these distances, while hot-bands are not affected by 
optical thickness. Consequently, the discrepancy between the $Q$(H$_2$O) values derived from the 
two bands is not explained, requiring further investigation which is beyond the scope of this paper. 
Note that H$_2$O 5~\micron{} lines observed with both MRS and NIRSPec have consistent intensities. So the 
discrepancy observed in Figure~\ref{fig:H2O-annuli}a cannot be attributed to comet variability.

The steep increase of  $Q$(H$_2$O) with increasing $\rho$ inferred from the 5~\micron{} lines is 
likely related to water release from icy dust particles. The much larger water production rate
($\sim$ 2$\times$ 10$^{29}$~molecules~s$^{-1}$) deduced from OH 18-cm data acquired with a 
large beam size (Nan\c{c}ay radio telescope, 3.6$\times$ 19 arcmin, Crovisier, personal communication)
further argues for an extended source of water in the coma of C/2017 K2 (PanSTARRS).

Maps shown in Figure~\ref{fig:H2O-maps} reveal a slightly asymmetric H$_2$O coma, with higher 
apparent $Q$(H$_2$O) and higher $T_{\rm rot}$ in the quadrant opposite to the Sun direction. One 
interpretation is a higher number density of subliming icy grains in this direction, possibly pushed 
anti-sunward by radiation pressure or drifted to the night side of the coma due to the rocket force 
from their sublimation. Modeling \citep{Fougere2012} showed that the addition of a substantial 
(distributed) source of gas from icy grains provides additional rotational excitation resulting in a 
higher $T_{\rm rot}$ than expected for adiabatic expansion from the nucleus, and a slower 
decay of $T_{\rm rot}$ with increasing $\rho$. 

\begin{deluxetable*}{ccccccc}
\tablenum{2}
\setlength{\tabcolsep}{8pt} 
\tablecaption{Volatile Inventory in comet C/2017 K2 (PanSTARRS) Measured with NIRSpec\label{tab:nirspec-qs}}
\tablewidth{0pt}
\tablehead{
\colhead{Molecule} & \colhead{$T$\subs{profile}} & \colhead{$Q$} & \colhead{$Q$/$Q$\subs{H2O}} & \colhead{Range in Comets} \\
\colhead{ } & \colhead{ } & \colhead{($10^{26}$ molecule s$^{-1}$)}  & \colhead{(\%)} & \colhead{(\%)}
}
\startdata
H$_2$O & H$_2$O & 760 $\pm$ 15 & 100 & $\cdot \cdot \cdot$ \\
CO$_2$ & CO$_2$ & 112 $\pm$ 2 & 14.7 $\pm$ 0.3 & $4-30$\\
       & H$_2$O & 82 $\pm$ 3 & 10.8 $\pm$ 0.3 & $\cdot \cdot \cdot$\\
CO & CO & 62 $\pm$ 1 & 8.2 $\pm$ 0.3 & $0.3-26$ \\
CH$_4$ & CH$_4$ & 14.8 $\pm$ 0.4 & 1.95 $\pm$ 0.05 & $0.15-2.7$ \\
CH$_3$OH & CH$_3$OH & 22.2 $\pm$ 0.7 & 2.93 $\pm$ 0.08 & $<0.13-6.1$ \\
$^{13}$CO$_2$ & $^{13}$CO$_2$ & 1.14 $\pm$ 0.05 & 0.151 $\pm$ 0.006 & $\cdot \cdot \cdot$ \\
C$_2$H$_6$ & H$_2$O & 2.3 $\pm$ 0.4 & 0.31 $\pm$ 0.04 & $0.1-2.7$ \\
HCN & H$_2$O & 1.7 $\pm$ 0.3 &  0.23 $\pm$ 0.05  & $0.03-0.5$ \\
H$_2$CO & H$_2$O & 2.2 $\pm$ 0.2 & 0.29 $\pm$ 0.03 & $<0.02-1.4$ \\
OCS & H$_2$O & 0.44 $\pm$ 0.03 & 0.058 $\pm$ 0.004 & $0.04-0.40$ \\
CN & H$_2$O & 0.66 $\pm$ 0.02 & 0.087 $\pm$ 0.003 & $\cdot \cdot \cdot$ \\
\enddata
\tablecomments{$T$\subs{profile} indicates which rotational temperature profile was used 
to calculate $Q_{x}$: the para-H$_{2}$O profile derived from MIRI or the CO, 
CH$_{4}$, CH$_{3}$OH, CO$_{2}$, or $^{13}$CO$_{2}$ profiles. 
$Q$ is the terminal production rate retrieved from $Q$-curves derived from NIRSPec 
data (see Section~\ref{sec:subsec-trace-results-q-and-t}). Opacity corrections are not included 
and, for water, results from the 5~\micron{} band are used.
The range of molecular abundances for each species measured in comets to date is indicated \citep{Biver2022}.}
\end{deluxetable*}

\vspace{-0.85 cm}
\section{Trace volatile species} 
\label{sec-trace-volatiles}

A wealth of molecular emission is superimposed on the continuum in the NIRSpec spectrum (Figure~\ref{fig:fig-nirspec-three})
of comet C/2017 K2 (PanSTARRS), including from H$_2$O (analyzed above, Section~\ref{sec:sec-watermodelling}), 
CO, CO$_2$, $^{13}$CO$_2$, OCS, CN, H$_2$CO, CH$_3$OH, CH$_4$, C$_2$H$_6$, HCN, NH$_2$, and 
OH$^*$ (prompt emission after UV photolysis of water). CN band emission (multiple lines) near
4.9~\micron{} is clearly detected. These ro-vibrational lines blended 
with the hot band water lines. Detection of CN emission in comet C/2017 K2 (PanSTARRS) at IR 
wavelengths is not totally surprising, as \citet{2023PSJ.....4....8F} present detection of these features 
in comet C/2021 A1 (Leonard). The CN emission also blends with water hot bands and the OCS band, hence
determination of the OCS abundance requires careful examination of the CN contribution.

\subsection{Analysis of Volatile Species}\label{subsec:trace-analysis}

We determined contributions from continuum and gaseous emissions across the NIRSpec
bandpass by adapting procedures previously applied to high-resolution near-infrared spectroscopy 
of comets using ground-based instruments such as iSHELL at the NASA-IRTF and NIRSPEC at the 
W.~M.~Keck Observatory \citep[e.g.,][]{2012JQSRT.113..202V, DiSanti2017, 2023PSJ.....4....8F}. These 
observatories utilize long (15\farcs0 to 24\farcs0) slit capabilities to sample molecular 
emission along a 1D spatial slice of the coma, extracting and averaging molecular 
emission in successive pixels equidistant from either side of the nucleus. For the 3D spatial-spectral 
maps enabled by the NIRSpec IFU capabilities, the logical progression of this technique is to 
extract azimuthally averaged spectra in 1-spaxel (0\farcs1) wide annuli with increasing radius 
from the photocenter, taken to be the position of the comet nucleus. 


\begin{figure*}[ht]
\figurenum{10}
\gridline{\fig{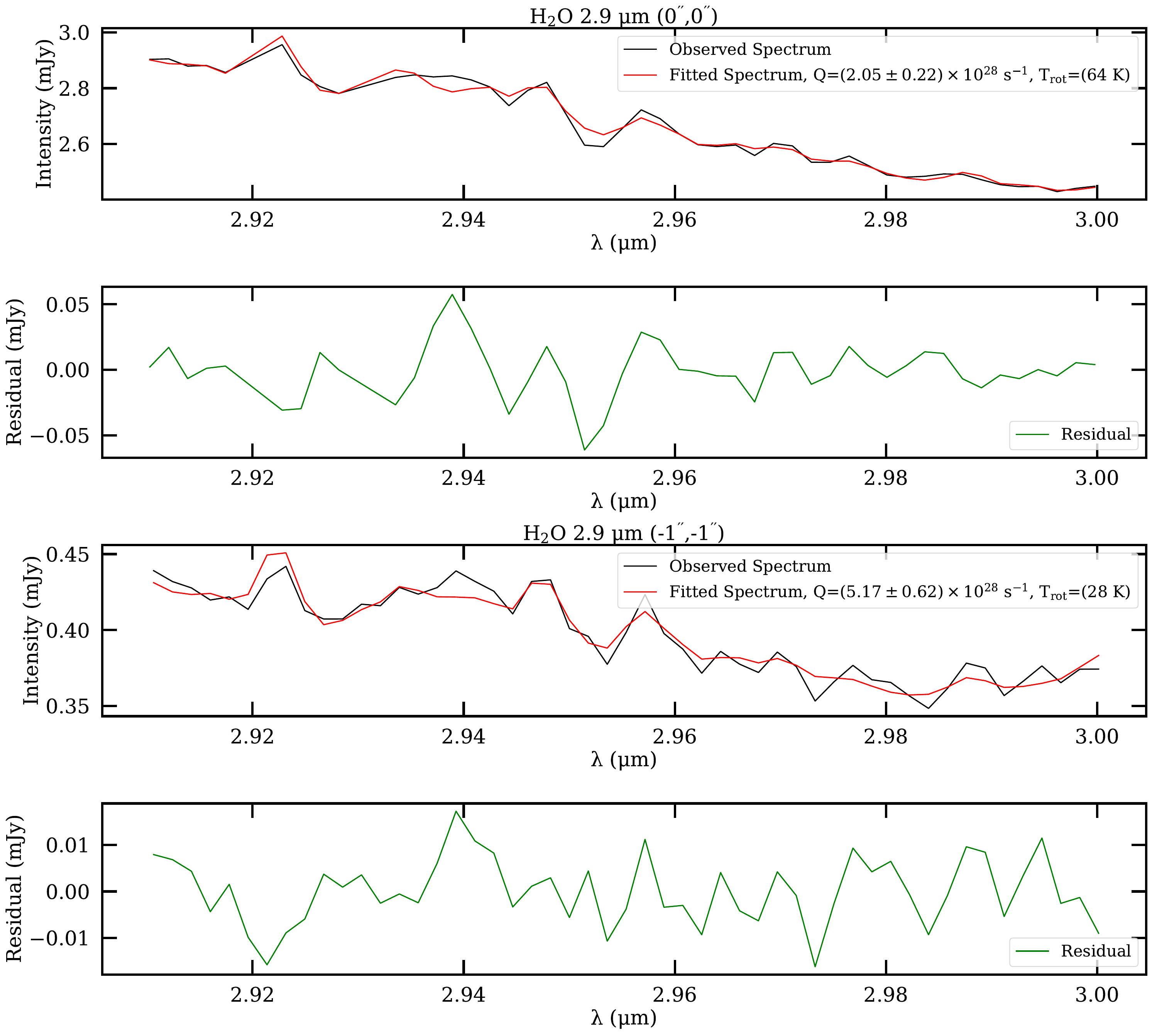}{0.61\textwidth}{(a)} }
\gridline{\fig{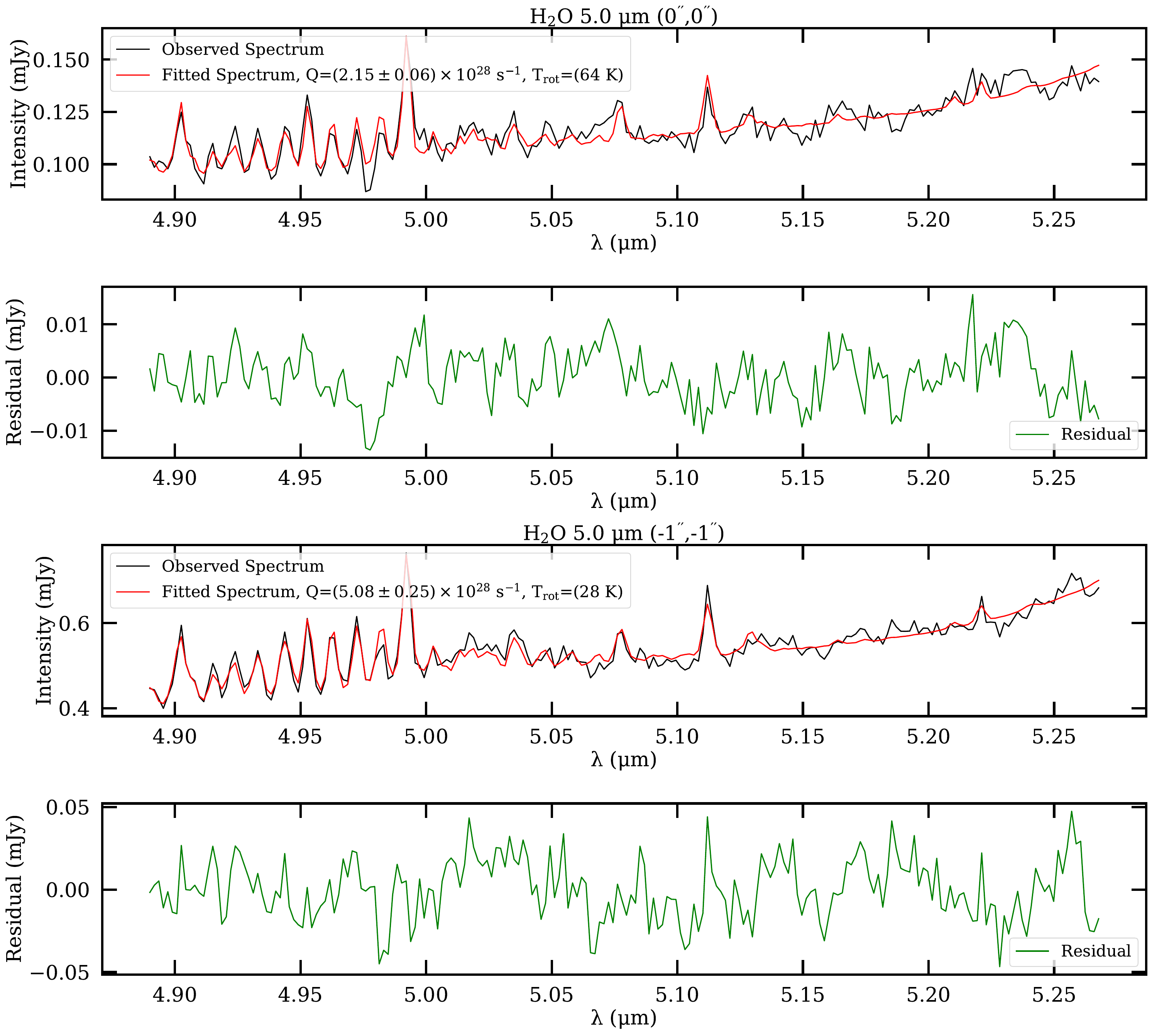}{0.61\textwidth}{(b)} }
\caption{\textbf{(a)--(b).} Model fits to H$_{2}$O 2.9~\micron{} and H$_{2}$O 5.0~\micron{} 
emission derived from the NIRSpec spectra of C/2017 K2 (PanSTARRS). (Top panel pair) The 
extracted on-nucleus spectrum.  (Bottom panel pair) The extracted off-nucleus spectrum (-1:-1) 
position (see Figure~\ref{fig:mrs-beamtile-positions} that describes the spatial location of the 
beams). Measurements and model fits are shown in black and red, respectively, while the 
residual is shown in green. Derived apparent production rate (for H$_{2}$O) or abundance 
relative to H$_{2}$O (for all other species) at each position is indicated. The derived or assumed 
(in parentheses) rotational temperature are listed in the inset. The complete
figure set (10 images), all species H$_{2}$O, CO$_{2}$, $^{13}$CO$_{2}$, CO, CH$_{4}$, CH$_{3}$OH, 
HCN, OCS, and C$_{2}$H$_{6}$ are available in the figure set of the online-journal. 
\label{fig:nathan10-extractions}
}
\end{figure*}
\clearpage


We used the NASA PSG to simultaneously fit continuum (described by a cubic spline) and 
molecular emission in discrete wavelength regions where each volatile has strong, relatively 
unblended emission, as is particularly important at the relatively 
low ($\lambda/\Delta\lambda\sim 1,000$) resolving power 
of JWST compared to ground-based platforms ($\lambda/\Delta\lambda\sim 10^4$). 
These are show in Figure~\ref{fig:nathan10-extractions}. The NASA PSG uses 
the Optimal Estimation method to retrieve molecular production rates and rotational temperatures 
using state-of-the-art synthetic fluorescence models for each targeted species, with detailed 
information provided in the NASA PSG Handbook \citep{2022fpsg.book.....V}.

Molecular emission in the innermost coma is affected by the NIRSpec PSF, and CO$_{2}$ 
emission is likely optically thick \citep{2015SSRv..197...47B}. The NASA PSG includes a 
correction for optical depth effects, which is most accurate at small solar phase angles 
\citep[for more details, see][]{2022fpsg.book.....V, 2023PSJ.....4..172R}.
Given the modest ($\sim25^{\circ}$) phase angle for comet C/2017 K2 (PanSTARRS) 
during our observations, and for consistency with the H$_{2}$O MRS analysis, we calculated 
molecular production rates using NASA PSG models without correcting for opacity (optically 
thin). To prevent the application of the NASA PSG opacity correction, we divided the observed 
molecular flux by a factor of 1,000 to approximate coma column densities consistent with 
optically thin conditions. As molecular production rate is linearly proportional to flux, the retrieved 
NASA PSG production rates were multiplied by 1,000 to recover the optically thin values. 
Newer, revised NASA PSG options now available offer alternative switches (e.g., FlourThin options);
however, full exploration of these evolving capabilities is beyond the scope of this manuscript.

Rotational temperature analysis is feasible for molecules having intrinsically bright lines 
and for which a broad range of excitation energies is sampled. In addition to para-H$_{2}$O (MRS), 
these conditions were satisfied for CO, CH$_{4}$, CH$_{3}$OH, CO$_{2}$, and $^{13}$CO$_{2}$ 
in the NIRSpec spectrum of comet C/2017 K2 (PanSTARRS). We calculated molecular 
production rates for the remaining species (whose rotational temperature profile could not 
be well-constrained) by assuming the profile of rotational temperature vs.\ nucleocentric distance
measured for para-H$_{2}$O with MRS. As the dominant coma volatile, H$_{2}$O sets the 
temperature profile in the inner coma and is an appropriate substitute when rotational 
temperatures cannot be derived for other species \citep[e.g.][]{2012ApJ...750..102G, 2018AJ....156..251R}. 

Given the larger spatial scale ($0\farcs13$) of the MRS pixels, we interpolated the spatial
dependence of the MRS H$_{2}$O rotational temperature and applied the appropriate value 
to each annulus extracted from the NIRSpec spectra. Owing to the likely high opacity of CO$_{2}$ in 
the inner coma, we also calculated its molecular production rate assuming the $^{13}$CO$_{2}$ 
rotational temperature profile.

\subsection{Production Rate and Rotational Temperature Profiles}
\label{sec:subsec-trace-results-q-and-t}

\begin{figure}[h!]
\figurenum{11}
\begin{center}
\includegraphics[trim=0.15cm 0.20cm 0.25cm 0.00cm, clip, width=0.48\textwidth]{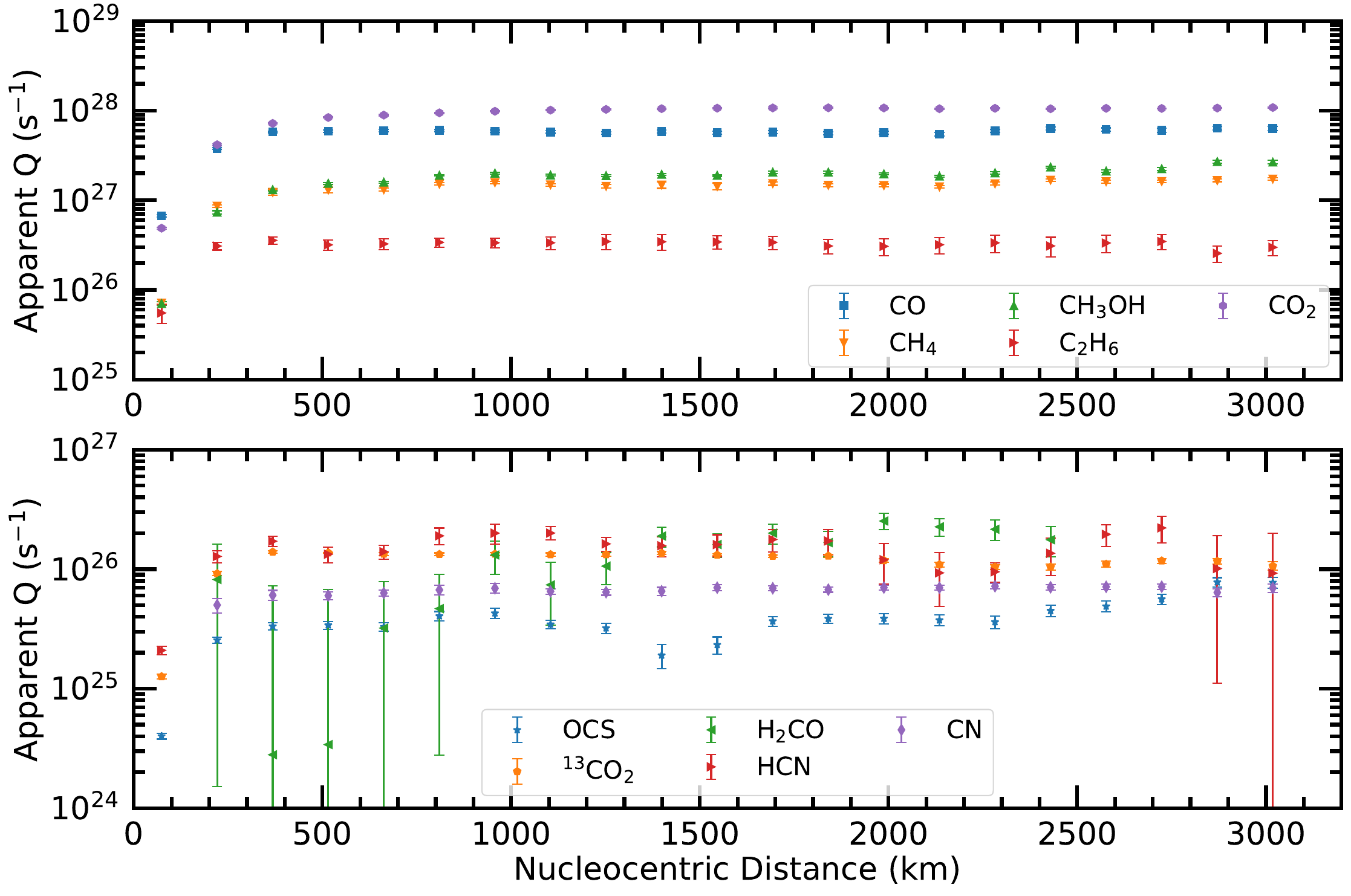}
\caption{Apparent molecular production rates ($Q$) as a function of nucleocentric distance for CO, 
CH$_{3}$OH, CO$_{2}$, CH$_{4}$, and C$_{2}$H$_{6}$ (upper panel) and for OCS, 
H$_{2}$CO, $^{13}$CO$_{2}$, HCN, and CN (lower panel) derived from NIRSpec observations
of comet C/2017 K2 (PanSTARRS) and modeled with the NASA PSG.  }
\label{fig:fig-nirspec-qs}
\end{center}
\end{figure}

\begin{figure}[h]
\figurenum{12}
\begin{center}
\includegraphics[trim=0.15cm 0.20cm 0.25cm 0.25cm, clip, width=0.46\textwidth]{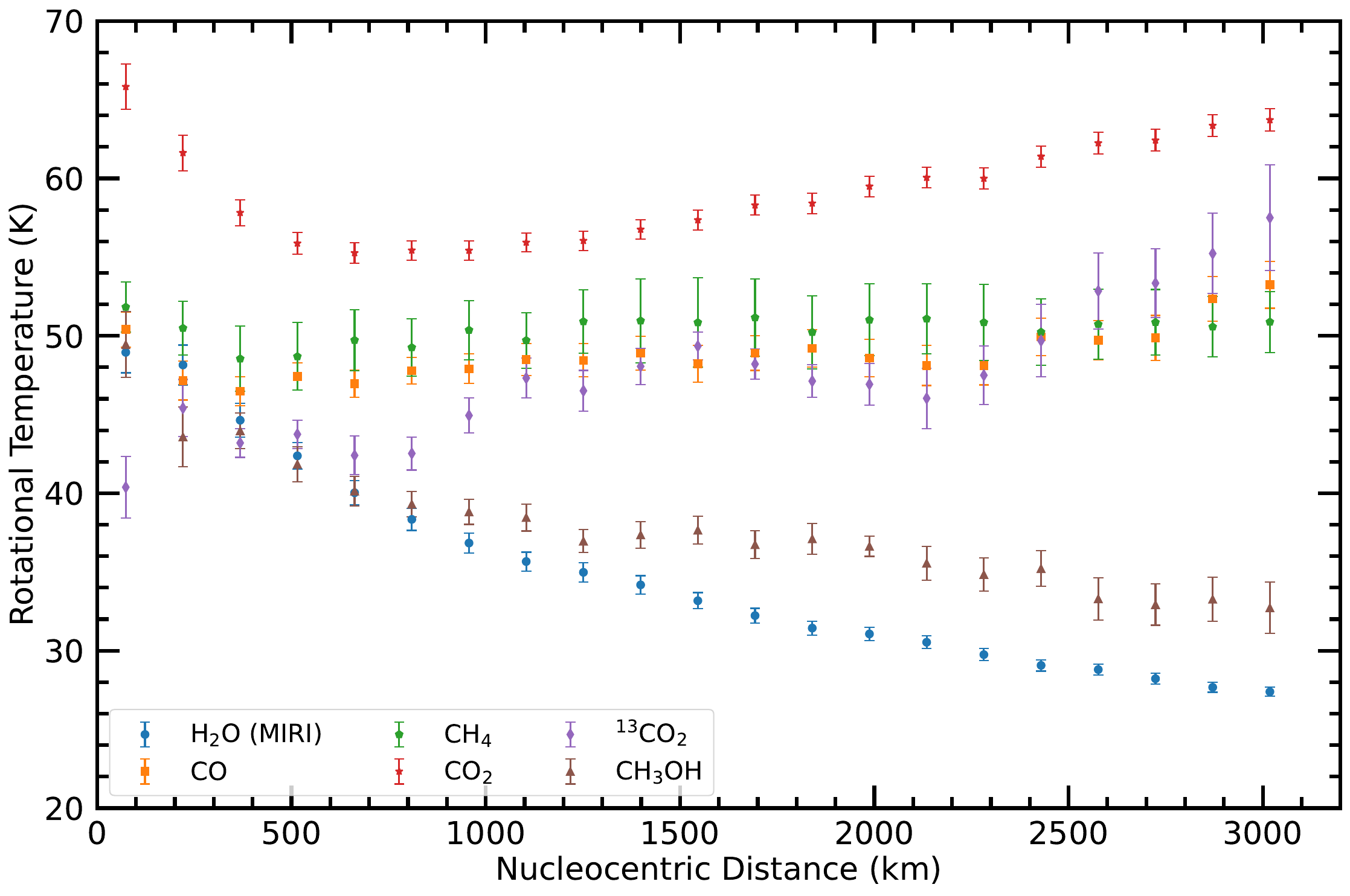}
\caption{Comet C/2017 K2 (PanSTARRS) rotational temperature profiles for H$_{2}$O ($\nu_{2}$ band 
para lines measured with MRS) along with CO, CO$_{2}$, $^{13}$CO$_{2}$, CH$_{4}$, and CH$_{3}$OH 
(measured with NIRSpec).}
\label{fig:fig-nirspec-temps}
\end{center}
\end{figure}

Figure~\ref{fig:fig-nirspec-qs} shows the molecular production 
rates retrieved by the NASA PSG for trace species measured with NIRSpec. As expected, 
the apparent $Q$'s are suppressed in the innermost annuli owing to opacity and the PSF, and increase 
with increasing nucleocentric distance until reaching a ``terminal value'' in coma regions 
where opacity effects no longer dominate. We calculate this terminal value to be the 
weighted average of five annular extracts once the 
``$Q$-curve'' \citep[e.g.,][]{1998Icar..135..377D, 2005PhDT.......258B, 2016ApJ...820...34D} has 
leveled \cite[see also][Figure 11 where this methodology is explained]{2011Icar..216..227V}.  
Table~\ref{tab:nirspec-qs} provides our molecular production rates and mixing ratios for 
each species. Figure~\ref{fig:fig-nirspec-temps} compares the rotational temperature profiles
derived for CO, CH$_{4}$, CH$_{3}$OH, CO$_{2}$, and $^{13}$CO$_{2}$ compared 
against that for para-H$_{2}$O from MRS.

The rotational temperatures measured for CO, CH$_{4}$, CH$_{3}$OH, $^{13}$CO$_{2}$, 
and H$_{2}$O are consistent within their respective $2\sigma$ uncertainties in the innermost 
coma, similar to results found from previous high-resolution ground-based studies of comets 
at near-infrared wavelengths \citep[e.g.,][]{2012ApJ...750..102G}. However, CO$_{2}$ is considerably 
higher because it is affected by opacity in the innermost coma and our models are unable to provide 
a good fit for spectra in the first few annular extracts. Therefore, we also calculated 
$Q$($^{12}$CO$_{2}$) using the temperature profile of $^{13}$CO$_{2}$, which was less affected by opacity
(Figure~\ref{fig:fig-12c-13c}).

However, the rotational temperatures for CO, CH$_{4}$, CO$_{2}$, and $^{13}$CO$_{2}$ 
show a consistent flat dependence with increasing nucleocentric distance, whereas CH$_{3}$OH 
shows a decreasing trend similar to (although shallower than) H$_{2}$O. This can be understood 
in terms of the differing radiative lifetime ($\tau_{R}$) for each molecule. In the inner coma, 
collisions with H$_{2}$O and electrons create a thermal equilibrium, gradually breaking 
down to fluorescence equilibrium in the outer coma, with the speed of departure from local 
thermodynamical equilibrium depending on the radiative vs.\ collisional rates of the particular 
molecule and band \citep{1994A&A...287..647B}. In the collisional coma, the gas is thermalized 
and T$_{\rm{rot}}$(K) of the molecules provides a direct diagnostic of the kinetic temperature. For molecules 
with smaller dipole moments (and thus longer radiative lifetimes) such as CO, CH$_{4}$, CO$_{2}$, 
and $^{13}$CO$_{2}$, rotational temperatures should remain relatively flat to greater nucleocentric 
distances before significant non-LTE effects begin to occur \citep{2022ApJ...929...38C}. H$_{2}$O 
and CH$_{3}$OH, with considerably larger dipole moments, radiatively cool much more efficiently.

\begin{figure*}[ht]
\figurenum{13}
\gridline{\fig{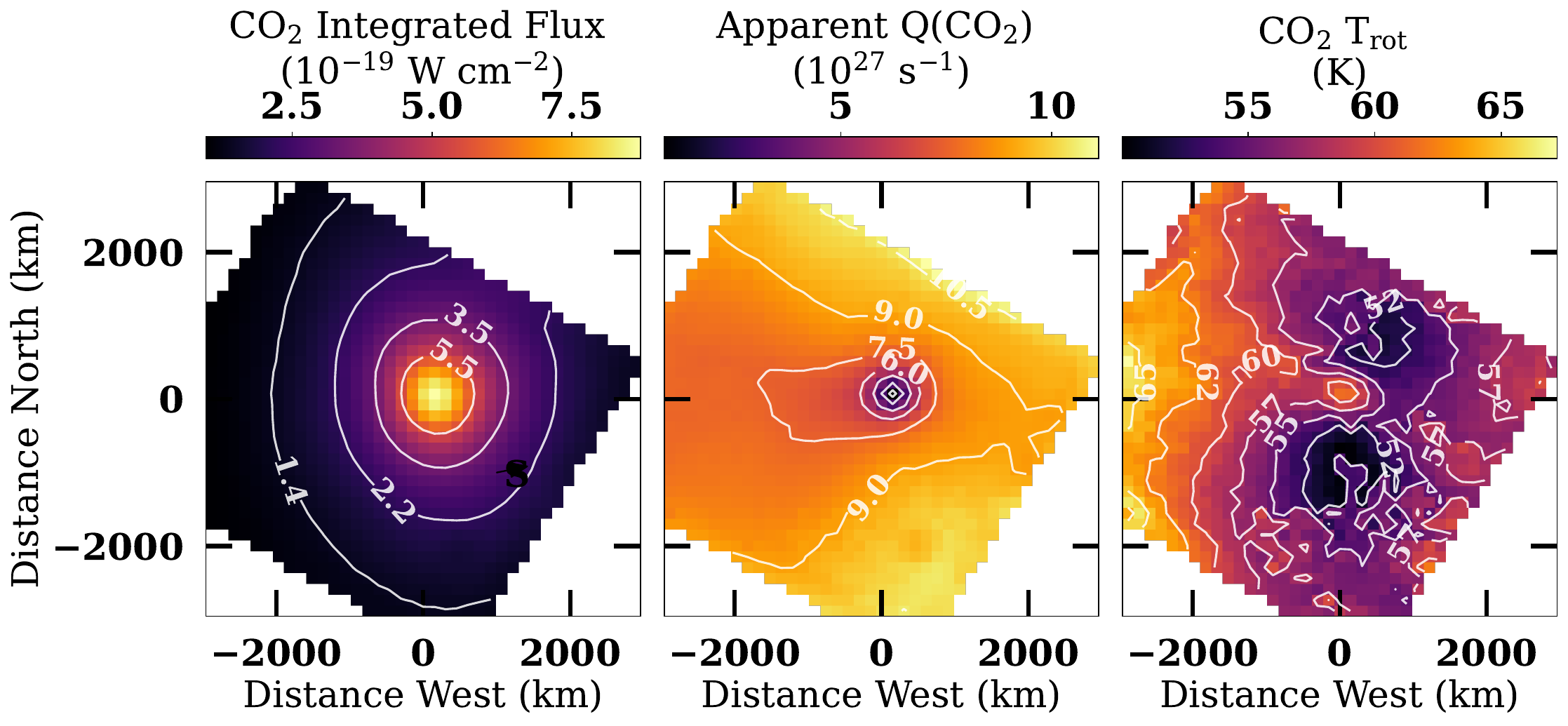}{0.48\textwidth}{(a)}
          \fig{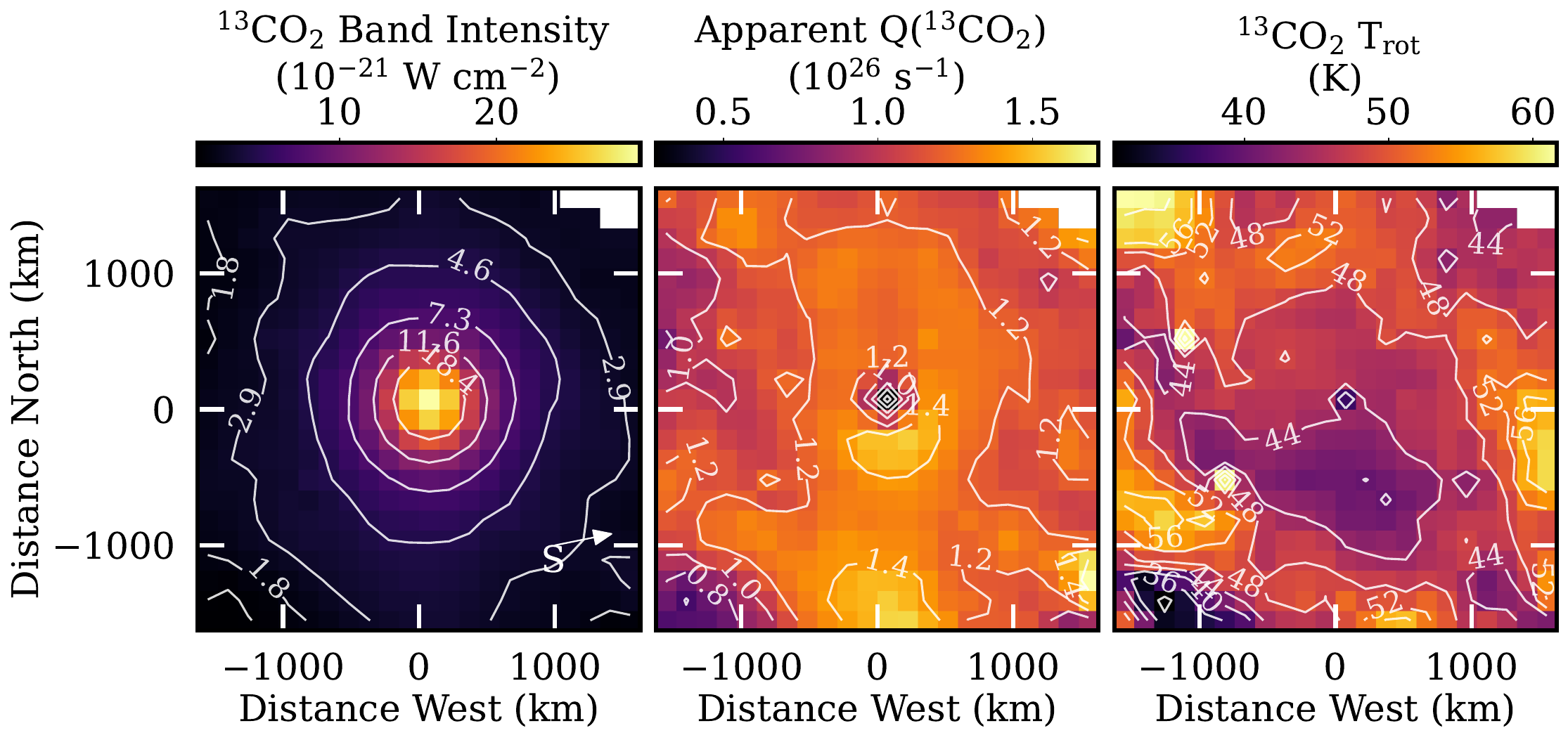}{0.48\textwidth}{(b)}
	}
\gridline{\fig{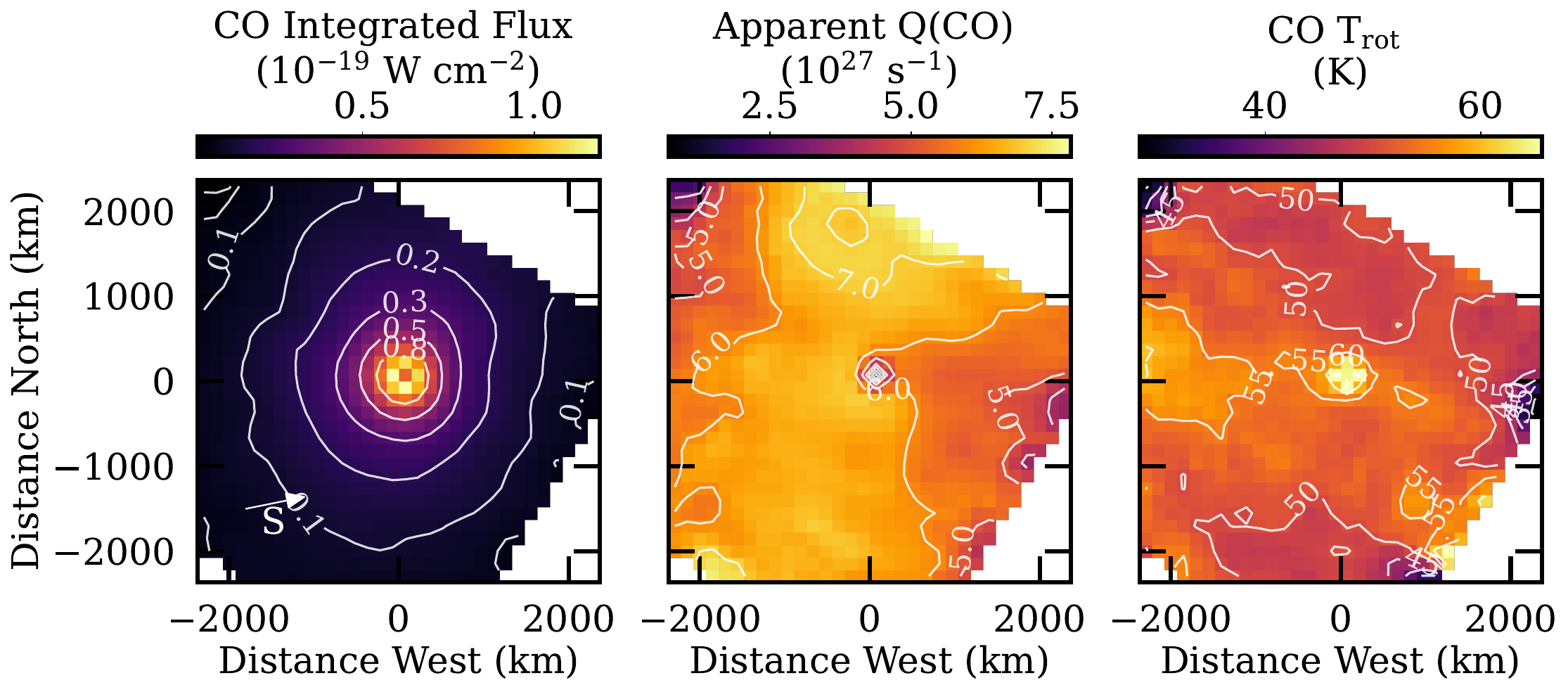}{0.48\textwidth}{(c)}
          \fig{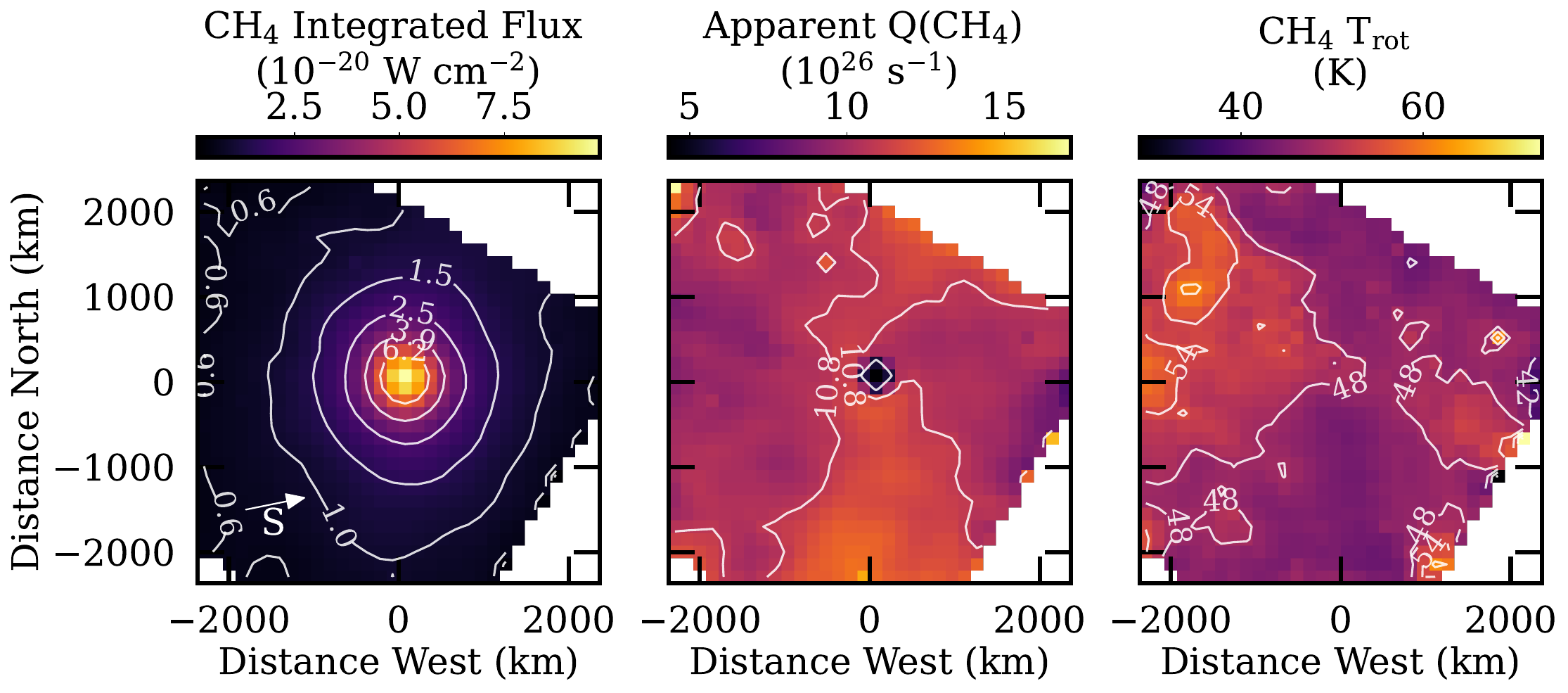}{0.48\textwidth}{(d)}
	}
\gridline{\fig{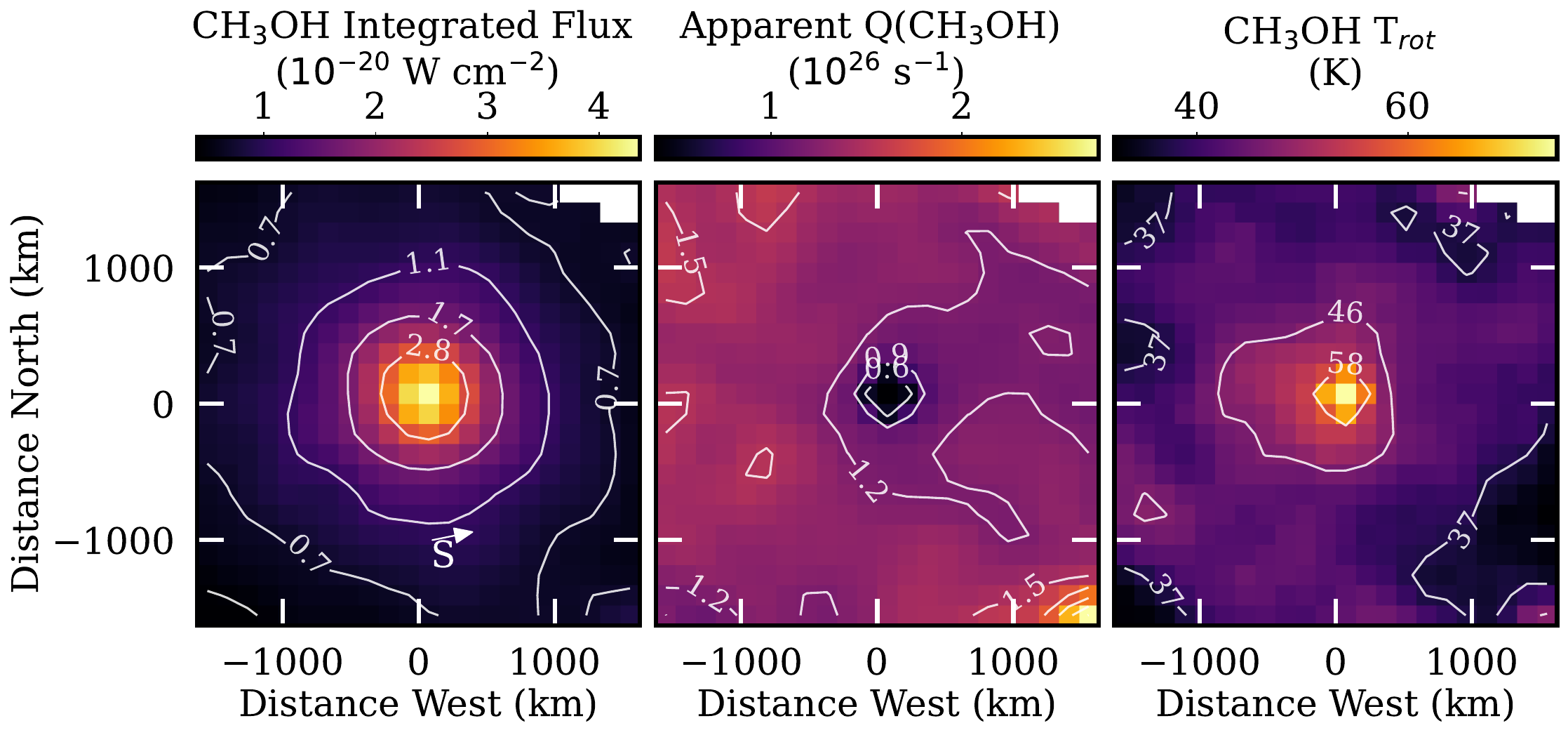}{0.48\textwidth}{(e)}
          \fig{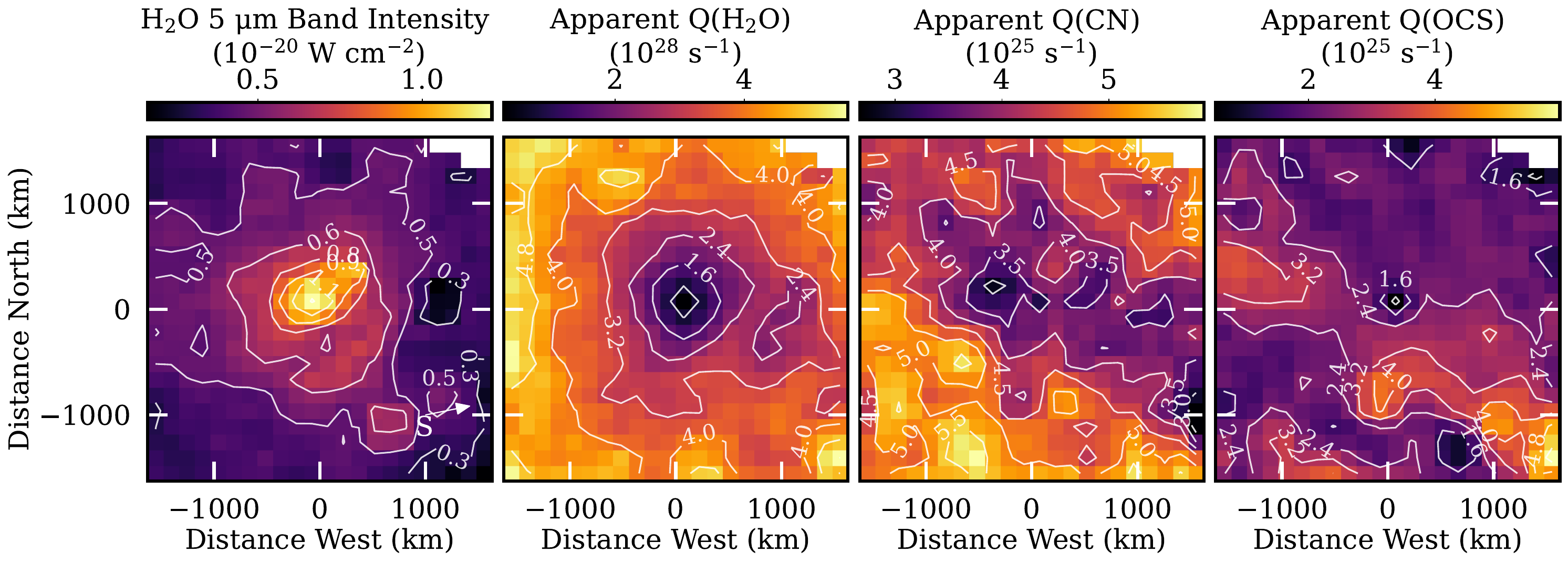}{0.52\textwidth}{(f)}
	}
\caption{\textbf{(a)--(e).} Maps of continuum-subtracted band intensity, apparent production 
rate, and rotational temperature for CO$_{2}$, $^{13}$CO$_{2}$, CO, CH$_{4}$, and 
CH$_{3}$OH. \textbf{(f)} Map of continuum-subtracted intensity for H$_{2}$O, along with 
apparent production rate for H$_{2}$O, CN, and OCS.
\label{fig:maps}}
\end{figure*}

\begin{figure}[h!]
\figurenum{14}
\begin{center}
\includegraphics[trim=0.05cm 0.15cm 0.25cm 0.25cm, clip, width=0.45\textwidth]{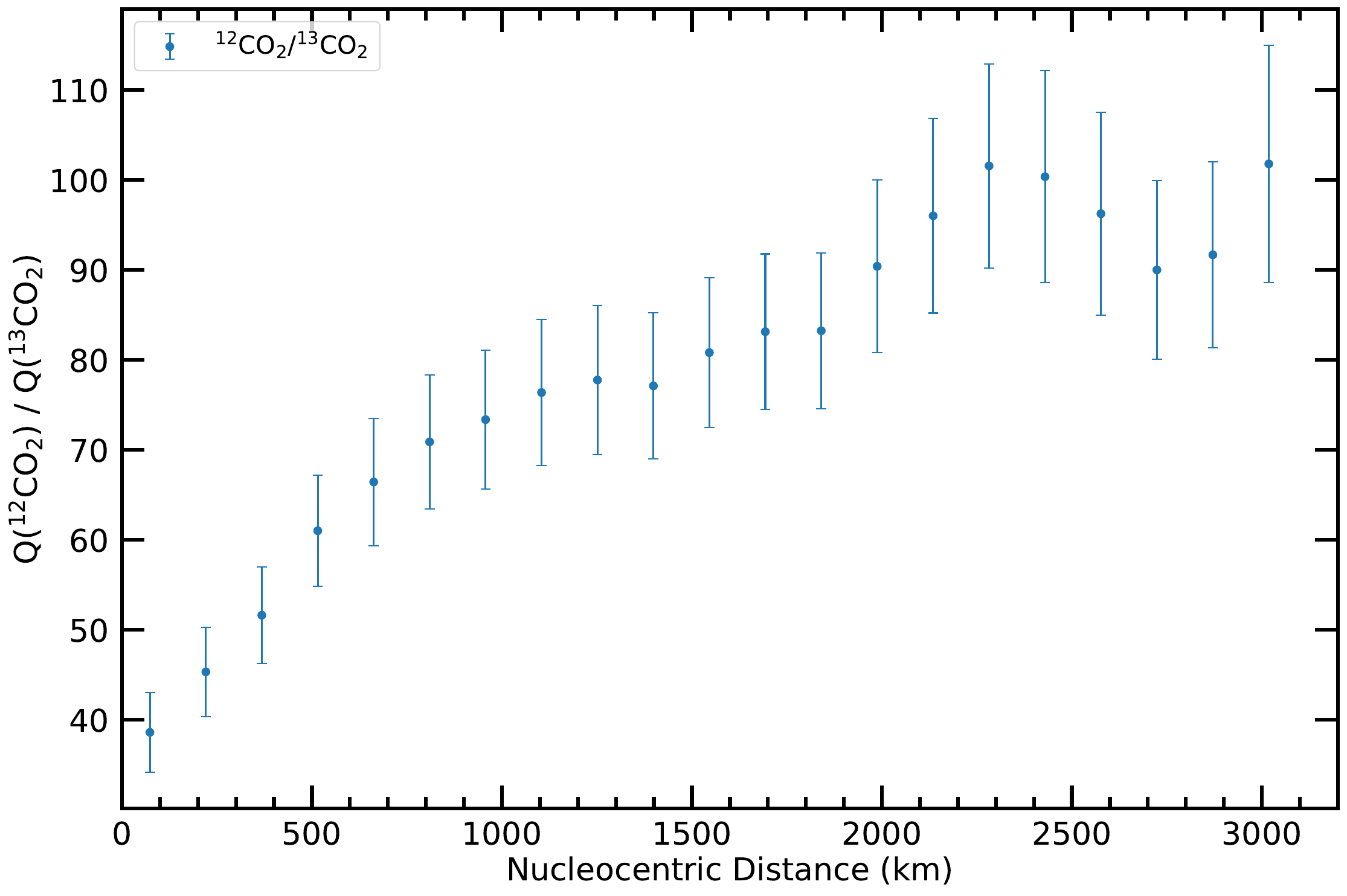}
\caption{The radial distribution of the $^{12}$CO$_{2}$ to $^{13}$CO$_{2}$ ratio in the inner coma
of comet C/2017 K2 (PanSTARRS) as a function of nucleocentric distance.}
\label{fig:fig-12c-13c}
\end{center}
\end{figure}

Further insight into the outgassing dynamics and physics of the coma can be gleaned by 
analyzing the spaxel-by-spaxel distribution of molecular abundances and (for a subset of these) 
their rotational temperatures. Previous work \citep{2011JGRE..116.8012V, 2016Icar..266..152D} has shown that 
molecules having common or distinct spatial distributions are reflective of the association or segregation of 
ices in the nucleus. Figures~\ref{fig:maps}a through \ref{fig:maps}e show maps of continuum-subtracted measured 
band intensities, apparent production rates, and rotational temperature for CO$_{2}$, $^{13}$CO$_{2}$, CO, 
CH$_{4}$, and CH$_{3}$OH. 

CO and CH$_{4}$ show remarkably similar abundance and temperature spatial distributions, 
perhaps from jets perpendicular to the Sun-comet line. 
Although CO$_{2}$ shows a similar spatial distribution to CO and CH$_{4}$, its temperature 
distribution is significantly warmer in the anti-sunward direction. This may be explained by less 
efficient adiabatic cooling owing to reduced collision rates in the anti-sunward coma, which 
has been previously observed in coma thermal studies of comet 46P/Wirtanen 
\citep{2021A&A...648A..49B, 2023ApJ...953...59C}.
CH$_{3}$OH shows a relatively smooth spatial distribution, with a rotational temperature 
that quickly decreases with increasing nucleocentric distance. The H$_{2}$O hotbands 
show a spatial distribution consistent with that measured by MRS H$_{2}$O 
maps, Figure~\ref{fig:H2O-maps}. CN (as expected for a product species) shows considerable 
enhancement off-nucleus. Finally, OCS (although at lower signal-to-noise) shows a 
relatively isotropic distribution. Collectively, these results demonstrate a complex 
coma for comet C/2017 K2 (PanSTARRS).

\subsection{Comparison with Ground-Based Near-Infrared Measurements}\label{sec:sec-ejeta}

\citet{2025AJ....169..102E} measured C/2017 K2 (PanSTARRS) using the NASA IRTF 3-m telescope
on 2022 August 19, 20, and 21, detecting (common to our JWST study) H$_{2}$O, CO, C$_{2}$H$_{6}$, 
CH$_{4}$, CH$_{3}$OH, and HCN and deriving upper limits for OCS. Owing to the use
of a unique echelle setting each date to obtain a full compositional makeup with IRTF, 
H$_{2}$O was only detected on August 20, with Q(H$_{2}$O) = $(3.65 \pm 0.66)\times10^{28}$~s$^{-1}$. 

This is roughly a factor of two lower than our Q(H$_{2}$O) = $(7.6 \pm 0.2)\times10^{28}$~s$^{-1}$ 
measured with NIRSpec, yet consistent with Q(H$_{2}$O) measured with MIRI. Similarly, 
their abundances relative to H$_{2}$O are roughly twice as high as ours. However, comparisons 
between the absolute molecular production rates are more nuanced: our studies are 
consistent within 1$\sigma$ uncertainty for Q(CO), Q(CH$_{4}$), and Q(HCN), yet disagree 
for Q(CH$_{3}$OH) and Q(C$_{2}$H$_{6}$). 

Our JWST study utilized the same models from the NASA PSG to analyze H$_{2}$O hot 
band emission near 2.9~$\mu$m and 5.0~$\mu$m as \cite{2025AJ....169..102E}, 
eliminating the possibility of significant differences in the fluorescence models. Similarly, 
the fact that measured Q's for CO, CH$_{4}$, and HCN are consistent between the 
two studies indicates that differences in methodology are not a contributing factor. 
As noted in Section~\ref{sec:results_waterband_analysis}, an investigation 
reconciling the $Q$(H$_{2}$O) measured at various bands and facilities is the subject of a future work.
Our finding of distinct outgassing activity for gases associated with H$_{2}$O versus CO$_{2}$ 
from JWST observations is consistent with results from the Rosetta mission to 
comet 67P/Churyumov-Gerasimenko \citep[e.g.,][]{2016Icar..277...78F, 2023MNRAS.526.4209R}.

\subsection{Trace volatiles compared with other comets}\label{sec:sec-trace-2-other-comets}

Molecular abundances in comet C/2017 K2 (PanSTARRS) can be compared against their 
respective mean or median among the comet 
population \citep{2016Icar..266..152D, 2022PSJ.....3..247H, 2020AJ....160..184S}: CO, 
CH$_{3}$OH, HCN, and CO$_{2}$ are consistent, C$_{2}$H$_{6}$, OCS, and H$_{2}$CO 
are depleted, and CH$_{4}$ is enriched. The simultaneous measure of CO$_{2}$ and 
$^{13}$CO$_{2}$ affords a tantalizing opportunity to investigate the $^{12}$C/$^{13}$C ratio. 
However, caution must be used when considering the significant opacity of the CO$_{2}$ band 
in the inner coma. With these comments in mind, we compared our measured $Q$($^{12}$CO$_{2}$) and 
$Q$($^{13}$CO$_{2}$) radial profiles for C/2017 K2 (PanSTARRS) and show these in Figures~\ref{fig:fig-12c-13c}. 
At nucleocentric distances were $Q$'s are no longer affected by the PSF (and for $^{12}$CO$_{2}$, 
opacity), our values are consistent with the terrestrial $^{12}$C/$^{13}$C = 89, similar to that found for 
other comets, including 67P/Churyumov-Gerasimenko \citep{2015SSRv..197...47B, 2022A&A...662A..69M}.

\begin{deluxetable*}{ccccc}
\tablenum{3}
\setlength{\tabcolsep}{20pt} 
\tablecaption{Modeled Ice Sublimation Rates, Active Areas, and Active Fractions of Comet C/2017 K2 (PanSTARRS)\label{tab:activity}}
\tablewidth{0pt}
\tablehead{
\colhead{Molecule}
& \colhead{$Q$}
& \colhead{$Z$}
& \colhead{$A$}
& \colhead{$f$} \\
& \colhead{(molecules s$^{-1}$)}
& \colhead{(molecules s$^{-1}$ m$^{-2}$)}
& \colhead{(km$^2$)}
& \colhead{(\%)}
}
\startdata
H$_2$O & $7.6\times10^{28}$ & $3.9\times10^{20}$ & 190 & $>$86 \\
CO$_2$ & $1.1\times10^{27}$ & $1.2\times10^{21}$ & 9.3 & $>$4 \\
CO & $6.2\times10^{27}$ & $4.2\times10^{21}$ & 1.5 & $>$0.7 \\
\enddata
\tablecomments{$Q$ is the measured molecule production rate (see Table~\ref{tab:nirspec-qs}), $Z$ is the ice sublimation 
rate per unit area, $A$ is the effective active area required for the production rate, and $f$ is the effective nuclear 
active fraction given a radius of $<$4.3 km.}
\end{deluxetable*}

\section{Activity}\label{sec:activity}

The production rate of gas produced by sublimation depends on the quantity of the 
ice and its temperature.  Simple models for comets generally assume spherical nuclei 
with surfaces in instantaneous equilibrium considering absorbed sunlight, thermal 
emission, and the latent heat of sublimation.  We use such a model 
\citep{1979M&P....21..155C} to compute the sublimation rates of water, 
CO$_2$, and CO for a nucleus with a 5\% visual Bond albedo and 
95\% infrared emissivity at the observed heliocentric distance of C/2017 K2 (PanSTARRS).  
The sublimation rates are compared to the measured production rates and used to compute the 
effective active surface area of the ice.  The ratio of the active area to the nucleus area defines the 
nuclear active fraction.  Most comets have water ice active fractions $\lesssim10$\%, but 
some comets have high active fractions, $>50$\%, and even exceeding 100\%, a phenomenon 
referred to as hyperactivity.

Table~\ref{tab:activity} lists the model sublimation rates, calculated active 
areas, and active fractions.  Our measured nucleus radius is an upper limit, 
therefore the active fractions are given as lower limits.  With a water ice active 
fraction of $>$86\%, comet C/2017 K2 (PanSTARRS) is a hyperactive comet.

\section{The Dust Inventory}  \label{sec:sec-dust}

Harker-thermal model code \citep{2007Icar..190..432H, 2002ApJ...580..579H, 2023PSJ.....4..242H} 
was used to fit the JWST MRS 7-to-27~\micron{} SEDs of comet C/2017 K2 (PanSTARRS). This model code constrains
the relative mass fractions that are quoted for the submicron- to micron-size portion of the particle differential 
size distribution. The model employs dust properties  (via optical constants) and  relative mass fractions of five 
primary dust compositions commonly used to interpret remote sensing observations of these materials in comet
comae. These include (as model free parameters): 
amorphous silicates including Mg:Fe = 50:50 amorphous olivine (\textit{AO50}) and Mg:Fe = 50:50 amorphous pyroxene 
(\textit{AP50}), and crystalline silicates of Mg = 90 crystalline olivine \textit{CO} and Mg =100 
crystalline ortho-pyroxene (\textit{CP}), as well as the particle differential size distribution parameters and the 
particle (dust grain) porosity. 

Our thermal models are similar to many forward models in the literature 
\citep[e.g.,][]{2005Icar..179..158M, 2005Sci...310..274S, 2015ApJ...809..181W, 2021PSJ.....2...25W, 2023PSJ.....4..242H}. 
The model flux is the sum over the particle differential size distribution of fluxes of individual 
discrete composition particles computed for their radiative equilibrium temperatures with
incident sunlight \citep{2024come.book..577E}. The absorptivities ($Q_{\rm{abs}}$) of the 
Mg:Fe = 50:50 amorphous silicates and amorphous carbon particles are computed with 
Mie plus effective medium theory \citep[mixing in vacuum with Bruggeman mixing formulae,][]{1983asls.book.....B} 
for a fractal porosity (the fractal porosity prescription is of the form 
$f = 1 - (a/0.1\, \micron)^{D-3}$, where $D$ is the fractal dimension parameter that ranges from
$D = 3$, solid to $D = 2.5$, fractal porous as described in \citet{2023PSJ.....4..242H} ) 
and with optical constants from, respectively, \citet[][and references therein]{1995A&A...300..503D} and
\citet{1983PhDT........18E}. For solid crystalline particles, $Q_{\rm{abs}}$ are computed using 
continuous distribution of ellipsoids (CDE) using optical constants for \textit{CP} (Mg = 100) from 
\cite{1998A&A...339..904J}.  CDE also is used for \textit{CO} and we employ our hot crystal model that 
was developed to fit comet C/1997~O1~(Hale-Bopp) \citep{2007Icar..190..432H}. 
Specifically, to generate radiative equilibrium temperatures for forward modeled fluxes at the 
specific $r_{\rm{h}}$ (au) and submicron radii of C/1997~O1~(Hale-Bopp), Maxwell-Garnett effective 
medium theory \citep[e.g.,][and references therein]{1990ApJ...355..680L} is 
used to mix (``dirty'') 0.2 to 2.0~\micron{} optical constants 
for the crystallographically non-oriented\footnote{\url{https://www2.astro.uni-jena.de/Laboratory/OCDB/crsilicates.html}}  
(Stubachtal) olivine with amorphous carbon with 
mixing parameter 0.043, which has the effect of increasing $Q_{\rm{abs}} \simeq 74$ in the ultra-violet to 2~\micron\
spectral region, thereby warming the crystals to the deduced radiative equilibrium temperatures; these optical 
constants define the ``hot crystal'' model \citep[for details see][]{2007Icar..190..432H}. 

\begin{figure}[h!]
\figurenum{15}
\begin{center}
\includegraphics[trim=0.05cm 0.05cm 0.2cm 0.05cm, clip, width=0.45\textwidth]{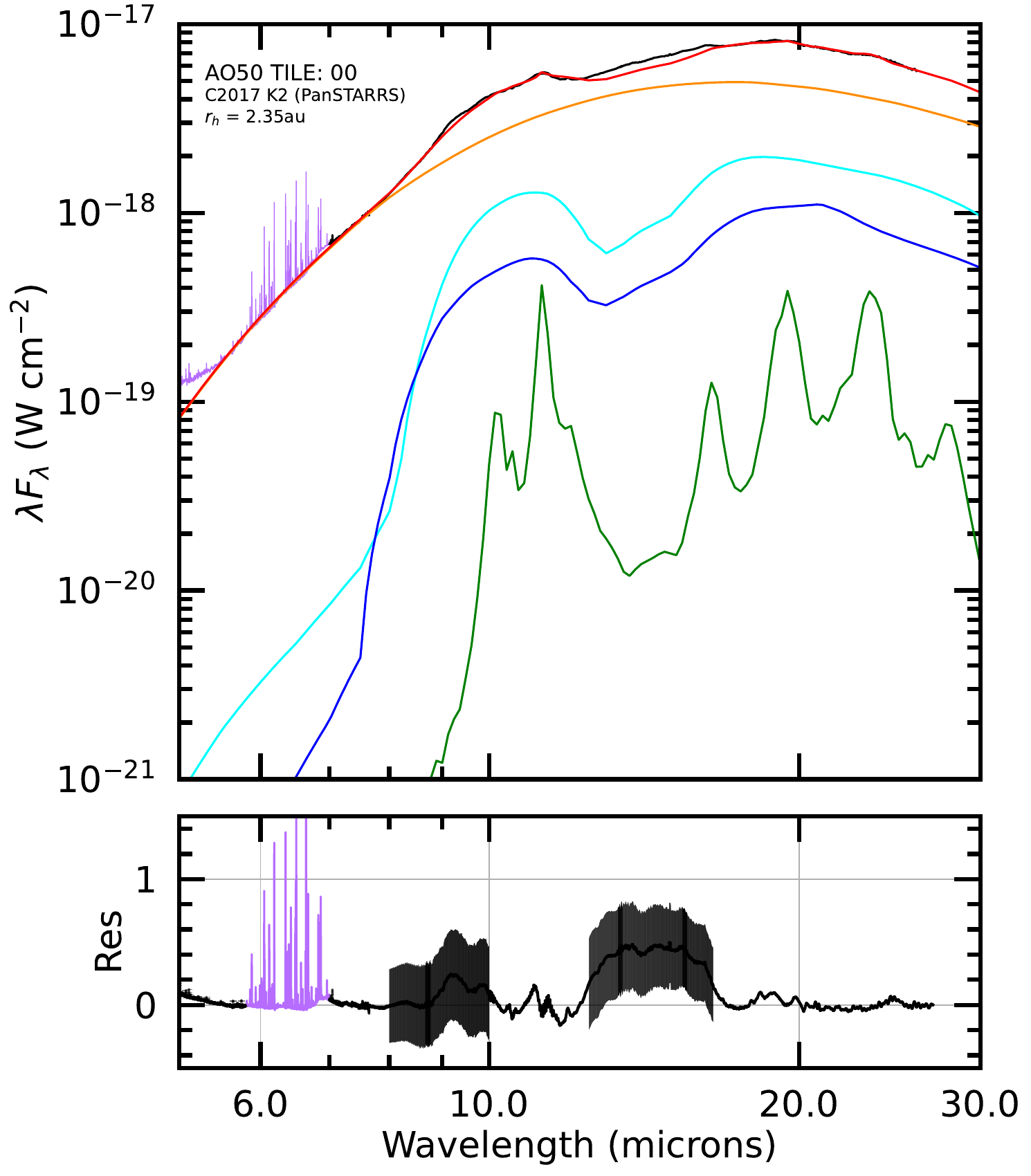}
\caption{(Top panel). The \textbf{`AO50'} thermal model fitted to the JWST MRS IR SED  in a 1\farcs0 diameter 
aperture centered on the photocenter (position 0:0) of comet C/2017 K2 (PanSTARRS). 
The models fitted to the observed 7.0 - 27~\micron{} SED constrain the dust mineralogy, assuming 
the coma refractory dust species have optical constants similar to that of amorphous carbon
(orange line), amorphous olivine (Mg:Fe = 50:50, cyan line), amorphous pyroxene (Mg:Fe = 50:50, blue line), 
Mg-crystalline olivine (Forsterite, green line), and when present, Mg-crystalline ortho-pyroxene 
(Enstatite, pink line). At position 0:0 enstatite was excluded from the best fit model.
The solid red line is the best-fit model composite spectra, superposed on the observed data (black solid curve) 
that was modeled ($\lambda \geq 7.0$~\micron) and the observed data that was omitted
from the thermal models (purple solid line) where strong water emission is dominant.
(Bottom Panel) The model residual (observed data - best fit thermal model) is in the bottom panel. 
The spectral region containing the $\nu_{2}$ water emission bands (purple, bottom panel) 
were not included in the model fit, and the grey vertical regions illustrated the point-point 
uncertainties (discussed in Section~\ref{sec:sec-dust-model-3cases}). Uncertainties were multiplied by 
40 in wavelength regions spanning 8.0 to 10.0~\micron{} and 12.5 to 16.5~\micron{} (grey vertical lines) 
that were shown by application of the Akaike Information 
Criterion \citep[AIC;][]{doi:10.1177/0049124104268644, 2007MNRAS.377L..74L} to yield the better 
fit for the \textbf{`AO50'} thermal model (see Section~\ref{sec-model-assessment}).
The complete figure set for all tile positions (7 images) is available in the online journal.
 }
\label{fig:fig-mrs-model00}
\end{center}
\end{figure}

The models constrain [$a_p, N, D$], where $a_p$(\micron)
is derived peak particle radii of the model Hanner grain size distribution \citep{1994ApJ...425..274H},  $N$ 
is the differential size distribution slope for the larger radii grains, and $D$ is the fractal dimension of
the parameter, as the amorphous porous grains are assumed to be hierarchical aggregates described 
by a factional porosity description. Specifically, in addition to the grain size distribution parameters the 
thermal model constrains the relative mass fractions of the amorphous silicates, crystalline silicates and 
amorphous carbon. The high signal-to-noise and broad spectral coverage afforded by JWST enabled some 
more nuanced modeling of the MRS spectra across the coma of comet C/2017 K2 (PanSTARRS)
to be conducted (Section~\ref{sec:sec-dust-model-3cases}).  Figure~\ref{fig:fig-mrs-model00} 
illustrates a best fit thermal model (as described in detail in Section~\ref{sec:sec-dust-ao50ac50}) 
from beam tile centered on the photocenter peak of the comet's surface brightness (tile position 0:0). 

The high spatial resolution of JWST spectral cubes, combined with the instrumental sensitivity to faint surface
brightness emission permit the study of dust properties of comet C/2017 K2 (PanSTARRS) to
be investigated by tiling synthetic aperture beams across the coma (Figure~\ref{fig:fig-three}).
In the thermal modeling fitting to the MRS SEDs, particular attention is paid to the far-infrared features to 
assess the relative abundances of Mg-rich crystalline olivine or Mg-rich crystalline pyroxene. Also, 
particular attention is paid to the short wavelength (Wein side) of the thermal emission, especially 
in a 1\farcs0 diameter aperture centered on the coma photocenter (tile position 0:0, Figure~\ref{fig:mrs-beamtile-positions}), 
because a non-negative residual, i.e., (data - model), is needed to properly assess the scattered light contribution 
(see Section~\ref{scattered-light-3to8}) over the range of wavelengths where the thermal emission onset
 occurs and to assess organic features in the 5 to 9~\micron{} region as discussed in Section~\ref{sec-pah-models-dw}.

A thermal model is fit to the observed spectra at each beam-tile position using a least-squares method 
as described in detail in \citep{2023PSJ.....4..242H}, where the compositional uncertainties are constrained 
using a Monte Carlo (bootstrap) analysis. The tabulated compositional uncertainties 
(Table~\ref{tab:compositon-modelparams}) and derived mass fractions (Table~\ref{tab:mod-massfractions}) 
are given for the inner 95\% confidence level for each parameter.


\subsection{Dust composition dependence} \label{sec:sec-dust-ao50ac50}

Thermal modeling of the coma in a 1\farcs0 diameter aperture centered on the coma photocenter 
(tile position 0:0, Figure~\ref{fig:mrs-beamtile-positions}) using all MRS data points 
at $\lambda \geq 7.0$~\micron{} results in the \textit{AC} component of the thermal model (which dominates 
the flux at the shortest wavelengths) producing too much flux at 5 to 7~\micron. We label this the first order thermal model. 
Hence, there are negative residuals at  $\lambda \leq 7.0$~\micron.  There are also residual emissions 
in the 14.7~\micron{} region that are not predicted by the thermal model, which also is the case for a 
half dozen Spitzer-observed comets  \citep{2023PSJ.....4..242H}.  Al-O bonds occur in this region of the spectrum in contrast to 
Si-O bonds of silicates. This residual emission in comet C/2017 K2 (PanSTARRS) is discussed in detail in 
Section \ref{sec:sec-residuals9314}.

In order to prevent the 14.7~\micron{} feature from biasing the fits, the MRS formal uncertainties in the region 
of the 14.7~\micron{} emission are enlarged by a factor of 40 to lessen the weight of these data points. 
The data with these revised uncertainties are fitted with the thermal model but 
the model still results in over-prediction of the short wavelength flux and a negative residual. The short-wavelength 
shoulder of the `10~\micron{}' silicate feature is sought to be fitted with these choice of uncertainties and the 
wavelength position of this shoulder is associated with amorphous pyroxene. For simplicity this set of models is 
called `favor AP50' or simply \textbf{`AP50'}. 

Models are also fitted to data with uncertainties multiplied by a factor of
40 for both the $\simeq 14.5$~\micron{} region (specifically 12.5 to 16.5~\micron) and the short wavelength 
shoulder of the `10~\micron{}' silicate feature (specifically 8.0 to 10.0~\micron). This set of models is called 
`favor AO50' or simply \textbf{`AO50'} because the far-infrared spectral shape of comet C/2017 K2 (PanSTARRS) is 
better fitted with amorphous olivine and the rise of the `10~\micron{}' silicate feature in the model is allowed, by the 
increased uncertainties, to occur at the longer wavelengths that are characteristic of amorphous 
olivine \citep[see][]{2023arXiv230503417E}.

 \subsection{Thermal model assessment} \label{sec-model-assessment}
 
We applied the Akaike Information Criterion \citep[AIC;][]{doi:10.1177/0049124104268644, 2007MNRAS.377L..74L} 
to compare these models (i.e., the first order thermal model, the \textbf{`AP50'} model, and the \textbf{`AO50'} model) for 
the dust emission in the coma comet C/2017 K2 (PanSTARRS).  AIC is a quantitative statistical framework 
with great utility in the assessment of  the relative suitability of two different model analyses of the same
spectrum, or model analyses of two renditions of the spectra \citep[e.g.,][]{2023PSJ.....4..242H}
which is the case for comet C/2017 K2 (PanSTARRS). 

There are aspects of the thermal emission observed in the SED of comet C/2017 K2 (PanSTARRS) that 
are not well represented by the thermal model. However, data points in those spectral regions are 
given less weight in the model-fitting via the $\chi^{2}$-minimization and then the resulting AIC value is significantly 
lower even with the logLikelihood's `partition function term'  that penalizes for the increase in the 
uncertainties, $\Sigma _i \left( 2~ln(\sigma_i)\right)$, \cite[see][Eq. 5.47]{2014sdmm.book.....I}. Often this 
partition function term is omitted from cited versions of the logLikelihood relation ($ln~\mathcal{L}$) and hence from the 
AIC equation as models fitted to the same data (same uncertainties), result in this term cancelling each 
other when two AIC values are subtracted. 
The expression for AIC \citep{2014sdmm.book.....I} is: 
$AIC=-2~ln\mathcal{L} + 2N_{param}$  where $-2~ln\mathcal{L} = \chi^2 + 2\sum_{i=1}^{N_{points}}ln \sigma_{i} + \frac{N_{points}}{2}ln2\pi$.
The best-fit model will be the model with the lowest 
AIC (AIC$_{min}$). The most probable model will be the model with the lowest AIC.
Taking position by position, AIC$_{min}$ always occurs for model \textbf{`AO50'}. 

For all the positions bean tile positions (e.g., Figures~\ref{fig:mrs-beamtile-positions}, \ref{fig:fig-three})
compared to the\textbf{`AO50'} models, the \textbf{`AP50'} models have relative 
probabilities less than $e^{-7399}$, and for the  first order thermal model the relative probabilities 
are less than $e^{-130341}$. In other words, the Signal-to-Noise ratios in JWST MRS data are so 
superb that the AIC assessments unambiguously show that the \textbf{`AO50'} models better fit the data. 
However, for completeness an example \textbf{`AP50'} thermal model SED decomposition
is provided in the Appendix~\ref{sec:appendix-0}. 

\subsection{Three model cases for detailed SED analysis} \label{sec:sec-dust-model-3cases}

The \textbf{`AO50'} model for the spectrum in 1\farcs0 diameter aperture centered on the coma photocenter 
(tile position 0:0, Figure~\ref{fig:mrs-beamtile-positions})  predicts slightly too much flux 
in the H$_{2}$O-emission spectral region. However, this model is successful enough to justify the 
fitting of the scattered light by a line of constant slope in reflectance of the 
combined NIRSpec and MRS data. At this juncture, three different cases for thermal modeling 
of the observed emission are considered. 

\begin{itemize}

\item Case A (\textbf{`AO50'}):~The thermal model is fitted to observed MRS data from 7 to 26~\micron{} 
that has the water lines (water model) removed. This is the model that is applied to all other 
positions besides 0:0.
 
\item Case B (\textbf{`AO50'}):~The scattered light ($\lambda \leq 5$~\micron) is extrapolated to the 
thermal infrared ($\lambda \gtsimeq 4.9$~\micron), subtracted from the observed MRS data, and 
the thermal model is fitted to these data for the wavelength range from 7.0 to 26~\micron{} with an 
estimate of the continuum under the 6 to 8.6~\micron{} region which is dominated by organic features in the 
residual (see Section \ref{sec-pah-models-dw}). 

\item Case C (\textbf{`AO50'}):~The scattered light ($\lambda \leq 5$~\micron) is extrapolated to the thermal 
infrared ($\lambda \gtsimeq 4.9$~\micron), subtracted from the observed MRS data, and the thermal 
model is fitted to these data for the wavelength range from 4.9 to 26~\micron{} with an estimate of 
the continuum under the 6 to 8.6~\micron{} region which is dominated by organic features in the residual 
(see Section \ref{sec-pah-models-dw}). 

\end{itemize}

\begin{figure}[h!]
\figurenum{16}
\begin{center}
\includegraphics[trim=0.25cm 0.25cm 0.25cm 0.05cm, clip, width=0.45\textwidth]{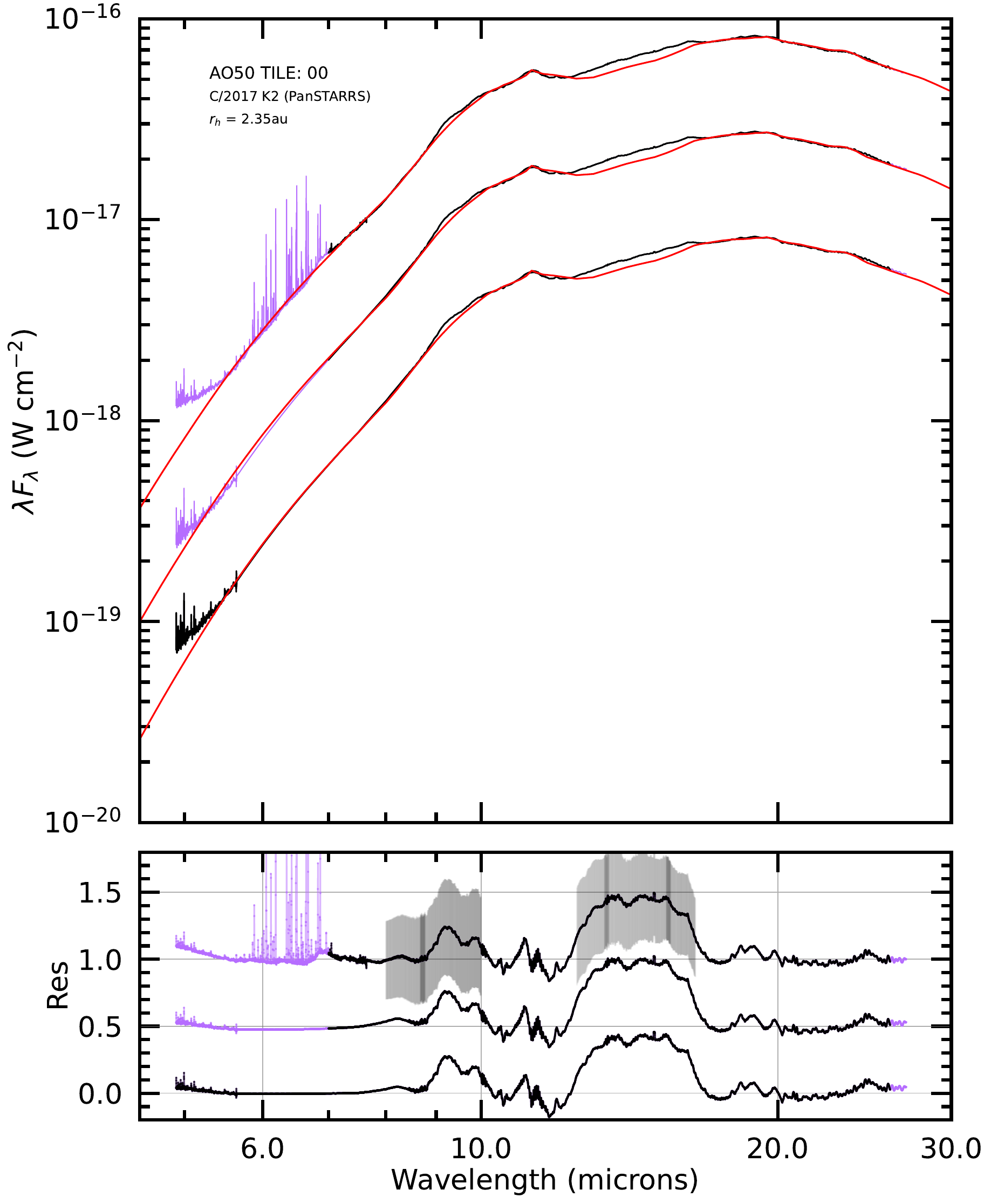}
\caption{The three  \textbf{`AO50'} model for Case~A, 
Case~B, and Case~C thermal model fits (Section~\ref{sec:sec-dust-model-3cases}). (Top panel) The top curve is Case~A,
the middle curve is Case~B, and the bottom curve is Case~C. The inset explains the details of each case and the red curve
in each instance is the best model fit. The purple denotes regions where water emission bands are present. (Bottom panel)
The residuals (fit data - the best fit thermal model). 
 }
\label{fig:fig-three-case-models}
\end{center}
\end{figure}

\begin{figure}[b]
\figurenum{17}
\begin{center}
\includegraphics[trim=0.25cm 0.25cm 0.25cm 0.25cm, clip, width=0.45\textwidth]{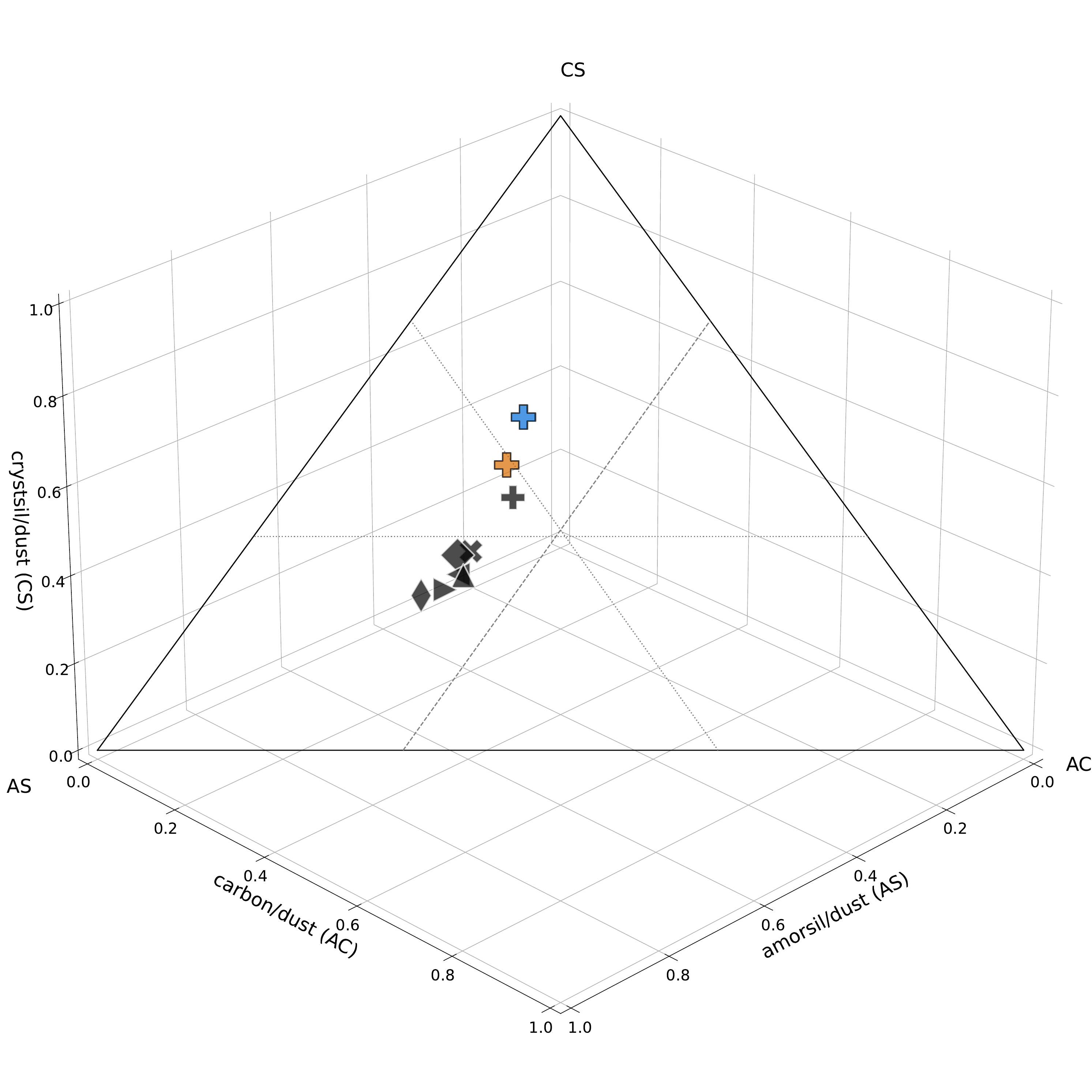}
\caption{The ternary diagram for the \textbf{`AO50'} model dust composition in the coma of comet C/2017 K2 
(PanSTARRS). The black cross is the dust composition derived for Case A; the orange cross is the 
dust composition derived for Case B; and the blue cross designates the dust composition derived from 
Case C. The filled black symbols represent the dust composition of the 7 MRS beam tile position derived 
from Case A treatment.
 }
\label{fig:fig-ternary}
\end{center}
\end{figure}

Models resulting from these three cases are illustrated in Figure~\ref{fig:fig-three-case-models}.
For the position 0:0, the dust compositions for \textbf{`AO50'} model for Case A, 
Case B, and Case C (plus symbols in Fig.~\ref{fig:fig-ternary}) differ by an increase in the mass fraction 
of \textit{CO} and hence an increase in the silicate crystalline mass faction for the submicron
to micron-size portion of the grain size distribution,  $f_{cryst} \equiv m_{cryst}$ / ($m_{amorphous} + m_{cryst}$)
where $m_{cryst}$ is the mass fraction  of the submicron crystalline materials, for thermal models fitted 
with scattered light subtracted, while the \textit{AC} mass fraction remains relatively 
unchanged between these three cases of \textbf{`AO50'} model. 

\subsection{Dust composition and size distribution across the coma} \label{sec:sec-ao50tile-reveal}

\begin{figure}[h]
\figurenum{18}
\begin{center}
\includegraphics[trim=1.5cm 1.5cm 1.5cm 2.0cm, clip, width=0.50\textwidth]{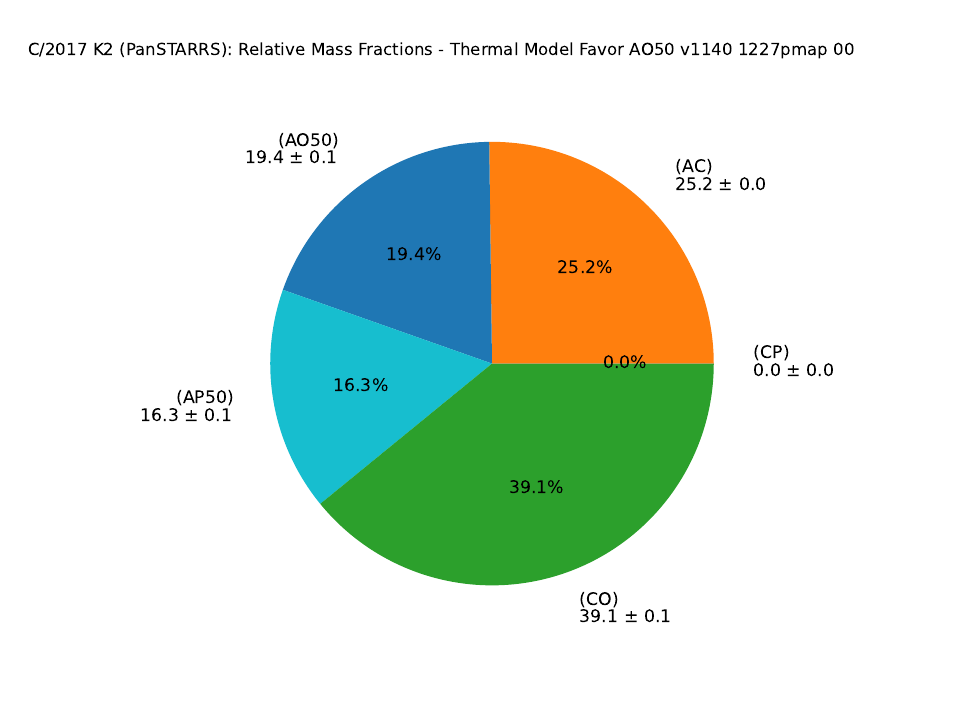}
\caption{The relative mass fraction 
for the submicron- to micron-size portion of the particle 
differential size distribution in the inner coma of comet C/2017 K2 (PanSTARRS) derived from the \textbf{`AO50'} 
Case C models. Tile position 0:0, comet photocenter beam position (see Figure~\ref{fig:mrs-beamtile-positions}. 
The compositional designation illustrated in the figure are: amorphous carbon (\textit{AC}), amorphous 
silicates (\textit{AO50, AP50}), and crystalline silicates (\textit{CO, CP}).  The \textit{AC} content of the coma grains 
is significant. The complete figure set (7 images) for each individual tile position (1\farcs0 diameter 
beam) is available in the online journal.
 }
\label{fig:fig-dustpie}
\end{center}
\end{figure}

The \textbf{`AO50'} set of Case A models shows that the composition of the dust in the inner coma is similar 
amongst the seven positions yet there are differences, Figures~\ref{fig:fig-ternary} and \ref{fig:fig-dustpie}.  
 Figure~\ref{fig:fig-ternary} demonstrates that all positions (black symbols) not 
centered on the nucleus have the same composition. All positions in the coma have four of the five 
compositions: \textit{AC, AO50, AP50,} and \textit{CO}. However, there is one position (+1:+1) that does has a small 
contribution from \textit{CP} ($\ltsimeq 0.06$). All positions have similar \textit{AC} values with (defined as mean $\pm$ standard 
deviation) of 0.247 $\pm$ 0.010, similar silicate-to-carbon ratios 3.049 $\pm$ 0.163, and $f_{cryst} = 0.384 \pm 0.065$ 
(Tables~\ref{tab:compositon-modelparams}, \ref{tab:mod-massfractions}) . The ratio of amorphous 
olivine to amorphous pyroxene has a wider range with $1.992 \pm 1.171$, Figure~\ref{fig:fig-dustpie}.
The coma center 0:0  position has a higher $f_{cryst} = 0.52$, whereas for other positions
$f_{cryst} = 0.361 \pm 0.034$. 

Towards the photocenter of the comet (position 0:0) the particle size distribution peaks at larger particle radii ($a_p$)
of 0.8~\micron, and has as stepper slope of 3.9. In the sunward direction, the  $a_{p}$ are all 0.5 to 0.6~\micron{} 
and the slope are 3.6 to 3.4 suggesting possibly fragmentation of the grains that make the size 
distribution more shallow. In the anti-sunward direction the trend of decreasing slope ($N$) is even more 
prominent. Investigation of grain fragmentation effects would require detailed 
dynamical \citep{2009Icar..203..571R, 2021MNRAS.504.4687F, 2024A&A...688A.177K}  
modeling which is beyond the scope of this manuscript.

\begin{figure*}[ht!]
\figurenum{19}
\begin{center}
\includegraphics[trim=0.20cm 0.25cm 0.20cm 0.20cm, clip, width=0.70\textwidth]{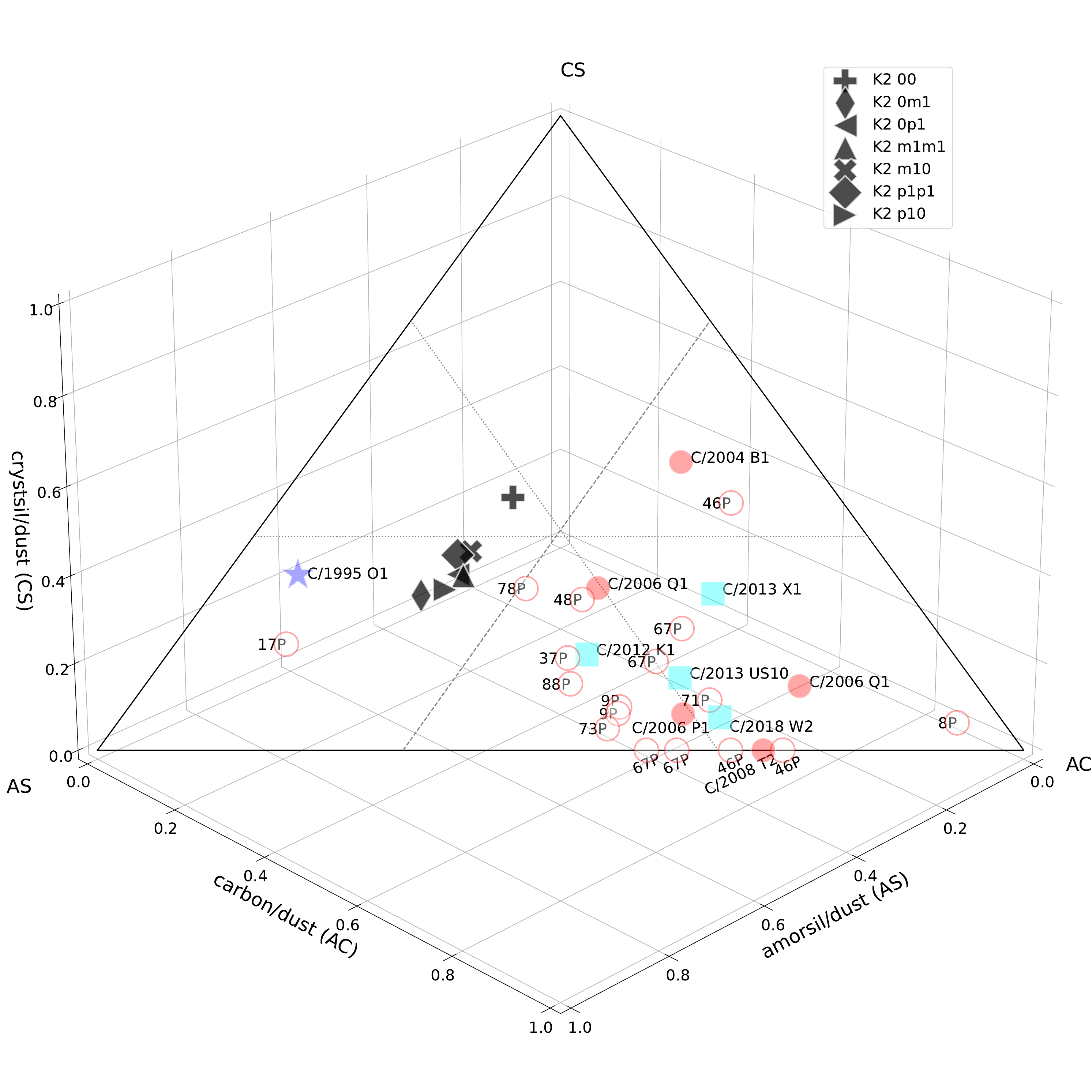}
\caption{The ternary diagram of comets showing the three-dimensional dependence
along the axes of amorphous carbon (\textit{AC}), amorphous silicates (\textit{AC}) and crystalline silicate (\textit{CS}) dust
composition. The filled black symbols represent the dust composition of the 7 JWST MRS beam tile position 
derived from Case A treatment of the JWST SEDs (see Section~\ref{sec:sec-dust-model-3cases}).
Filled-symbols are Oort Cloud Comets (OCCs) and open-symbols are Jupiter Family comets (JFCs). 
The purple filled star is that of OCC comet C/1995~O1~(Hale-Bopp) 
\citep{2002ApJ...580..579H, 2021PSJ.....2...25W}. The circles are values 
derived from thermal models of OCCs and JFCs from the Spitzer observed 
full-wavelength-coverage SEDs \citep{2023PSJ.....4..242H}.  The filled-cyan squares 
are values derived from SOFIA SEDs of OCCs \citep{2021PSJ.....2...25W}.
}
\label{fig:fig-tenary-comets-comparisions}
\end{center}
\end{figure*}

\begin{figure*}[ht!]
\figurenum{20}
\begin{center}
\includegraphics[trim=0.25cm 0.25cm 0.25cm 0.10cm, clip, width=0.70\textwidth]{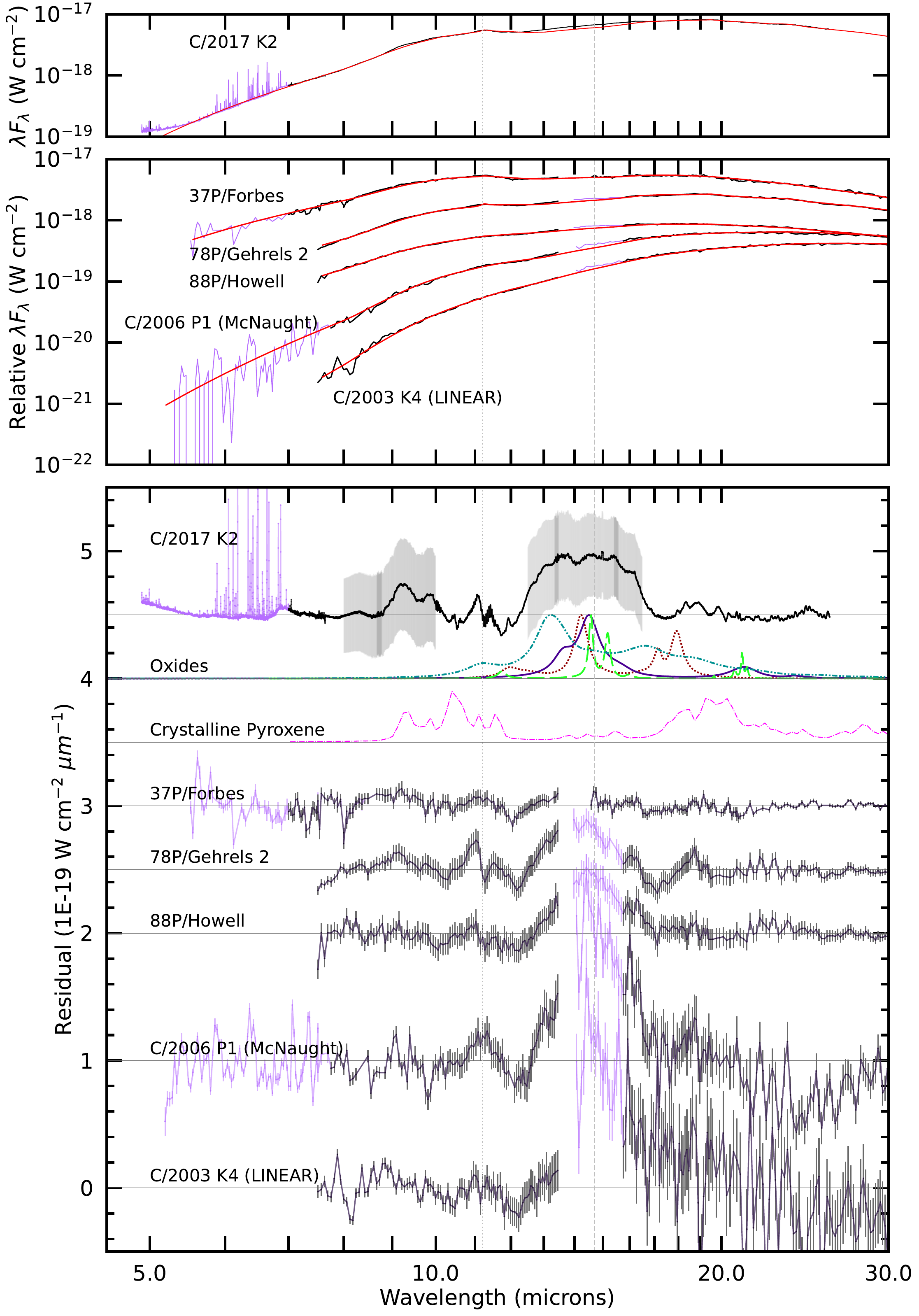}
\caption{The 14~\micron{} residual feature in comet C/2017 K2 (PanSTARRS) compared to 
other comets observed with Spitzer \citep{2023PSJ.....4..242H} that span this spectral range 
in the SED. The blue dotted vertical line in the middle panel is at 14.0~\micron. The SEDs in 
the middle panel are taken from the model compilations of Spitzer observed 
comets \citep{2023PSJ.....4..242H}, where the fluxes are relative, scaled to that
of comet C/2017 K2 (PanSTARRS) for comparison. The bottom panel shows the 
residual (observed SED - model SED) to highlight residual features present at 9.3~\micron{} 
and $\simeq 14.7$~\micron{}. The top residual plot is that of comet C/2017 K2 (PanSTARRS). 
Immediately below that black solid line (which is the residual) are 4 curves of various 
oxides (see text Section~\ref{sec:sec-residuals9314}) scaled to 0.5 at their peak of a blackbody function (F$^{bb}$) of 
F$^{bb}[T_{\rm{BB}}] \times Q_{\rm{abs}}$, where  $T_{\rm{BB}}$ is the derived blackbody 
temperature of the model fit at position `0:0' and $Q_{\rm{abs}}$ are the absorption efficiencies of common
characteristic minerals  commonly found in CAIs \citep{1996Sci...272.1316G}. The plum-colored dash-dot line shows that
dust with a crystalline pyroxene composition cannot produce the observed $\simeq 14$~\micron{} 
residual nor the features in the residual near 9.3~\micron{} and 10.0~\micron{} because the other 
resonances of \textit{CP} are absent from the residual at the contrast predicted by thermal model for 
\textit{CP} for C/2017 K2 (PanSTARRS) (scaled from the position +1:+1 \textbf{`AO50'} model Case A).
}
\label{fig:fig-spitzer-comps}
\end{center}
\end{figure*}


\subsection{Comparisons with other comets} \label{sec:sec-dust-comparison}

The relative abundances of amorphous carbon, amorphous silicates, and crystalline silicates in comet 
C/2017 K2 (PanSTAARS) reveal a coma that is generally more silicate rich than most of the comet population \citep{2023PSJ.....4..242H}.

The \textit{AC} content of the coma grains is significant, yet comet C/2017 K2 (PanSTARRS) is not ``AC-rich'' as is 
the OCC C/2013 US$_{\rm{10}}$~Catalina \citep{2021PSJ.....2...25W}. Comet 
C/2017 K2 (PanSTARRS) has a silicate-to-carbon ratio $>$1 and an average crystalline mass fraction 
$f_{cryst} \simeq 0.384$  and therefore shares compositional similarities to other OCCs including 
comet C/2006 Q1 (NEAT) at $r_{\rm{h}} = 2.8$~au (Si/C $= 1.40^{+0.42}_{-0.39}$) \citep[20080702.11,][ ]{2023PSJ.....4..242H}
and comet C/2012 K1 (PanSTARRS) at  $r_{\rm{h}} = 1.7$~au (Si/C $ = 1.20 \pm 0.27$) \citep{2021PSJ.....2...25W}.
In constrast, comet C/1995~O1~(Hale-Bopp) at  $r_{\rm{h}} = 2.8$~au \citep{1997Sci...275.1904C, 2002ApJ...580..579H, 2021PSJ.....2...25W} 
has a much higher Si/C $= 11.678\pm 0.008$.  Comet C/1995~O1~(Hale-Bopp) notably has higher contrast 
amorphous and crystalline silicate features due to the higher Si/C ratio and due to the differential 
size distribution having a greater relative abundance of smaller particles with the peak grain radii $a_p =0.2$~\micron,
which is smaller than the C/2017 K2 (PanSTARRS) size distribution that peaks at particle radii 
($a_p$) of $\sim$0.5~\micron{} to 0.6~\micron{}, depending on the position in the 
coma (see Table~\ref{tab:compositon-modelparams}).

Figure~\ref{fig:fig-tenary-comets-comparisions} presents the dust composition ternary diagram
derived for comets from thermal modeling of remote sensing observations of their SEDs. 
Comet C/2017~K2 (PanSTARRS) has a higher silicate to carbon ratio than most comets and
has has crystalline mass fraction ($f_{cryst}$) value that is 20\% larger than the Spitzer comet 
survey median $f_{cryst}$ of 0.31 \citet{2023PSJ.....4..242H}.

\subsection{Thermal model 14~\micron{} residuals} \label{sec:sec-residuals9314}

The signal-to-noise of the JWST MRS spectra afford an opportunity to explore subtle residuals in the SED of 
comet C/2017 K2 (PanSTARRS). These data present a transformative opportunity to explore the possibility that
the dust in the coma might contain species thought to be present but heretofore not detected due to sensitivity
limits. Materials containing CAIs are one such species that may be
contained within the refractory dust in comet coma. Stardust comet return samples have shown that terminal 
particles may include CAIs that are assemblages of high temperature minerals that are amongst
the oldest solar system materials \citep{2024A&A...687A..65W, 2024SSRv..220...79Z}. Three such 
Stardust CAIs are mineralogically comparable to type C CAIs, composed of nodules of mixtures 
of diopside and Al-, Ti-bearing diopside rimming with grains 
of spinel, anorthite and sometimes melilite, perovskite enclosed in Al-Si-rich glass and Al-rich 
diopside \citep{2014LPI....45.2282J, 2014AREPS..42..179B}. CAIs are relatively rare, only occurring 
in about 1\% of the Stardust sample materials \citep[e.g.,][and D.~Brownlee, priv. communication]{2017M&PS...52.1612J}.
The most common minerals seen in CAIs in meteorites are anorthite, melilite, perovskite, aluminous 
spinel, hibonite, calcic pyroxene, and forsterite-rich olivine \citep{1996Sci...272.1316G}.

Analysis of our thermal models  shows that dust with a crystalline pyroxene composition cannot produce 
the observed $\simeq 14$~\micron{} residual nor the features in the residual near 9.3~\micron{} and 
10.0~\micron{} because the other resonances of \textit{CP} are absent from the residual at the contrast 
predicted by thermal model for \textit{CP} for C/2017 K2 (PanSTARRS). The 14~\micron{} residual emission is
also seen in some Spitzer-studied comets, albeit at much lower fidelity. 
Figure~\ref{fig:fig-spitzer-comps}  compares  the 14~\micron{} residual of C/2017 K2 (PanSTARRS) with
five thermal modeled Spitzer comets \citep{2023PSJ.....4..242H}, 37P/Forbes that has no feature and 
78P/Gehrels~2, 88P/Howell, C/2006 P1 (McNaught), and C/2003 K4 (LINEAR) that all have this residual feature
near $\simeq 14$~\micron.

Using the laboratory result of NASA Stardust sample return analysis as a guide we tentatively explored whether 
CAIs could be responsible for residual emission.  The absorption efficiencies ($Q_{\rm{abs}}$) of common 
characteristic minerals commonly found in CAIs \citep{1996Sci...272.1316G}, and plausibly potential present in comet
coma materials, were computed using optical constants of four potential Al-, Ti-oxides -- MgAl$_{2}$O$_{3}$
natural Burma spinel \citep{2011A&A...526A..68Z}, CaTiO$_{3}$ perovskite \citep{1999A&AS..136..405H}, 
Al$_2$O$_3$ corundum \citep{2013A&A...553A..81Z}, and CaAl$_{12}$O$_{19}$ 
hibonite \citep{2002A&A...392.1047M} and for ellipsoidal shapes 
defined by their L-axes [Lx,Ly,Lz] \citep{1983asls.book.....B} or by their physical axial lengths (x,y,z).  Plotted under 
the C/2017 K2 (PanSTARRS) residual in Figure~\ref{fig:fig-spitzer-comps} are the ellipsoidal shapes for
each of the four Al-,Ti-oxides that place the strongest resonance nearest to
$\sim$14.5~\micron{}: [L$_{x}$,L$_{y}$,L$_{z}$]=[0.171,0.0697,0.697] for spinel (oblate 0.25x1x1), 
[0.426,0.640,0.650] for CaTiO$_{3}$ perovskite (oblate, 0.66x1x1), [0.147,0.699,0.699] for 
corundum and for hibonite (0.21x1x1).  Perovskite is the one mineral of the four minerals that has a 
single resonance and that is relatively insensitive to the shape of the ellipsoid. The spectral signatures 
of Al-, Ti-oxides are in the spectral range of the residuals present comet C/2017 K2 (PanSTARRS) as well
as that seen in certain Spitzer comets. 

The presence of `hot-temperature' refractory  materials (e.g., CAIs) in coma coma is clear evidence of
radial mixing in the early solar system \citep{2017ApJ...840...86D, 2019AJ....157..181V}.
Whether CAIs are present and with sufficient mass fractions to produce the 14~\micron{} residual feature emission
requires mode detailed future analysis coupled with additional high signal to noise JWST comet spectra from
7 to 26~\micron. However, these studies may provide an observational metric for 
constraining theoretical models radial transport of inner disk condensates out to the comet forming 
regime \citep[e.g.,][]{2002A&A...384.1107B, 2008SSRv..138...75W, 2020PNAS..11723426W} and 
demonstrate the potential JWST comet observations to transform our understanding of the early solar system 
environment, reveal materials that were the building blocks of planetesimals, and provide spectral 
templates to contextualize JWST studies of small bodies and planetary systems in active formation.



\begin{longrotatetable}
\begin{deluxetable*}{@{\extracolsep{0pt}}crcccccccccccccc}
\setlength{\tabcolsep}{3pt} 
\tablenum{4}
\tablewidth{0pc}
\tablecaption{Comet C/2017 K2 (PanSTARRS): Best-Fit Cases A, B, C (`AO50') Thermal Emission Model Parameters \label{tab:compositon-modelparams}}
\tablehead{
& \\[-1.75mm]
& &  & & & & \multicolumn{5}{c}{\underbar{$N_p (\times 10^{18}$)\tablenotemark{\,a} } }  \\
& & & & &  & \colhead{Amorphous} &
\colhead{Amorphous} & \colhead{Amorphous} & \colhead{Crystalline} & 
\colhead{Crystalline} \\
\colhead{Case} & \colhead{Tile Pos.} &  \colhead{N} & \colhead{M} & \colhead{$a_p$ \tablenotemark{$\dagger$}} & \colhead{D} & \colhead{Pyroxene} & \colhead{Olivine} &
\colhead{Carbon} & \colhead{Olivine} & \colhead{Orthopyroxene} & \colhead{$\chi^{2}_{\nu}$} & \colhead{$\chi^{2}$} & \colhead{$N_{pts}$} & \colhead{$\Sigma_{i=1}^{Npts} 2~ln~\sigma_i$} & \colhead{AIC \tablenotemark{$\ddagger$}}\\
& \colhead{(x:y)} && &  \colhead{($\mu$m)} & & \colhead{(AP50)} & \colhead{(A50)} & \colhead{(AC)} & \colhead{(CO)} & \colhead{(CP)}
}
\startdata
& \\[0.9mm]
A & 0:+1 & 3.3 & 13.2 & 0.5 & 2.727     & $2.5616^{+0.0229}_{-0.0225}$   & $4.4703^{+0.0146}_{-0.0144}$  & $8.2165^{+0.0035}_{-0.0035}$ & $2.3123^{+0.0148}_{-0.0154}$ & \nodata & 15.23 & 124909.83 & 8208 & -781465.1 & -648996.62 \\[2.9mm]
A & +1:+1 & 3.4 & 13.6 & 0.5 & 2.857   & $0.9884^{+0.0204 }_{-0.0202 }$ & $4.4445^{+0.0144}_{-0.0143 }$ & $6.1406^{+0.0035}_{-0.0035}$ & $2.1398^{+0.0118}_{-0.1197}$ & $0.5132^{+0.0229}_{-0.0230}$ & 7.15 & 58656.26 & 8208 & -785321.43 & -719106.52 \\[2.9mm]
A & -1:0 & 3.3 & 13.2 & 0.5 & 2.727    & $2.8555^{+0.0243}_{-0.0236}$   & $4.6875^{+0.0151}_{-0.0154 }$ & $9.2891^{+0.0039}_{-0.0040}$ & $3.0110^{+0.0161}_{-0.0163}$ & \nodata & 7.60 & 62363.40 & 8208 & -780149.63	& -710227.58\\[2.9mm]
A & 0:0 & 3.9 & 27.3 & 0.8 & 2.727       & $3.6608^{+0.0216}_{-0.0220}$   & $4.3598^{+0.0134}_{-0.0132}$  & $12.4860^{+0.0033}_{-0.0034}$ & $4.9653^{+0.0203}_{-0.0205}$ & \nodata & 20.50 & 168138.42 & 8208 & -772904.22 & -597207.15\\[2.9mm]
A & +1:0 & 3.6 & 18.0 & 0.6 & 2.857     & $2.6500^{+0.0189}_{-0.0192}$  & $3.6133^{+0.0132}_{-0.0130}$  & $6.8890^{+0.0032}_{-0.0034}$  & $2.3006^{+0.0199}_{-0.0126}$ & \nodata & 8.26 & 67742.59 & 8208 & -782255.88 & -706954.64 \\[2.9mm]
A & -1:-1 & 3.6 & 18.0 & 0.6 & 2.857 & $1.1583^{+0.0169}_{-0.0169}$  & $3.1896^{+0.0114}_{-0.0116}$   & $5.2681^{+0.0029}_{-0.0029}$ & $1.8500^{+0.0109}_{-0.0108}$ & \nodata & 6.72 & 55105.67 & 8208 & -784365.0 & -721700.68 \\[2.9mm]
A & 0:-1 & 3.6 & 18.0 & 0.6 & 2.857    & $3.9033^{+0.0192}_{-0.0195}$  & $2.9435^{+0.0131}_{-0.0130}$  & $6.4834^{+0.0033}_{-0.0033}$  & $2.2837^{+0.0123}_{-0.0124}$ & \nodata  & 9.85 & 80810.61 & 8208 & -782280.21 & -693910.95 \\[2.9mm]
B & 0:0 & 4.1 & 32.8 & 0.9 & 2.727     & $3.8122^{+0.0190}_{-0.0190}$  & $3.4701^{+0.0115}_{-0.0115}$  & $10.4251^{+0.0030}_{-0.0029}$ & $5.3481^{+0.0221}_{-0.0213}$ & \nodata & 21.31 & 174822.37	& 8208 & -772904.22 & -590523.20 \\[2.9mm]
C & 0:0 & 4.3 & 38.7 & 1.0 & 2.727       & $3.3569^{+0.0166}_{-0.0163}$   & $2.6199^{+0.0099}_{-0.0100}$  & $9.3398^{+0.0021}_{-0.0021}$ & $6.1203^{+0.0242}_{-0.0241}$ & \nodata & 21.99 & 242990.67 & 11058 & -1053772.96 & -800604.67 \\[2.9mm]
\enddata
\tablecomments{\, \nodata indicates that composition inclusion is not required by model to achieve fit.}
\tablenotetext{\dagger}{ \ Derived parameter.}
\tablenotetext{a}{\ Number of grains, $N_p$ at the peak ($a_p$) of the Hanner grain-size distribution (HGSD).}
\tablenotetext{\ddagger}{ \ Akaike Information Criterion. $N_{param}=8$ parameters: five compositions, and $D$, $N$, $M$. }
\end{deluxetable*}

\end{longrotatetable}


%


\begin{longrotatetable}
\begin{deluxetable*}{@{\extracolsep{0pt}}crcccccccc}
\setlength{\tabcolsep}{3pt} 
\tablenum{5}
\tabletypesize{\footnotesize}
\tablewidth{0pc}
\tablecaption{Comet C/2017 K2 (PanSTARRS):  Best-Fit Cases A, B, C (`AO50') Mass Fractions of Submicron Grains\label{tab:mod-massfractions}}
\tablehead{
&\\[0.2mm]
& \colhead{Tile} &\colhead{Total\tablenotemark{$\dagger$}} & \colhead{Amorphous} & \colhead{Amorphous} & \colhead{Amorphous} & \colhead{Crystalline} & \colhead{Crystalline} & \colhead{Silicate/Carbon}  \\
\colhead{Case} & \colhead{Pos.} &  \colhead{Mass}  & \colhead{Pyroxene} & \colhead{Olivine} & \colhead{Carbon} & \colhead{Olivine} & \colhead{Orthopyroxene} & \colhead{Ratio}  &\colhead{$f_{\rm{cryst}}$ }\\
& \colhead{(x:y)} & \colhead{( kg $\times 10^{7}$)} & \colhead{( $f$(ap50) $\times 10^{-1}$)} & \colhead{( $f$(ao50) $\times 10^{-1}$ )} & \colhead{( $f$(ac) $\times 10^{-1}$ )} & \colhead{( $f$(co) $\times 10^{-1}$  )} & \colhead{( $f$(cp) $\times 10^{-1}$ )} & &\colhead{ ( $\times 10^{-1}$) } 
}
\startdata
& \\[0.9mm]
A & 0:+1     & $1.9948^{+0.0036}_{-0.0037}$ &  $1.7349^{+0.0145}_{-0.0140}$ & $3.0275^{+0.0135}_{-0.0133}$ & $2.5294^{+0.0049}_{-0.0047}$ & $2.7083^{+0.0129}_{-0.0132}$  & \nodata &  $2.9535^{+0.0074}_{-0.0077}$ & $3.6252^{+0.0153}_{-0.0159}$ \\[2.9mm]
A & +1:+1   & $2.0391^{+0.0057}_{-0.0057}$ & $0.8405^{+0.0167}_{-0.0165}$ & $3.7793^{+0.0019}_{-0.0019}$ & $2.3734^{+0.0068}_{-0.0067}$ & $2.4251^{+0.0118}_{-0.0123}$   & $0.5817^{+0.0245}_{-0.0246}$ & $3.2133^{+0.0120}_{-0.0120}$ & $3.9425^{+0.0024}_{-0.0024}$ \\[2.9mm]
A & -1:0   & $2.2930^{+0.0039}_{-0.0039}$ & $1.6824^{+0.0135}_{-0.0132}$ & $2.7618^{+0.0119}_{-0.0119}$ & $2.4877^{+0.0044}_{-0.0044}$ & $3.0681^{+0.0166}_{-0.0118}$    & \nodata  & $3.0198^{+0.0072}_{-0.0071}$ & $4.0841^{+0.0135}_{-0.0136}$ \\[2.9mm]
A & 0:0       & $3.9374^{+0.0064}_{-0.0065}$ & $1.6282^{+0.0091}_{-0.0093}$ & $1.9390^{+0.0078}_{-0.0077}$ & $2.5242^{+0.0043}_{-0.0042}$ & $3.9086^{+0.0101}_{-0.0110}$   & \nodata & $2.9617^{+0.0066}_{-0.0067}$ & $5.2283^{+0.0109}_{-0.0110}$ \\[2.9mm]
A & +1:0     & $2.5685^{+0.0034}_{-0.0035}$ & $2.1241^{+0.0140}_{-0.0141}$ & $2.8963^{+0.0013}_{-0.0013}$ & $2.5100^{+0.0037}_{-0.0036}$ & $2.4696^{+0.0099}_{-0.0103}$   & \nodata & $2.9841^{+0.0057}_{-0.0058}$ & $3.2971^{+0.0120}_{-0.0124}$ \\[2.9mm]
A & -1:-1 & $1.8982^{+0.0031}_{-0.0031}$ & $1.2563^{+0.0172}_{-0.0174}$ & $3.4594^{+0.0158}_{-0.0161}$ & $2.5971^{+0.0046}_{-0.0045}$ & $2.6872^{+0.0120}_{-0.0118}$   & \nodata & $2.8504^{+0.0067}_{-0.0068}$ & $3.6299^{+0.0143}_{-0.0143}$ \\[2.9mm]
A & 0:-1    & $2.6460^{+0.0035}_{-0.0036}$ & $3.0371^{+0.0138}_{-0.0138}$ & $2.2903^{+0.0121}_{-0.0120}$ & $2.2930^{+0.0034}_{-0.0033}$ & $2.3796^{+0.0100}_{-0.0102}$    & \nodata &  $3.3611^{+0.0063}_{-0.0064}$ & $3.0876^{+0.0120}_{-0.0123}$ \\[2.9mm]
B & 0:0       & $3.6315^{+0.0066}_{-0.0064}$ & $1.7699^{+0.0085}_{-0.0085}$ & $1.6110^{+0.0069}_{-0.0070}$ & $2.2000^{+0.0040}_{-0.0041}$ & $4.4192^{+0.0103}_{-0.0102}$   & \nodata & $3.5455^{+0.0084}_{-0.0083}$ & $5.2283^{+0.0105}_{-0.0107}$ \\[2.9mm]
C & 0:0       & $3.2745^{+0.0067}_{-0.0067}$ & $1.5858^{+0.0077}_{-0.0077}$ & $1.2376^{+0.0060}_{-0.0062}$ & $2.0055^{+0.0041}_{-0.0041}$ & $5.1710^{+0.0102}_{-0.0101}$   & \nodata & $3.9862^{+0.0103}_{-0.0101}$ & $6.4682^{+0.0097}_{-0.0096}$ \\[2.9mm]
 \enddata
\tablecomments{\, Asymmetric uncertainties indicate values constrained to a confidence level of 95\%; table entry of  \nodata  indicates zero mass fraction of the particular composition returned by best-fit model.}
\tablenotetext{\dagger}{ \ Derived parameter for emission within a 1\farcs0 diameter beam toward the particular spatial location in coma.}
\end{deluxetable*}
\end{longrotatetable}


\clearpage

\section{Polycyclic Aromatic Hydrocarbons}  \label{sec-pah-models-dw}

The detection of polycyclic aromatic hydrocarbon (PAHs) in the coma of comets from remote sensing observations has been 
controversial  \citep{1997P&SS...45.1539J, 2007Icar..187...69L, 2009ApJ...696.1075B, 2023JApA...44...89V}.
However, JWST observations of comets obtained at high-signal-to-noise opens a new domain of discovery 
space in the 3 to 8~\micron{} wavelength regime were in other astrophysical environments strong emission from 
PAHs are evident. JWST observations of comet C/2017 K2 (PanSTARRS) affords an opportunity to critically 
assess whether emission from PAHs species is present in comet comae, complimented by robust thermal 
model fits to the long wavelength ($\gtsimeq 9$~\micron) SED which constraints the particle size and physical properties 
of coma dust. With JWST, the fidelity of possible detection PAH and analysis is enhanced compared to other 
observations of comets, such as those observed Spitzer \citep{2016PASP..128a8009K, 2023PSJ.....4..242H}. Furthermore, 
the JWST  comet C/2017 K2 (PanSTARRS) spectrum enables a detailed study of the continuum in 
the 8~\micron{} region were both scattered light and thermal emission overlap, tightly constraining the 
interpretation of spectral residuals at wavelengths attributed to PAH emission. 
The \textit{relative strengths} of the features are critical to the PAH model. Hence, a robust 
determination of the continuum under the features requires the scattered light model and the thermal model, which 
was iteratively determined by estimating the PAH emission, subtracting it, re-fitting the thermal model
and finally re-deriving the PAH residual that was fitted.

\begin{figure}[h!]
\figurenum{21}
\begin{center}
\includegraphics[trim=0.05cm 0.15cm 0.05cm 0.15cm, clip, width=0.45\textwidth]{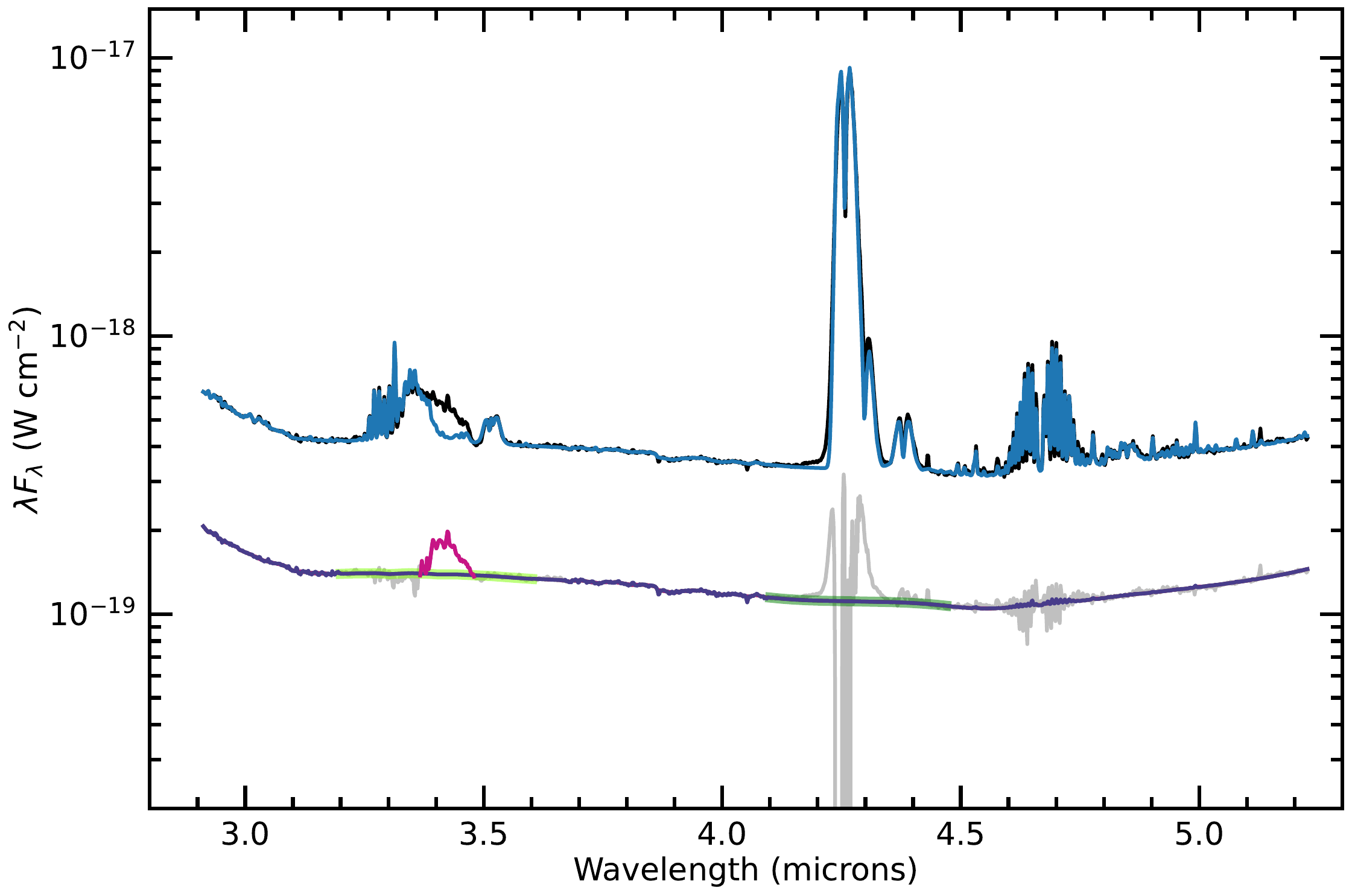}
\caption{The 3.0 to 5.0~\micron{} spectral region of comet C/2017 K2 (PanSTARRS) derived from NIRSpec data 
of the central coma (position 0:0, see Figure~\ref{fig:fig-nirspec-three}). (Top spectrum): The spectra decomposed 
using a NASA Planetary Spectrum Generator \citep[NASA PSG;][]{2018JQSRT.217...86V} model which
incorporates CH$_{3}$OH adopting a temperature of 40~K.  The black solid line is the observed spectrum 
while the NASA PSG model is given by the blue solid line, both scaled up by factor of 2 for offset clarity.  
(Bottom spectrum): The lower purple line is the result of median filtering (medfilt) of the residual (observed - model). 
The lower chartreuse and lower green lines are the continuum derived by more complex steps than just medfilt as 
that are described in the text (Section~\ref{sec:residual_3to8}). The medium violet red is the residual emission 
centered near 3.42~\micron{} that cannot be accounted for by the PSG model. This residual
emission is fitted by models with small heavily hydrogenated PAHs fluorescing in the solar radiation 
field (see Section~\ref{sec:pah_model}).
}
\label{fig:fig-PSG34mu}
\end{center}
\end{figure}

Comet C/2017 K2 (PanSTARRS) is observed to have CH$_{3}$OH present in relatively high
abundance.  Examination of the spectrum in the 3.4~\micron{} region, Figure~\ref{fig:fig-PSG34mu}, 
and its residuals from a NASA PSG model suggest that gas-phase CH$_{3}$OH  $\nu_{9}$ band, 
which is not included in our spectral models, contributes to a portion of the observed residual. However, 
organic species (including PAHs) may also be emitting in this region as well. Decomposition of the 
observed emission in the 3 to 8.6~\micron{} spectrum to search of residual emission from PAHs requires an 
accurate continuum model. Derivation of this continuum model is discussed in the Appendix.

\subsection{Residual Features 3 to 8~\micron} \label{residual-feature-3to8}

\begin{figure*}[ht!]
\figurenum{22}
\gridline{
    \fig{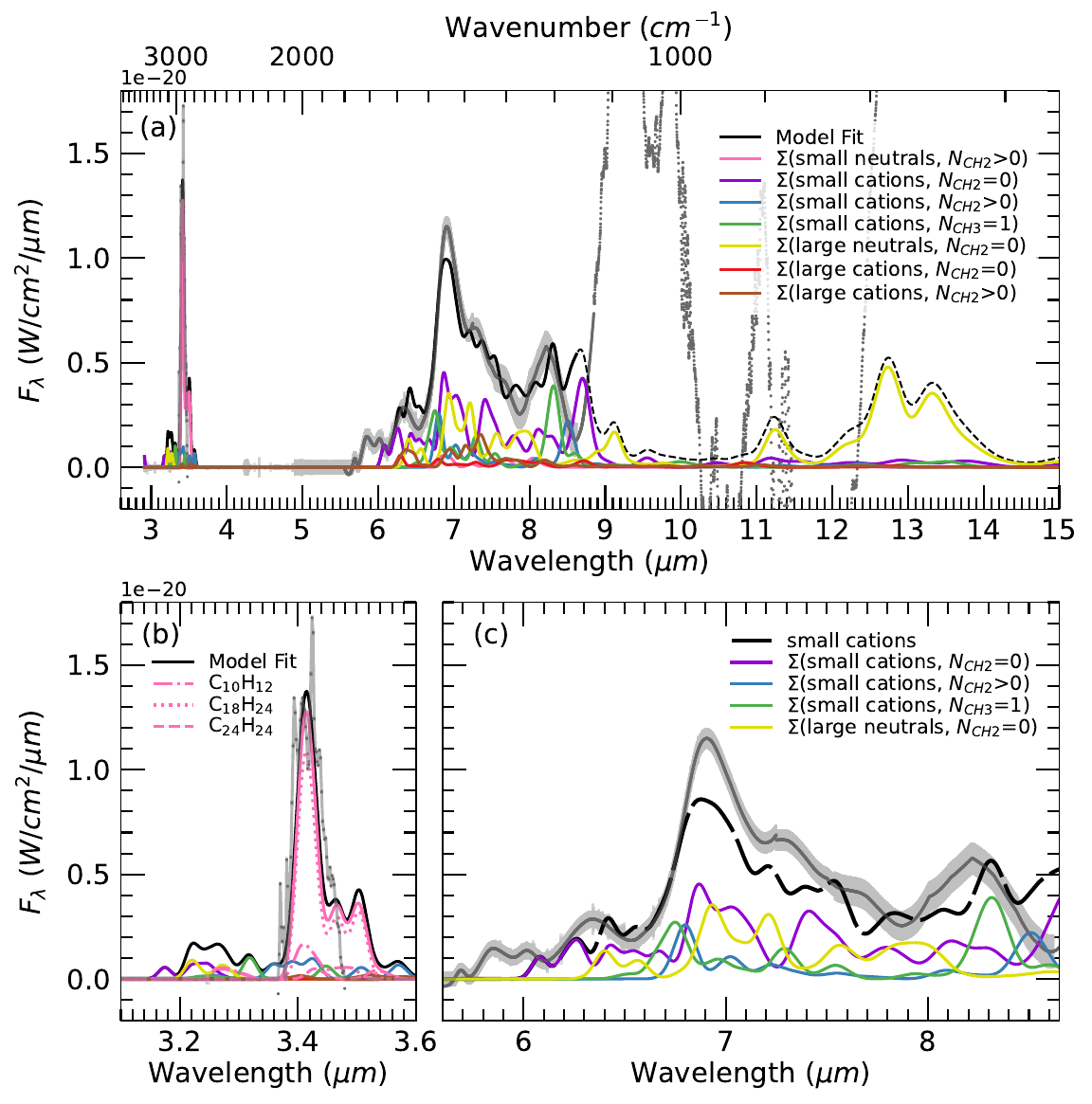}{0.70\textwidth}{ }
   }
\vspace{-1.0cm}
\gridline{
    \fig{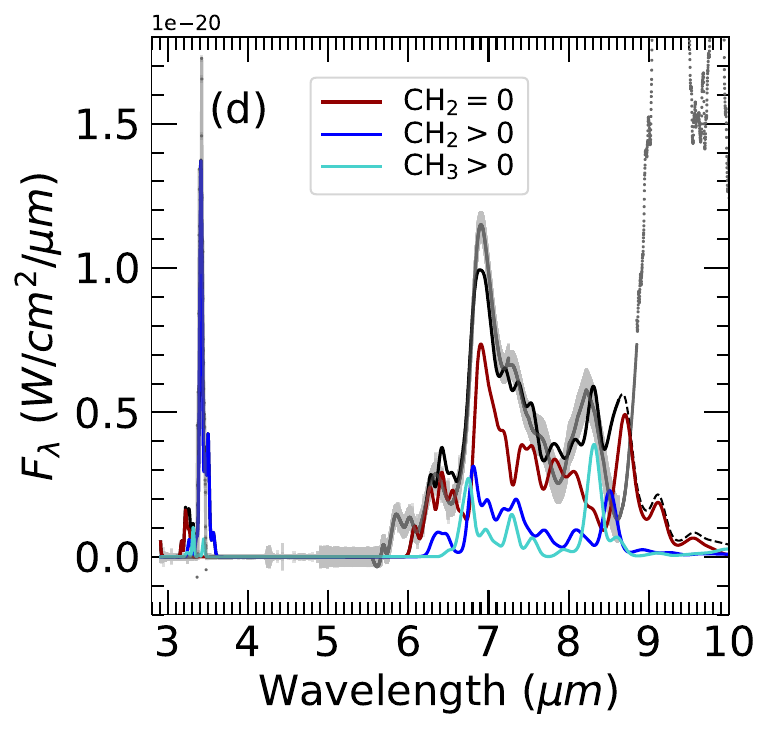}{0.33\textwidth}{ }
    \fig{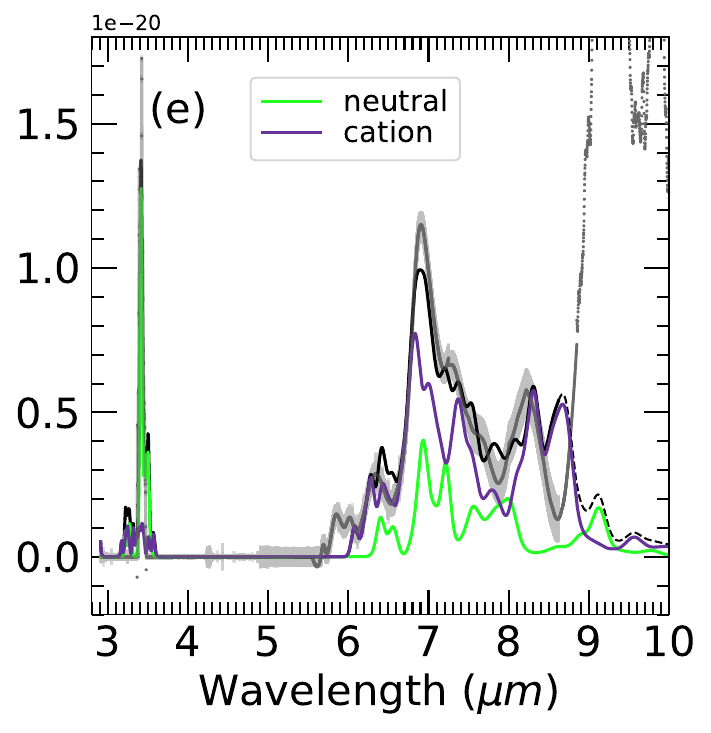}{0.305\textwidth}{ }
    \fig{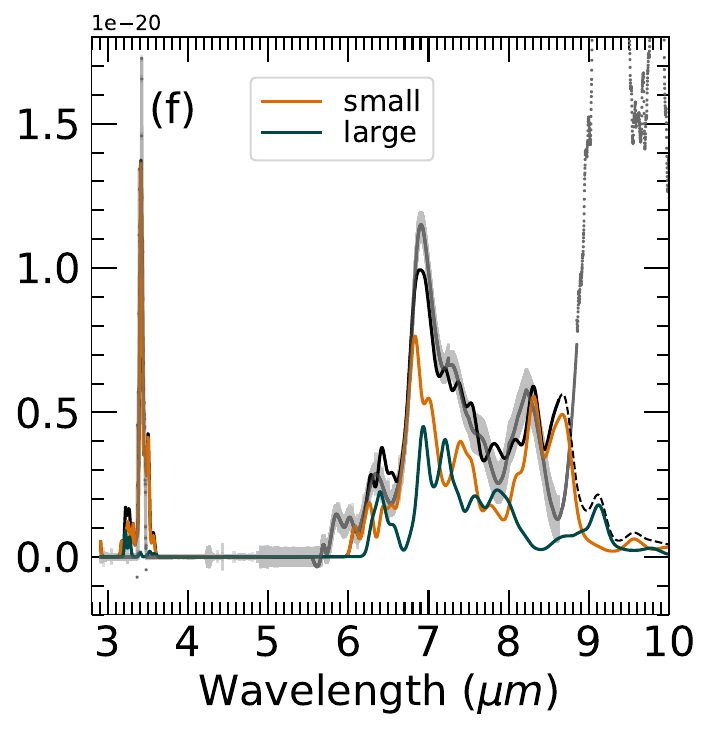}{0.305\textwidth}{ }
}
\vspace{-0.65cm}
\caption{PAH model fitted to $F^{\rm{PAH extract}}_{\rm{Residual}}$ for comet C/2017 K2 (PanSTARRS). 
(a) Model fit to 3.2--3.6~\micron{} and 5.5--8.62~\micron{} ({\it black solid}) spectral regions, and model predicted 
outside of that range ({\it black short dashed}), shown with breakdown into seven categories of PAHs 
n the model: small ($N_{C}\leq 40$) neutral PAHs with CH$_2$ groups ($N_{\rm CH_2}$$>$0) ({\it pink}); 
small cations ($N_{\rm CH_2}$$=$0) ({\it purple}), small cations with CH$_2$ groups ($N_{\rm CH_2}$$>$0) ({\it blue}), 
and small cations with metylene side groups ($N_{\rm CH_3}$$>$0)  ({\it green}); large neutrals ({\it yellow}); 
large cations ($N_{\rm CH_2}$$=$0) ({\it red}); and large cations with CH$_2$ groups ($N_{\rm CH_2}$$>$0) ({\it brown}). 
(b) Model fit of the 3.42~\micron{} region where primarily three small neutrals with CH$_2$ groups
generate the observed peak, with fully hydrogenated C$_{18}$H$_{24}$ ({\it pink dotted}) dominating the 
flux density compared to heavily hydrogenated C$_{10}$H$_{12}$ ({\it pink dash-dot}) and fully hydrogenated 
C$_{24}$H$_{24}$ ({\it pink dashed}). (c) Model components of small cations and large neutrals with 
their sum ({\it black long-dash}) that dominate the 6.85--8.62~\micron{} emission: 
small cations ($N_{\rm CH_2}$$=$0) ({\it purple}), 
and large neutrals ($N_{\rm CH_2}$$=$0) ({\it yellow}). Small cations with groups 
($N_{\rm CH_2}$$>$0, $N_{\rm CH_3}$$>$1) ({\it blue, green}) contribute near 6.85~\micron {}, 
and ($N_{\rm CH_3}$$>$1) to the  8.25~\micron{} feature. Small cations and large neutrals
contribute to the 6.3~\micron{} feature. (d) Model breakdown into functional 
groups: $N_{\rm CH_2}$$=$0, $N_{\rm CH_2}$$>$0, $N_{\rm CH_3}$$>$0 ({\it brown, blue, cyan}). 
e) Model breakdown into neutrals ({\it chartreuse}) and cations ({\it deep purple}). (f) Model breakdown into 
small ({\it orange}) and large PAHs ($N_{C}$$>$40) ({\it dark teal}). 
}
\label{fig:fig-K2-pah-model}
\end{figure*}

The $F^{\rm{PAH extract}}_{\rm{Residual}}$ (for a definition see Appendix~\ref{sec:appendix-A}) 
shows spectral features, with peaks occurring at 
3.42~\micron{}, 6.35~\micron{}, 6.92~\micron{}, 8.25~\micron{}, as well a weaker emissions 
spanning 6.9~\micron{} to 7.88~\micron{} with a lower contrast feature at 7.28~\micron{} 
(2920--2925~cm$^{-1}$, 1575~cm$^{-1}$, 1445~cm$^{-1}$, 1212~cm$^{-1}$, and 1374~cm$^{-1}$).
Residual emission in the  C/2017 K2 (PanSTARRS) spectra can be explained by PAHs. 
The presence of a strong emissions near 6.85~\micron{} (Figure~\ref{fig:fig-K2-pah-model}) 
suggests either aliphatic hydrocarbons \citep{2016MNRAS.462.1551Y, 2017ApJ...835..291S} 
or heavily- to fully-hydrogenated PAHs \citep{2013ApJS..205....8S}, and the presence of the 
3.42~\micron{} feature in emission in the C/2017 K2 (PanSTARRS) data strongly support this contention. 
The hydrogenation of PAHs, i.e., the addition of peripheral hydrogen atoms, converts aromatic (sp2) 
carbons to aliphatic (sp3) carbons and generates aliphatic C\textendash H stretch bands 
of \textendash CH$_2$\textendash groups (methylene) and significantly weakens the original 
aromatic bands, where each \textendash CH$_2$\textendash \ has two C\textendash H stretches in 
contrast to the one C\textendash H if the carbon were sp3 \citep{2013ApJS..205....8S}. While some 
authors refer to these aliphatic bonds as \textendash H side-groups \citep{2020ApJ...892...11B}, we refer 
to them as CH$_2$ groups \citep{2013ApJS..205....8S}. No other species can be hot enough to 
emit at 3.4~\micron{} \citep{2007ApJ...657..810D}. The 6.3~\micron{} feature is weaker than the 
6.9~\micron, which is unusual compared to ISM and proto-planetary 
disks \citep{2017ApJ...835..291S, 2024A&A...685A..75C}. Also unusual is the lack 
of emissions near 3.3~\micron{} as well as the lack of a secondary weaker peak at 3.5~\micron{} 
typically associated with a 3.4~\micron{} peak \citep[Figure~3]{2013ApJS..205....8S}.

In comet C/2017 K2 (PanSTARRS), the 3.42~\micron{} peak is narrow (2020--2025~cm$^{-1}$, 
FWHM = 100~cm$^{-1}$) and does not have the typical three-peak structure of meteoritic insoluble 
organic matter (IOM) of CH$_{3}$ at 2960~cm$^{-1}$, CH$_{2}$ at 2930~cm$^{-1}$, 
and CH$_{3}$+CH$_{2}$ at 2860~cm$^{-1}$ (3.378~\micron{}, 3.413~\micron{}, and 3.497~\micron{}) 
\citep{2019M&PS...54.1632K}, further supporting the concept of macromolecules rather than 
solid state organics are producing the observed features. 

The NASA AMES PAH database v3.20\footnote{\url{https://www.astrochemistry.org/pahdb/}} and the 
AmesPAHdbPythonSuite tools \citep{2014ApJS..211....8B, 2018ApJS..234...32B, 2020ApJS..251...22M} 
were used to identify and model potential PAHs species in the residual  comet C/2017 K2 (PanSTARRS) spectra, 
employing excitation in solar radiation field characterized by a 5770~K blackbody ($\simeq 3.5$~eV) 
which is sufficient for these materials to fluoresce. In our modeling, we assumed that there 
are no absorption features, the PAHs species are limited in size ($N_{C}\leq 100$~atoms), and 
PAHs are either cations (PAH$^{+}$) or  neutrals (PAH$^{0}$), a natural expected consequence (\citep[e.g.][]{2003ApJ...596L.195G})
in the coma of comets where PAHs and ices are expected to be admixed. A model fit was attained using 
the `NNLS' method \citep{2018ApJ...858...67B}, using the 1~$\sigma$ measurement uncertainties. 
Multiple models were run with varying search criteria (size, charge, composition) and the models'  
relative probabilities assessed using the Akaike Information Criterion with the result that 
$N_{C}\leq 100$~atoms yielded the most probable model fit for the MRS wavelength range and 
second most probable model for the NIRSpec and combined NIRSpec and MRS data. The 
most probable PAH model included a very large compact PAH cation C$_{216}$H$_{36}{^+}$ that 
contributed less out-of-band emissions near 3.42~\micron{} but which is not considered a likely PAH 
candidate because PAHs with $N_{C}\leq 100$~atoms are anticipated for the ISM \citep{2012ApJ...754...75R}, 
and thus the PAH model with $N_{C}\leq 100$ is considered here. Out of 2550 PAHs considered in the model, 
22 PAHs are in the model fit.

\subsection{PAH Model}
\label{sec:pah_model}

The PAH model is shown in Figure~\ref{fig:fig-K2-pah-model}(a) with its decomposition into seven 
categories of size, charge, CH$_2$ groups or 
CH$_3$ side groups. Figure~\ref{fig:fig-K2-pah-model}(b) shows the model in the region of the 
3.42~\micron{} feature. Figure~\ref{fig:fig-K2-pah-model}(c) highlights the major contributors to the 
6.9~\micron{} feature including large neutrals, small cations with 
CH$_{2}$ groups and with CH$_{3}$ methyl side groups, but mostly small cations without CH$_{2}$ groups. 
Figures~\ref{fig:fig-K2-pah-model}(d--f) show the model breakdown into tuples showing side groups, 
charge state, and size.  We find that small PAHs, which we define as number of carbon atoms $N_C\leq 40$, 
which are neutral and with aliphatic CH$_{2}$ groups \citep{2013ApJS..205....8S} are required to produce 
the isolated 3.42~\micron \ emission feature. 
On the one hand, small neutral PAHs that produce the 3.42~\micron \ feature do not contribute significantly 
to the 6--9~\micron{} region. On the other hand, features in the 6--9~\micron{} region requires emission 
from small cations, large neutrals, and large cations. In particular, the 6.85~\micron{} (1460~$\rm{cm}^{-1}$) 
short wavelength rise of the 6.9~\micron{} feature requires small cations with CH$_{2}$ groups,
where the CH$_{2}$ groups are expected from prior studies \citep{2016MNRAS.462.1551Y, 2020A&A...637A..82D}. 
In contrast to expectations based on prior studies, we find small cations and large neutrals, both 
without CH$_{2}$ groups, contribute to the strong 6.9~\micron{} feature and its long wavelength shoulder 
(Figure~\ref{fig:fig-K2-pah-model}(c)). These small cations are linear two- to four-ring chains 
and ``zigzag'' PAHs \citep[e.g.,][]{2024ApJ...968..128R}. 

The PAH model is fitted to the residual (Figure~\ref{fig:fig-K2-pah-model}a) that is at 
wavelengths shorter than where silicates emit and we note that there are no strong features predicted 
by this PAH model at longer wavelengths, e.g., in the vicinity 11--12.5~\micron{}. 

Edge defects in PAHs with CH$_{2}$ groups as well as in solid-state hydrogenated amorphous carbons 
(HACs) affect the 3.4~\micron{} feature profile \citep{1998ApJ...503L.183D}.  In an effort to fit JWST 
spectra of the Orion Bar (ISM), new computations of edge defects \citep{2024ApJ...968..128R}, 
i.e., the steric repulsion between two adjacent hydrogen atoms, show the trend for PAH cations to move 
the ``6.25~\micron{} skeletal mode" to longer wavelengths (6.3 to 6.35~\micron). These new theoretical 
spectra of PAHs with edge defects will be available for model fitting in the next release (v4.00) of the Ames PAH 
database and so the larger PAHs in the model fit for C/2017 K2 (PanSTARRS), which currently include 
some armchair and zigzag PAHs, may change with this next database release. Likely, similarly structured 
PAHs as those fitted to Orion photodissociation regions (PDRs) are present in the coma of comet C/2017 K2 (PanSTARRS). 

\begin{figure}[h!]
\figurenum{23}
\begin{center}
\includegraphics[width=0.45\textwidth]{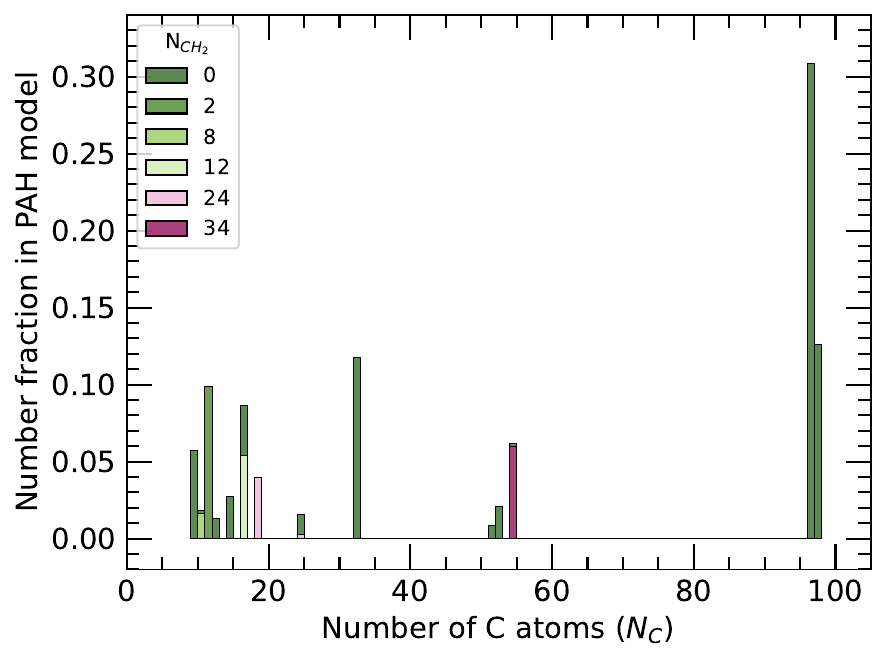}
\caption{Number fraction of PAH molecules in the PAH model for comet C/2017 K2 (PanSTARRS) versus 
number of carbon atoms ($N_{C}$) and colored by number of CH$_2$ groups ($N_{\rm CH_2}$). The number 
fractions are grouped by ($N_{C}$, $N_{\rm CH_2}$) and summed.}
\label{fig:fig-K2numberfrac-vs-NC-3umresidual}
\end{center}
\end{figure}

The number fraction of PAHs in the model fit can be characterized by the number fraction versus number 
of carbon atoms ($N_{C}$) and the number of CH$_{2}$ groups as shown in 
Figure~\ref{fig:fig-K2numberfrac-vs-NC-3umresidual}. 

The small PAHs have sizes that range from smaller than ISM to classic ISM PAHs such as 
coronene C$_{24}$H$_{12}$ except that in the PAH model for comet C/2017 K2 (PanSTARRS) this 
species is fully hydrogenated as C$_{24}$H$_{24}$.  Two- to four-ring PAHs in the PAH model for K2 are 
commensurate with: (a)~mass spectroscopy by Rosetta's ROSINA for 67P/Churyumov-Gerasimeko that found 
2-ring PAHs C$_{10}$H$_{8}$ to fully hydrogenated C$_{10}$H$_{18}$ \citep{2022NatCo..13.3639H}; 
(b)~mass spectroscopy of aromatic nano globules in Bennu samples that reveal  2- to 4-ring PAHs and a 
dominance of 4-ring C$_{14}$H$_{10}$ and its alkylated homologues \citep{2024LPICo3036.6422C}; 
(c)~2- to 4-ring PAHs found in Ryugu samples (predominantly 4-ring) \citep{2023Sci...382.1411Z}; and 
(d)~mass spectroscopy of Stardust samples that found 2- to 4-ring PAHs \citep{2010M&PS...45..701C}. 

The PAH models yield in a suite of PAH molecules. Descriptive parameters of the cometary 
organic molecules include the Hydrogen Deficiency Index (HDI) and the CH$_{2}$:CH$_{3}$ 
ratio \citep{2022NatCo..13.3639H}. The HDI is defined as, = C + 1 + N/2 - H/2 - X/2, where C is the 
number of C atoms in the molecule, N the number of N atoms, H the number of H atoms and X the 
number of halide atoms. In comet C/2017 K2 (PanSTARRS), X is zero. Thus, HDI = C + 1 + N/2 - H/2. 
The lower the HDI, the more abundant are peripheral H atoms.  In mass spectroscopy, HDI can be called the 
Double Bond Equivalent \citep{2020PSJ.....1...55D}. The HDI compares the number of 
H atoms in a given molecule to the maximum number of H atoms. It attempts to quantify 
information on the molecule’s chemical structure as it corresponds to the total number 
of multiple C\chemone C bonds and cycles in the molecule. These descriptive parameters 
enable comparisons of PAHs species modeled from JWST spectra of comet C/2017 K2 (PanSTARRS) 
to complex organic molecules measured via ROSINA mass spectrometry of the coma of comet 
67P/Churyumov-Gerasimeko and to those studied via rotational spectroscopy \citep{2022NatCo..13.3639H} 
for regions in the ISM including cold environments. 

\begin{figure}[t!]
\figurenum{24}
\begin{center}
\includegraphics[trim=0.75cm 1.7cm 0.1cm 0.1cm, clip, width=0.49\textwidth]{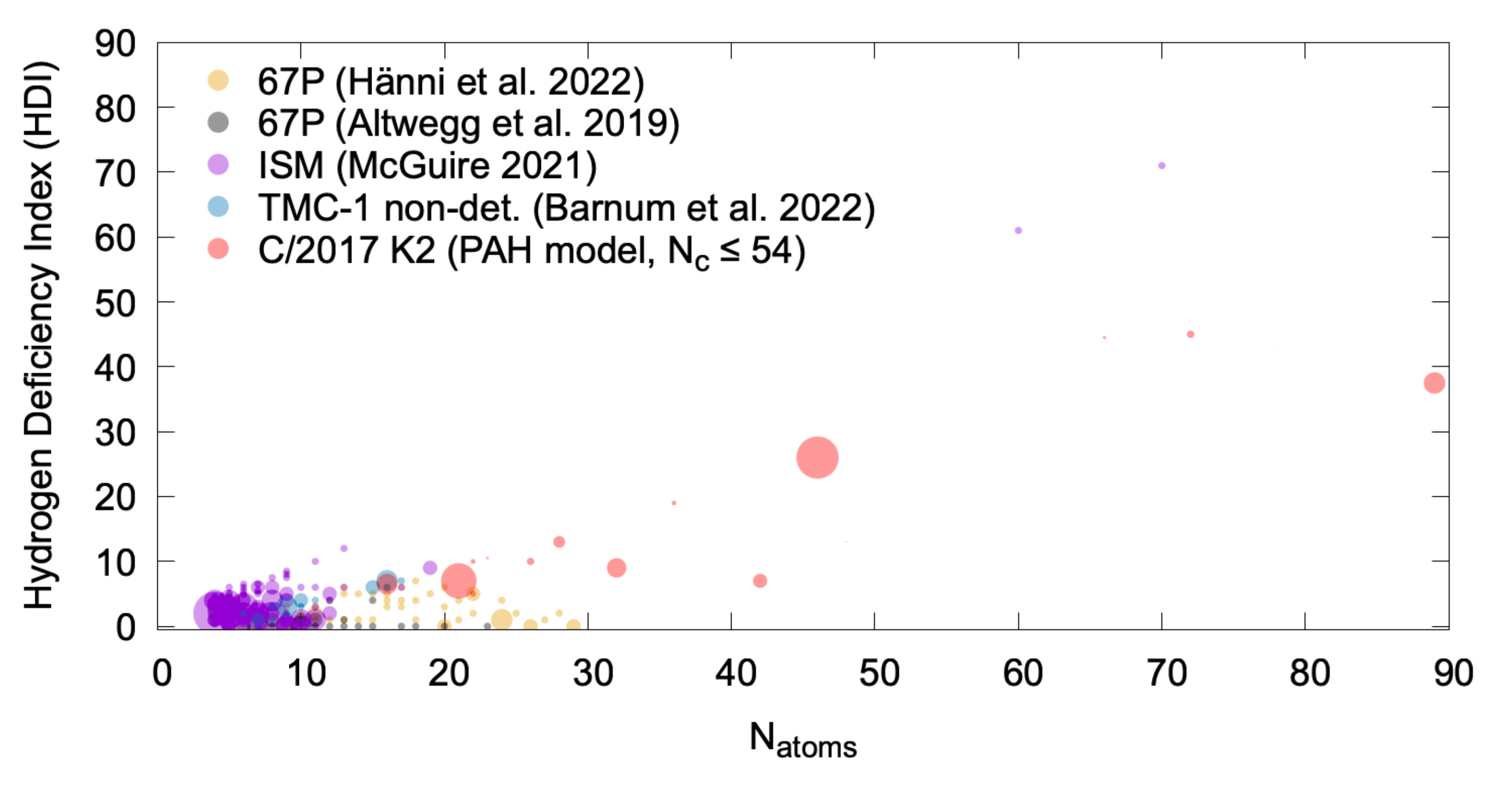}
\caption{Hydrogen Deficiency Index (HDI) versus $N_{\rm{atoms}}$. PAH molecules in the 
PAH model for comet  C/2017 K2 (PanSTARRS)
with $N_C\leq 54$ are plotted here ({\it red}, symbol size is 10 times the 
number fraction in PAH model) in the HDI vs.\ $N_{\rm{atoms}}$ relation to set PAHs in 
comet comet C/2017 K2 (PanSTARRS) into the context of complex organic molecules assessed 
for comet 67P/Churyumov-Gerasimeko ({\it yellow, grey}), for the ISM ({\it purple}), and for non-detections in the 
Taurus Molecular Cloud 1 (TMC-1) ({\it blue}), where symbol size represents the number of 
molecules sharing the HDI value and number of atoms, with the smallest circles representing 
one molecule. Organic molecules identified by ROSINA in 67P/Churyumov-Gerasimeko include linear, heterocycles, 
and aromatic molecules of $N_{C}\geq 4$. PAHs in comet C/2017 K2 (PanSTARRS) are more hydrogenated than the
trend for the ISM and less hydrogenated than 67P/Churyumov-Gerasimeko. Adapted from \citep{2022NatCo..13.3639H}, 
using their data and their data compilations from \citep{2019ARA&A..57..113A, 2021zndo...5046939M, 2022JPCA..126.2716B}. 
}
\label{fig:fig-hanni2022-K2}
\end{center}
\end{figure}

Figure~\ref{fig:fig-hanni2022-K2} shows the HDI versus $N_{\rm{atoms}}$, limited to $N_{C}\leq 40$ 
and correspondingly $N_{\rm{atoms}}\leq 50$ to better show the overlap with the other sources. 
Comet C/2017 K2 (PanSTARRS) in the relation HDI vs $N_{\rm{atoms}}$ lays between the ISM and 
comet 67P/Churyumov-Gerasimeko. Our PAH model suggests larger $N_{\rm{atoms}}$ (correlated with larger $N_{C}$) and 
less hydrogenation species are extant the coma of comet /2017 K2 (PanSTARRS). ROSINA did not 
measure higher-mass PAHs than C$_{10}$H$_{18}$ (fully hydrogenated naphthalene) because their 
fragmentation signatures were not sufficiently prominent and specific to allow a clear identification. Hence, 
the contribution of PAHs larger than naphthalene to ROSINA's mass spectra is unclear \citep{2022NatCo..13.3639H}. 

\citet{2022NatCo..13.3639H} make succinct arguments for comet 67P/Churyumov-Gerasimeko's heterocyclic and 
aromatic molecules being inherited materials and not products of processing on the comet. From the census 
of organic molecules measured by ROSINA, the trend is for ISM molecules to be smaller and more hydrogen 
deficient than cometary species. These latter trends imply that cometary organics are the outcome of a long 
sequence of material processing starting in the ISM. By comparison of comets 67P/Churyumov-Gerasimeko, 
1P/Halley, Stardust return samples from comet 81P/Wild~2, and the IOM and soluable organic matter 
(SOM) with meteoritic samples, \citet{2022NatCo..13.3639H} postulate that comets have a shared
pre-solar history of Solar System organics and subsequent material processing. 

JWST spectrum strongly suggests (i.e., the analysis of the residual spectral features in the 3 to 8.6~\micron{} region)
that PAHs are present in the inner coma of comet C.2017 K2 (PanSTARRS). These species could be 
affixed to the ice- or dust-ice mixture sublimed from the nucleus or driven off by nucleus activity. Laboratory 
experiments show PAH cations form readily by UV irradiation of water-ice--embedded 
PAHs \citep{2010A&A...511A..33B, 2012ApJ...756L..24G, 2014ApJ...784..172H, 2015ApJ...799...14C}.
In the coma of C/2017 K2 (PanSTARRS), spatially extended H$_{2}$O production rates are presumed to 
arise from the transport of ice grains into the coma. PAH emission features are extant in the MRS beam
positions of nucleus.

Heavily hydrogenated small neutral PAHs are linked to the 3.4~\micron{} feature and 
similar PAHs contribute to emission in the 6.9~\micron{} region in  comet C/2017 K2 (PanSTARRS). 
The presence of small PAH cations and extended ice sublimation are commensurate with 
small PAH cations residing in and being released in the coma from water ice grains in the inner coma. 
The PAH model for comet C/2017 K2 (PanSTARRS) may be updated with 
the next version of the PAH database, which will include a broader range of PAH species, larger PAH species, and 
also theoretical computations focused on edge effects for armchair and zigzag PAHs \citep{2024ApJ...968..128R}. 

\section{Conclusion} 
\label{sec-summary}

The JWST study of comet C/2017 K2 (PanSTARRS) demonstrates the discovery potential to advance 
cometary science through high spatial resolution  moderate spectral resolution data-cubes
produced by the JWST NIRSpec and MRS IFUs. These instruments provide spectral access to key 
diagnostics of coma materials produced by comet nucleus sublimation activity including molecular emission 
bands, organic signatures, and dust features in the spectral energy distribution (SEDs) of active comets. 

In comet C/2017 K2 (PanSTARRS) at 2.35~au pre-perihelion our analysis of JWST NRISpec and MRS observations
show:

\begin{itemize}

\item The nucleus of comet C/2017 K2 (PanSTARRS) likely is $\ltsimeq 4.2$~km in radius based on NIRSpec
and MRS (short) azimuthally averaged radial surface brightness profiles and WebbPSF simulations.

\item Emission lines from H$_2$O are detected in both NIRSpec and MRS spectra of C/2017 K2 (PanSTARRS),
including faint H$_2$O ro-vibrational lines from the $\nu_3$-$\nu_2$ and $\nu_1$-$\nu_2$ hot bands in the 4.5
to 5.2~$\mu$m range, and strong lines from the $\nu_2$ fundamental vibrational band, with some minor 
contribution from hot bands (e.g., $\nu_3+\nu_2-\nu_3$, 2$\nu_2-\nu_2$) in region 5.5 to 7.25~$\mu$m.
The effective H$_{2}$O OPR of C/2017 K2 (PanSTARRS) is larger than 2.75

\item A wealth of trace volatile species (molecular emission) is superimposed on the continuum in the 
NIRSpec spectrum of C/2017 K2 (PanSTARRS), including  CN (band emission centered at 4.9~\micron{}), 
H$_2$CO, CH$_3$OH, CH$_4$, C$_2$H$_6$, HCN, NH$_2$, and OH$^*$ (prompt emission after UV photolysis 
of water) are clearly detected.

\item In the inner coma, the CO and CH$_{4}$ show remarkably 
similar abundance and temperature spatial distributions, perhaps from jets perpendicular to the Sun-comet line. 
CO$_{2}$ shows a similar spatial distribution to that of CO and CH$_{4}$; however, its temperature 
distribution is significantly warmer in the anti-sunward direction. This may be explained by less efficient adiabatic 
cooling owing  to reduced collision rates in the anti-sunward coma.

\item  At the epoch of the JWST observations of comet C/2017 K2 (PanSTARRS) the CH$_{3}$OH and 
H$_{2}$O show significant increases in production, yet the CO, CH$_{4}$, and HCN production rates
are in agreement with \cite{2025AJ....169..102E}.  Combined with the JWST derived spatial distribution
of these species in the coma indicate that the comet underwent an H$_{2}$O-driven 
outburst that did not propagate to molecules associated with CO$_{2}$ (distinct outgassing activity for 
gases associated with H$_{2}$O versus CO$_{2}$). This dichotomy was also seen in the Rosetta study
of comet 67P/Churyumov-Gerasimenko \citep[e.g.,][]{2016Icar..277...78F}.

\item Comet C/2017 K2 (PanSTARRS) is a hyperactive comet with a water ice active fraction of $>$86\%. 

\item The particle differential size distribution and relative abundances of amorphous carbon, amorphous 
silicates, and crystalline silicates in comet C/2017 K2 (PanSTARRS) are commensurate with other
comets, with average (spatially averaged across the coma) amorphous carbon mass fraction values of 
$\simeq 0.247 \pm 0.010$,  silicate-to-carbon ratios $\simeq 3.049 \pm 0.163$, 
and crystalline mass fraction of the sub-micron grains, $f_{cryst} \simeq 0.384 \pm 0.065$.

\item Thermal models  of the JWST MRS SEDs shows that dust with a crystalline pyroxene composition 
cannot produce the observed $\simeq 14$~\micron{} residual nor the features in the residual near 9.3~\micron{} and 
10.0~\micron{}. Possibly these residuals may be due materials similar in composition to minerals seen in 
calcium-aluminum-rich inclusions (CAIs) as suggested by the Stardust sample return collections.

\item Analysis of residual spectral features in the 3 to 8.6~\micron{} JWST spectrum of comet 
C/2017 K2 (PanSTARRS) strongly support the contention that polycyclic aromatic hydrocarbons (PAHs) 
are present in the inner coma. Organic species are found to be present in the coma 
despite this comet not being super carbon-rich.

\end{itemize}

\noindent As this initial case study of C/2017 K2 (PanSTARRS) highlights, JWST can provide provide 
a wealth of observational constraints to models exploring the physio-chemical conditions in the disk 
realms from which a comet nuclei agglomerated and the nature of individual comets themselves.

\vspace{0.25cm}
\noindent \textbf{Acknowledgments:} The authors acknowledge the efforts and critiques provided by the anonymous referees which
guided improvements to the manuscript. The authors also would like to thank Christiian Boersma 
for their assistance with the AMES-PAHdbPythonSuite tools
used in the manuscript analysis, as well  as conversations with Nora H{\"a}nni regarding 
cometary organics, PAHs, and the Hydrogen Deficiency Index and efforts by
Geronimo Villanueva to adapt and improve aspects of the NASA PSG for use with JWST
spectra.  This work is based in part on observations made with the NASA/ESA/CSA James Webb 
Space Telescope Cycle 1 GO program 1556 whose analysis was supported via
StSci grant JWST-GO-01566.001. All of the data presented in this paper were obtained from the Mikulski 
Archive for Space Telescopes (MAST) at the Space Telescope Science Institute, doi: 10.17909/qj24-7k98.
The specific observations analyzed can be accessed via STScI is operated by the Association of Universities
for Research in Astronomy, Inc., under NASA contract NAS5–26555. Support to MAST for these data is provided by the 
NASA Office of Space Science via grant NAG5–7584 and by other grants and contracts. 

\facilities{MAST, JWST \citep{2023PASP..135f8001G} } 
\software{Astropy \citep{2018AJ....156..123A}, Numpy \citep{2020Natur.585..357H},
SciPy \citep{2020NatMe..17..261V}, Photoutiles \citep{2024zndo..13942182B},
SUBLIME \citep{2022ApJ...929...38C}, NASA Planetary Spectrum
Generator \citep{2018JQSRT.217...86V}, NASA AMES PAH database v3.20
and the AMES-PAHdbPythonSuite \citep{2014ApJS..211....8B, 2018ApJS..234...32B, 2020ApJS..251...22M}.
}

\clearpage
\appendix

\section{SED decomposition \textbf{`AO50'} vs. \textbf{`AP50'} } \label{sec:appendix-0}

The Akaike Information Criterion \citep{doi:10.1177/0049124104268644, 2007MNRAS.377L..74L}
which is a quantitative statistical framework, was used to quantitatively compare to thermal models, 
the \textbf{`AO50'} and the \textbf{`AP50'} in the spectral decomposition analysis (see Section~\ref{sec-model-assessment})
For completeness, and to enable comparison with (Figure~\ref{fig:fig-mrs-model00})  the \textbf{`AP50'} spectral 
model for the comet photocenter is provided here. We find AP50 overshoots wavelengths $\ltsimeq 7.0$~\micron{} and
creates a negative residual.

\begin{figure}[h!]
\figurenum{25}
\begin{center}
\includegraphics[trim=0.05cm 0.05cm 0.2cm 0.05cm, clip, width=0.50\textwidth]{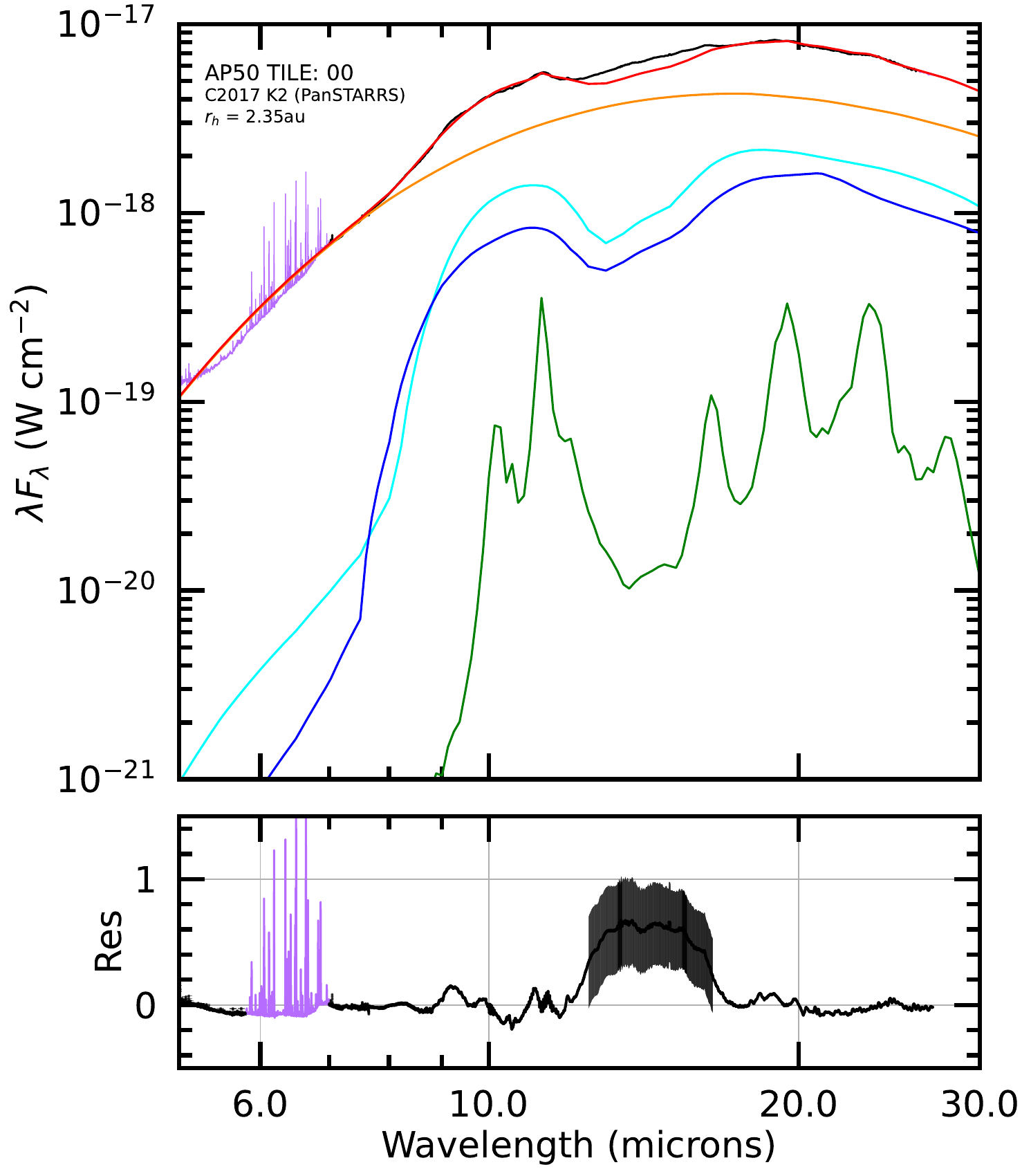}
\caption{(Top panel). The \textbf{`AP50'} thermal model fitted to the JWST MRS IR SED  in a 1\farcs0 diameter 
aperture centered on the photocenter (position 0:0) of comet C/2017 K2 (PanSTARRS). 
The models fitted to the observed 7.0 - 27~\micron{} SED constrain the dust mineralogy, assuming 
the coma refractory dust species have optical constants similar to that of amorphous carbon
(orange line), amorphous olivine (Mg:Fe = 50:50, cyan line), amorphous pyroxene (Mg:Fe = 50:50, blue line), 
Mg-crystalline olivine (Forsterite, green line), and when present, Mg-crystalline ortho-pyroxene 
(Enstatite, pink line). At position 0:0 enstatite was excluded from the best fit model.
The solid red line is the best-fit model composite spectra, superposed on the observed data (black solid curve) 
that was modeled ($\lambda \geq 7.0$~\micron) and the observed data that was omitted
from the thermal models (purple solid line) where strong water emission is dominant.
(Bottom Panel) The model residual (observed data - best fit thermal model) is in the bottom panel. 
The spectral region containing the $\nu_{2}$ water emission bands (purple, bottom panel) 
were not included in the model fit, and the grey vertical regions illustrated the point-point 
uncertainties (discussed in Section~\ref{sec:sec-dust-model-3cases}). Uncertainties were multiplied by 
40 in wavelength regions spanning 12.5 to 16.5~\micron{} (grey vertical lines). AP50 
overshoots wavelengths $\ltsimeq 7.0$~\micron. 
 }
\label{fig:fig-mrs-AP50-model00}
\end{center}
\end{figure}

\vspace{-0.50cm}
\section{The Residual 3 to 8.6~\micron{} continuum model} \label{sec:appendix-A}

Decomposition of the observed emission in the 3 to 8.6~\micron{} spectrum to search of residual emission from 
PAHs requires an accurate continuum model. Identification of residual emission feature requires 
the generation of a continuum corrected

\begin{equation}
F^{PAH}_{\rm{Residual}} \equiv F_{\rm{obs}}  - F_{\rm{thermal\, model}} - F_{\rm{scattered\, light}}
\end{equation}

\noindent that properly co-joins NIRSpec and MRS spectra in the region from 3 to 8~\micron. Deriving the continuum 
required masking water emission and other molecular emission lines and determining the contribution 
of scattered light, and generating a residual thermal model, one where the observed thermal emission SED is corrected 
for the contribution from a scattered light component (Figure~\ref{fig:fig-K2-reflectance-3panel}). 
To ascertain the dust continuum beneath the PAH emissions, particular attention was given to choosing the best-fitting 
thermal model, which requires extending a thermal model fit down to 4.9~\micron, fitting the thermal after subtracting
the scattered light and a first estimate of the PAH emissions (6 to 8.6~\micron{}) and then lessening the weights 
of specific regions of the SED: this thermal model is the so-called \textbf{‘AO50’} Case C 
model and is discussed in detail in Section~\ref{sec:sec-dust-model-3cases}ff. This process is iterative.

\section{The Scattered light 3 to 8~\micron} \label{scattered-light-3to8}

\begin{figure}[ht!]
\figurenum{26}
\begin{centering}
\includegraphics[trim=0.20cm 0.15cm 0.15cm 0.05cm, clip, width=0.40\textwidth]{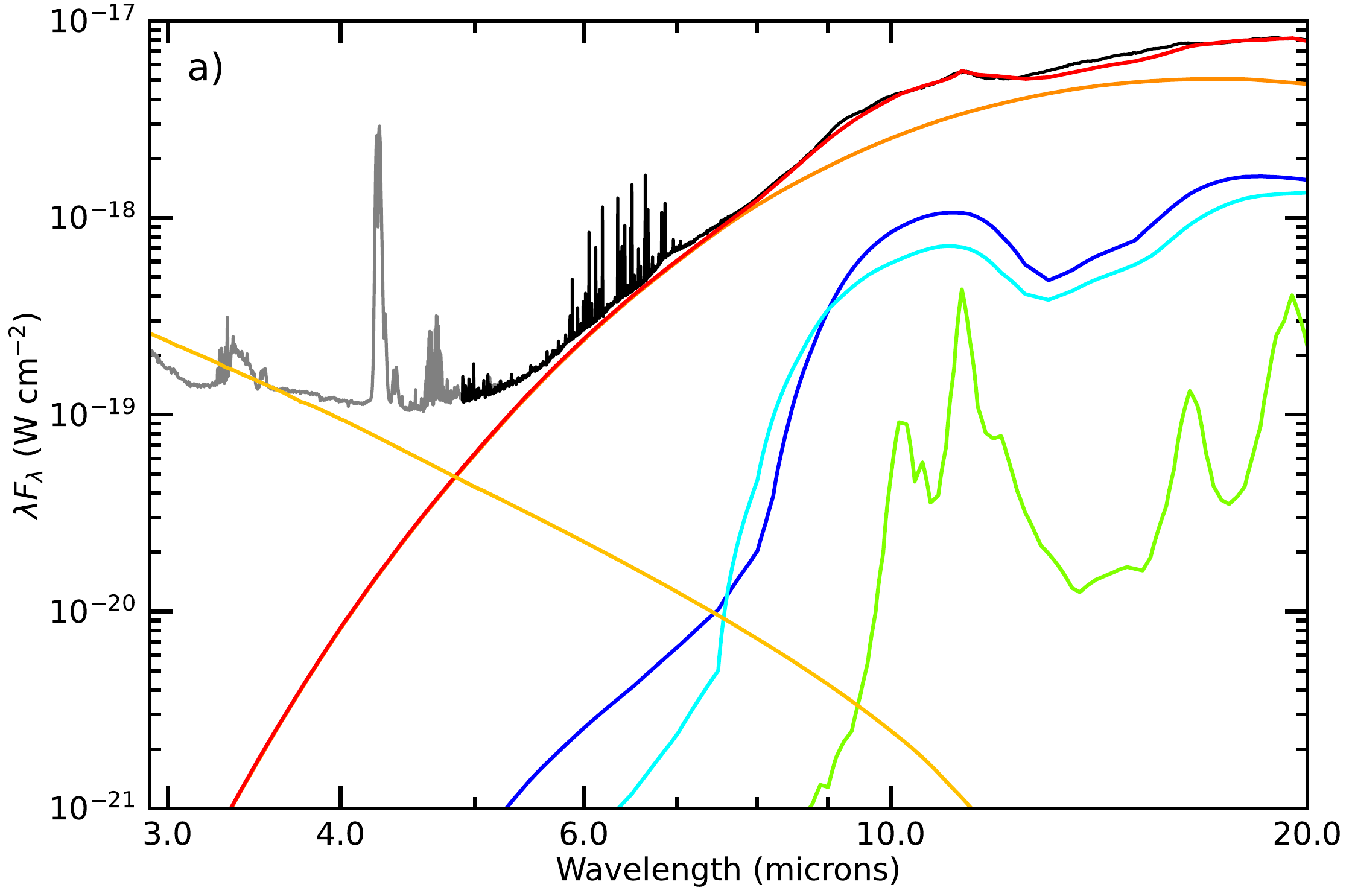}
\includegraphics[trim=0.20cm 0.15cm 0.15cm 0.05cm, clip, width=0.40\textwidth]{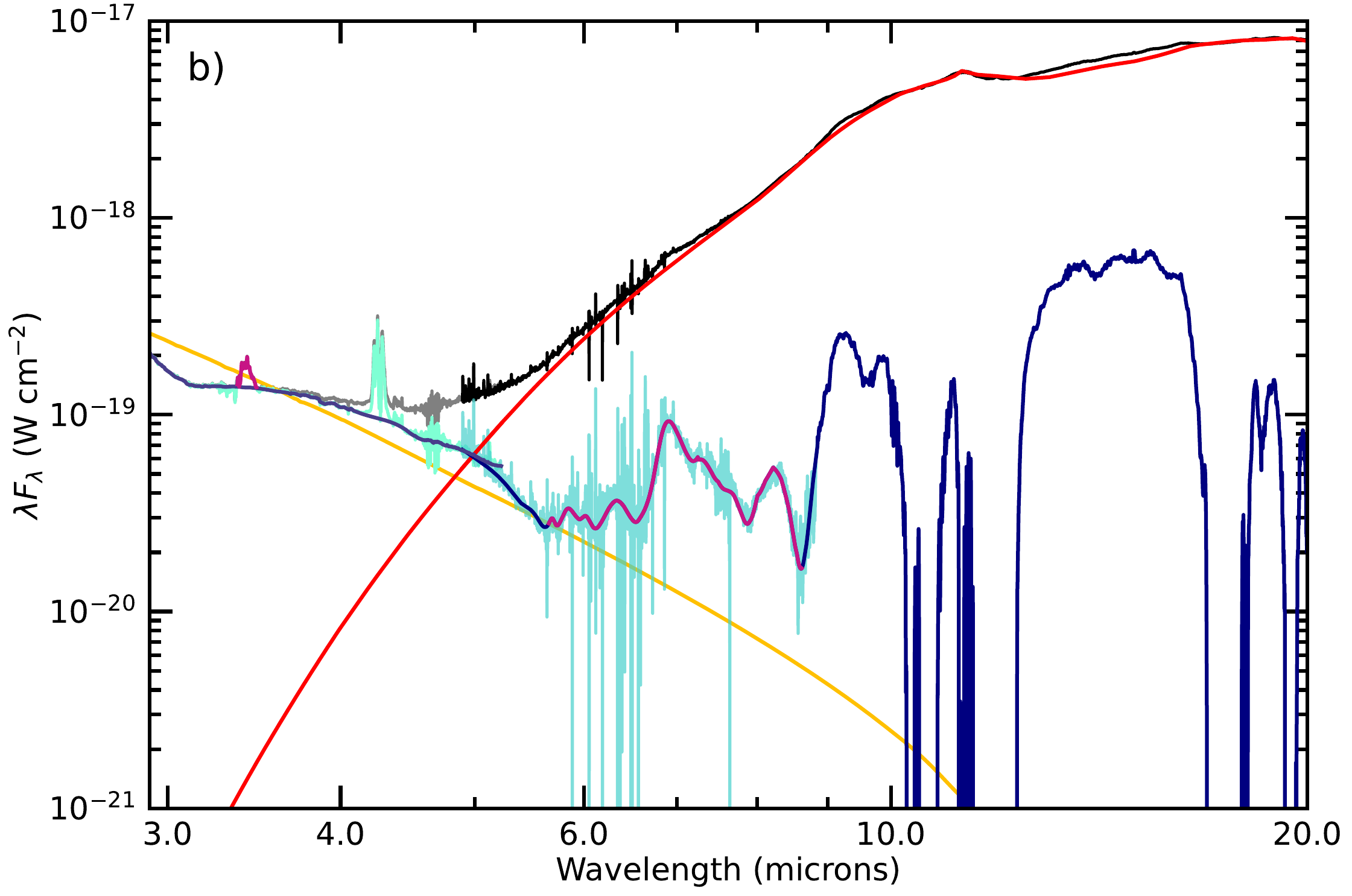}
\includegraphics[trim=0.20cm 0.15cm 0.15cm 0.05cm, clip, width=0.40\textwidth]{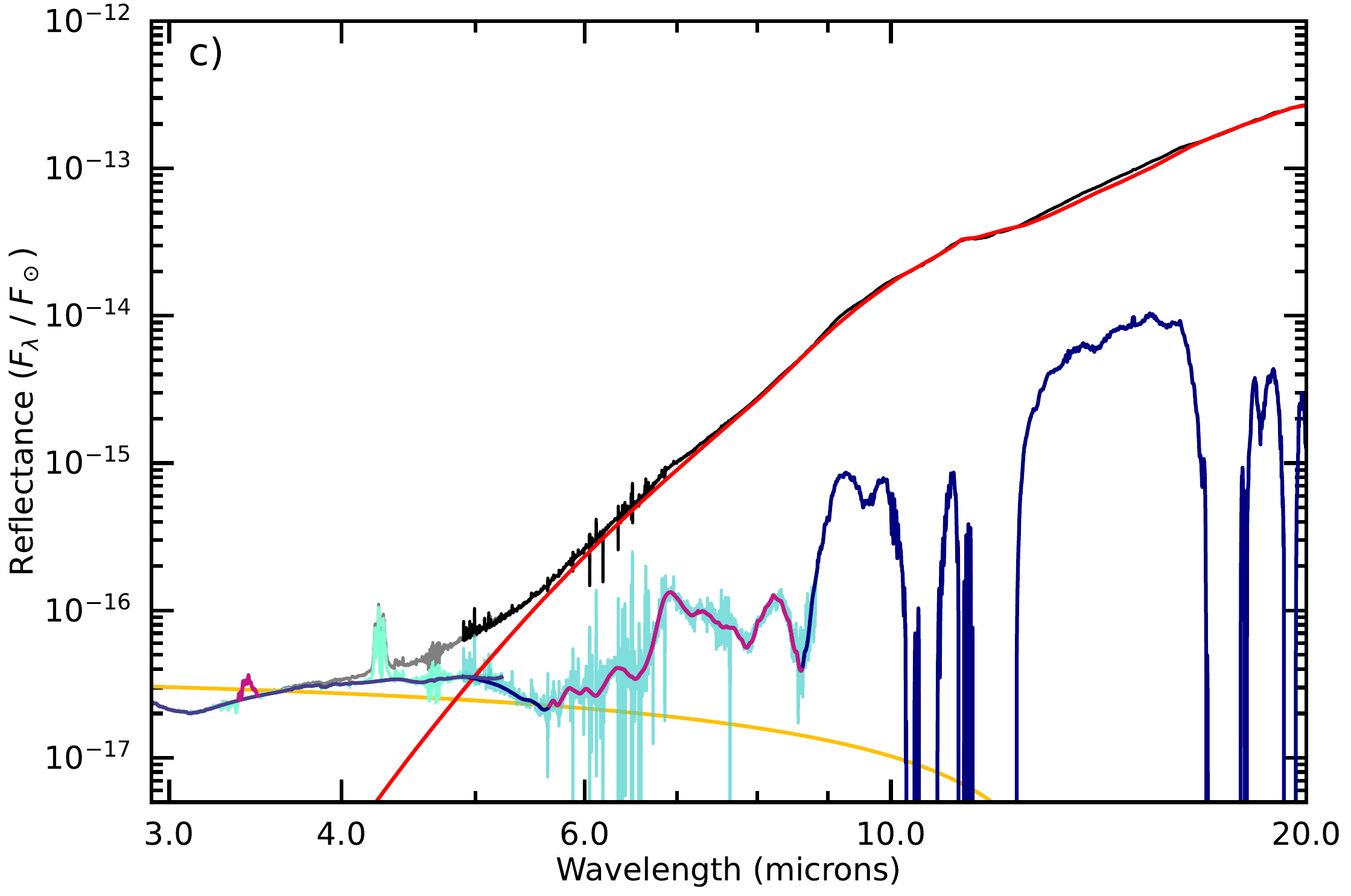}
\caption{The \textbf{`AO50'} Case C modeling of the scattered light, removal of water lines, and
NASA PSG contributions from volatiles and extraction of residual from the SEDS in comet C/2017 K2 (PanSTARRS)
from beam position 0:0 (a 1\farcs0 diameter beam centered on the photocenter of the comet).
(Top panel) The observed JWST NIRSPEC (gray line) + MRS (black line) and the \textbf{`AO50'} Case C 
model with the components of the model shown and extrapolated to NIRSpec wavelengths, and 
the scattered light (slope) assessed for NIRSpec and extrapolated to MRS. (Middle panel) 
The observed JWST NIRSPEC (gray line) + MRS (black line) and the \textbf{`AO50'} Case C `F(total)'
and the scattered light (slope) extrapolated to MRS, with the also the MRS residual show in medium turquoise 
line and the residual median filtered result plotted in navy-colored line on top of medium turquoise. (Bottom panel) Same
as the middle panel, but in reflectance. The MRS medium turquoise ``hash'' arises from imperfect 
modeling of the H$_{2}$O lines and the NIRSpec ``hash''
is not poor signal-to-noise but rather arrises from imperfect molecular modeling of the lines
(likely due to likely due to optical depth effects in regions of CO$_{2}$ and HCN emission).
}
\label{fig:fig-K2-reflectance-3panel}
\end{centering}
\end{figure}

To determine the contribution of scattered light to the observed SED, a reflectance spectrum 
(F$_{\lambda}$ / F$_{\odot}$) was derived by dividing the observed NIRSpec and MRS spectra described above 
by a spectrum of the Sun solar reference E490 \citep{e490}.
The  C/2017 K2 (PanSTARRS) coma is blue (decreasing reflectance with increasing wavelength), with a 
mean (linear) spectral slope of  $-1.00 \pm 0.03$\% per 100~nm between 3.6 and 5.45~\micron{} (normalized at 3.6~\micron).  
The wavelength dependence of the scattered light of the coma deviates from that of a straight 
line (i.e., it is not linear over this wavelength range, Figure~\ref{fig:fig-K2-reflectance-3panel}). The scattered 
light behavior of comet comae in the 3--6 \micron{} region is unknown. Modeling the scattered light with 
aggregate particles is beyond the scope of this work. 

Reflectance spectra of other comets often are red colored, for example 238P/Read \citep{2023Natur.619..720K}  
spectral slope of $2.18 \pm 0.02$\% per 100~nm between 1.0 and 2.55~\micron{} (normalized at 2.0~\micron) 
or 103P/Hartley~2 which has a spectral slopes of $1.41 \pm 0.02$\% inside a jet emerging from a small lobe of  
nucleus facing the Sun to $5.10 \pm 0.08$\% per 100~nm for the outer coma \citep{2014Icar..238..191P}. In general, 
red slopes are well reproduced by refractory components such as amorphous carbon \citep{2006AJ....132.1346C}. 

Comet 17P/Holmes experienced a substantial outburst, releasing dust and ice particles into its coma 
with a scattered light negative (blue) slope that became increasingly negative over a few days,
spanning $-2.7 \pm 0.1$\% per 100~nm to $-3.5 \pm 0.1$\% per 100~nm \citep{2009AJ....137.4538Y}.
The temporal behavior of the scattered light slope could either arise from an increase in the relative abundance of icy grains 
or the decrease of the average grain size. However, in comet 17P/Holmes over this period the gas outflow 
was observed to weaken with time, thus \citep{2009AJ....137.4538Y} favor a hypothesis that
the change in slope was due to a decrease in the particle sizes in the size distribution. 

Models presented to explain the nominal coma of comet 67P/Churyumov-Gerasimeko show that 
the scattered light slope at 2~\micron{} cannot reproduce the blue slope for 1~\micron{} grains of 
amorphous carbon and Mg:Fe = 50:50 olivine even for a steep differential size 
distribution \citep{2017MNRAS.469S.842B, 2017MNRAS.469S.443B}. 
Similarly, models for the outbursting coma of 67P/Churyumov-Gerasimeko that incorporate 25\% 
fractal aggregates can produce a slope of $-1.0$\% per 100~nm for particle sizes 
$\ltsimeq 0.21$~\micron{} but the aggregate particles 
at 1.3~au are hot (590~K). \citet{2017MNRAS.469S.842B, 2017MNRAS.469S.443B} carbon-olivine 
model's nominal volume fraction of silicates for 67P/Churyumov-Gerasimeko grains is 0.34 and a porosity of 0.5.
This assumption produces a steep size distribution with slopes of between 4.0 and 4.5 and minimum particle sizes
$\ltsimeq 0.2$~\micron{}  for a power law differential size distribution yielding a scattering color of $-2$\% per 100~nm.

In the coma center, the size distribution has significantly fewer numbers of particles as small as 
0.5~\micron{} and is a mixture of Mg:Fe amorphous silicates, Mg-crystalline olivine, and amorphous 
carbon (Section~\ref{sec:sec-ao50tile-reveal}). We assert that the blue slope color in 
comet C/2017 K2 (PanSTARRS) inner coma is strongly affected by the presence of the ice particles. 
The presence of of ice in the inner coma also is traced by water distribution in the coma 
(Section~\ref{sec:results_waterband_analysis}). These assertions are also in agreement with the 
high water production (hyperactivity, see Section~\ref{sec:activity}) of the comet, which can be explained 
by water ice in the coma \citep[e.g.,][]{2021PSJ.....2...92S}.  However, detailed modeling 
the scattered light in comet C/2017 K2 (PanSTARRS) is beyond the scope of this work. 

\section{The Residual Spectra Determination from 3 to 8~\micron} \label{sec:residual_3to8}

Emission lines from water are present in the short-wavelength range of MRS spectra. (Section~\ref{sec:sec-watermodelling}). 
First, the water line model was subtracted from the MRS 5.5 to 7.25~\micron{} data. 
Imperfectly modeled water lines resulted in some under and over subtracted lines. These were identified by a 
line-to-continuum threshold of $\gtsimeq 5$, subsequently masked, and then median filtered
to produce a continuum spectrum. To correct the MRS spectra for scattered light, the spectral
reflectance was converted to flux density and subtracted from the MRS flux density.  Here, we adopt a 
functional form for the scattered light between 3.6 and 5.45~\micron{}  (where the NIRSpec 
and MRS continua overlap) as a line of constant spectral slope of $-1.00 \pm 0.03$\% per 100~nm. A thermal 
model (Section~\ref{sec:sec-dust}) was fitted to the observed flux minus the scattered light spectral 
energy distribution (SED) down to 4.9~\micron. This modeling is necessary to accurately estimate the 
thermal emission in the 4.9 to 8.5~\micron{} region in the SED which is dominated by the amorphous carbon 
component. The resultant thermal model provides the most robust assessment of the ``baseline'' under 
the PAH emissions, the latter is necessary to assess the relative strengths of identified residual spectral features. 

However, as discussed in Section~\ref{sec:sec-dust} the thermal model did not well fit the shoulder of 
the 10~\micron{} silicate feature because of emission at 9.3~\micron{} and at 8.55~\micron. Emission 
at $\simeq 8.55$~\micron{} is too short to be attributed to silicates and therefore are presumed to 
be associated with the PAH emissions. The shape of 9.3~\micron{} emission is not overall well fitted 
by orthopyroxene, despite the coincidence in the expected peak wavelength of 9.3~\micron. At longer 
wavelengths ($\lambda \simeq 18$~\micron) the modeled resonances of crystalline pyroxene and 
their relative strengths do not match the observed data (Figure~\ref{fig:fig-spitzer-comps}). The measured 
uncertainties in the region of the silicate feature shoulder were multiplied by 40 to lessen their 
weight in the thermal model fit. The application of the Akaike Information 
Criterion \citep{doi:10.1177/0049124104268644, 2023PSJ.....4..242H} clearly demonstrates that the thermal 
model better fits these data weighted in this manner (Section~\ref{sec:sec-ao50tile-reveal}).

The NIRSpec continuum was derived after subtraction of the NASA PSG molecular emission 
model (Section~\ref{sec-volatiles}). Slight over-subtractions of the molecular emissions shortward of 
3.4~\micron{} require the baseline under the 3.4~\micron{} feature to be established by fitting a 
spline and subsequently a positive slope 2$^{nd}$-order polynomial that span between the ice band 
($\simeq 3.01$~\micron) and 3.6~\micron, Figure~\ref{fig:fig-PSG34mu}. The thermal model 
and the scattered light are also subtracted from the NIRSpec data, because the thermal emission 
contributes to the observed SED from $\simeq 4.0$ to  5.0~\micron, Figure~\ref{fig:fig-K2-reflectance-3panel}.

Splicing the these latter continuum corrected spectra together in the region of overlap produces
$F^{\rm{PAH extract}}_{\rm{Residual}}.$  It has excess emissions (a hump) between 4 and 5~\micron{}
that could be aspects of the scattered light that are not well modeled by the adopting a constant slope 
trend for the scattered light or there could be emissions from very small grains \citep[VSGs,][]{2024A&A...685A..74P}
not accounted for in our decomposition methodology. 

\clearpage

\bibliographystyle{aasjournal}
\bibliography{c17k2-jw6}{}

\end{document}